\documentclass[12pt]{article}
\usepackage{amsmath}
\usepackage{graphicx}
\usepackage{enumerate}
\usepackage{natbib}
\usepackage{url} 

\usepackage{chngcntr}

\usepackage[T1]{fontenc}
\usepackage[latin9]{inputenc}
\usepackage{multirow}
\usepackage{tikz}
\usepackage{amsmath}
\usepackage{amsthm}
\usepackage{amssymb}
\usepackage{amstext}
\usepackage{bm}
\usepackage{booktabs} 
\usepackage{color}
\usepackage{enumerate} 
\usepackage[shortlabels]{enumitem}
\usepackage{epsfig,epsf,psfrag}
\usepackage{epstopdf}
\usepackage{float}
\usepackage{graphics}
\usepackage{graphicx}
\usepackage{latexsym}
\usepackage{lscape}
\usepackage{mathabx} 
\usepackage{mathtools}
\usepackage{mathrsfs}
\usepackage{multirow}
\usepackage{natbib}
\usepackage{rotate}
\usepackage{setspace}

\usepackage{sectsty}

\usepackage{subcaption}
\usepackage{xargs}[2008/03/08]
\usepackage{xcolor}

\usepackage[text={7in,9.5in},centering]{geometry}
\usepackage{ifthen}
\usepackage[hidelinks]{hyperref}
\hypersetup{
	colorlinks,
	linkcolor={red!50!black},
	citecolor={blue!50!black},
	urlcolor={blue!80!black}
}

\newtheorem{thm}{Theorem}
\newtheorem{lem}{Lemma}
\newtheorem{cor}{Corollary}
{
	\theoremstyle{remark}

	\newtheorem{assumption}{Assumption}
}

\newtheorem*{lem*}{Lemma}

\newcommand{\bea}{\begin{eqnarray*}}
\newcommand{\eea}{\end{eqnarray*}}
\newcommand{\be}{\begin{eqnarray}}
\newcommand{\ee}{\end{eqnarray}}
\newcommand{\beq}{\begin{equation}}
\newcommand{\eeq}{\end{equation}}

\newcommand{\bal}{\begin{equation}\aligned}
\newcommand{\eal}{\endaligned\end{equation}}
\newcommand{\bgt}{\begin{equation}\begin{gathered}}
\newcommand{\egt}{\end{gathered}\end{equation}}
\newcommand{\ed}{\end{document}}

\newcommand{\btab}{\begin{tabular}}
\newcommand{\etab}{\end{tabular}}

\newcommand{\bi}{\begin{itemize}}
\newcommand{\ei}{\end{itemize}}
\newcommand{\bfi}{\begin{figure}}
\newcommand{\efi}{\end{figure}}
\newcommand{\ben}{\begin{enumerate}}
\newcommand{\een}{\end{enumerate}}
\newcommand{\bay}{\begin{array}}
\newcommand{\eay}{\end{array}}

\def\vs{\vspace{.1cm}}

\def\bco{\iffalse}

\newcommand{\no}{\noindent}
\newcommand{\bc}{\begin{center}}
\newcommand{\ec}{\end{center}}

\definecolor{DarkBlue}{rgb}{0,.08,.45}
\definecolor{DarkRed}{rgb}{.7,0,.4}
\definecolor{DarkGreen}{rgb}{0,0.4,0}

\newcommand{\bpmt}{\begin{pmatrix}}
	\newcommand{\epmt}{\end{pmatrix}}

\newcommand{\bsp}{\begin{split}}
	\newcommand{\esp}{\end{split}}

\newcommand{\bdes}{\begin{description}}
	\newcommand{\edes}{\end{description}}

\newcommand{\bass}{\begin{assumption}}
	\newcommand{\eass}{\end{assumption}}
\newcommand{\bthm}{\begin{thm}}
	\newcommand{\ethm}{\end{thm}}
\newcommand{\blem}{\begin{lem}}
	\newcommand{\elem}{\end{lem}}
\newcommand{\bcor}{\begin{cor}}
	\newcommand{\ecor}{\end{cor}}
\newcommand{\bpf}{\begin{proof}}
	\newcommand{\epf}{\end{proof}}

\def\vs{\vspace{.125cm}}

\def\vs{\vspace{.125cm}}

\def\i1{_{i1}}

\def\var{{\rm var}}

\def\cp{\citep}
\def\eps{\varepsilon}

\def \mbb {\mathbb}

\DeclareMathOperator*{\argmin}{argmin}

\def\expect{\mbb{E}}

\def\real{\mbb{R}}

\def\O{O}

\def\Op{O_P}

\def\p#1{\left(#1\right)}
\def\b#1{\left[#1\right]}
\def\br#1{\left\{#1\right\}}

\newcommand \mopi[1]{m_{i\oplus}(#1)}
\newcommand \hmopi[1]{\hat{m}_{i\oplus}(#1)}

\def \nuom{\gamma_{\nu_0,\nu_1}^{(\omega_1)}}
\def \pert {\mathcal{P}}

\def \Hom{H_{\omega_1}}
\def \Gom{G_{\omega_1}}
\newcommand{\mustarom}{\mu^\ast_{\omega_1}}
\def \mutilom{\tilde{\mu}_{\omega_1}}
\def \supom{\underset{\omega_1 \in \p{\Omega_1,\mathcal{F}_1,P_1}}{\sup\ }}

\def \gamz{\zeta_{0}(z)}
\def \hgamz{\hat{\zeta}_{0}(z)}

\def \ggamz{\zeta_{1}(z)}
\def \hhgamz{\hat{\zeta}_{1}(z)}

\newcommand \geodi[1]{\gamma_{\nu_{i0},\nu_{i1}}^{(i)}(#1)}
\newcommand \geodpti[1]{\nu_{i#1}}

\def \M{\mathcal{M}}
\def \T{\mathcal{T}}

\def \S{\mathcal{S}}

\newcommand{\argminsp}{\underset{\mu \in \mathcal{M}}{\argmin\ }}

\newcommand{\ltwoNorm}[2][]{%
	\ifthenelse{ \equal{#1}{} }
	{\ensuremath{\|#2\|}}
	{\ensuremath{\|#2\|_{#1}}}
} 
\usepackage{authblk}
\def\bco{\iffalse}

\newcommand{\single}{\renewcommand{\baselinestretch}{1.2}\normalsize}
\newcommand{\double}{\renewcommand{\baselinestretch}{1.6}\normalsize}

\usepackage{xr}
\begin{document}
\thispagestyle{empty} \single \bc {\bf \sc \Large Geodesic Mixed Effects Models for Repeatedly Observed/Longitudinal Random Objects}
\vspace{0.15in}\\
Satarupa Bhattacharjee$^{1}$  and  Hans-Georg M\"uller$^2$ \\
$^1$ \small{Department of Statistics, Pennsylvania State University}\\
$^2$\small{Department of Statistics, University of California, Davis}
\ec 


%

%
%
\begin{abstract}
	Mixed effect modeling for longitudinal data is challenging when the observed data are random objects, which are complex data taking values in a general metric space without linear structure. In such settings the classical additive error model and distributional assumptions are unattainable. Due to the rapid advancement of technology,  longitudinal data containing complex random objects, such as covariance matrices, data on Riemannian manifolds, and probability distributions are becoming more common.  Addressing this challenge, we develop a mixed-effects regression for data in geodesic spaces, where the underlying mean response trajectories are geodesics in the metric space and the deviations of the observations from the model are quantified by perturbation-maps or transports. A key finding is that the geodesic trajectories assumption for the case of random objects is a natural extension of the linearity assumption in the standard Euclidean scenario. Further, geodesics can be recovered from noisy observations by exploiting a connection between the geodesic path and the path obtained by global Fr\'echet regression for random objects.  The effect of baseline Euclidean covariates on the geodesic paths is modeled by another Fr\'echet regression step. We study the asymptotic convergence of the proposed estimates and provide illustrations through simulations and real-data applications.	
\end{abstract}
\no {KEY WORDS:\quad Random Effects; Random objects; Geodesics; Perturbation; Optimal transport; Fr\'echet regression; M-estimation}.
\thispagestyle{empty} \vfill
\noindent \vspace{-.2cm}\rule{\textwidth}{0.5pt}\\
{\small Research supported in part by grants NSF DMS-2310450.}

\newpage
\pagenumbering{arabic} \setcounter{page}{1} 

\double

\section{Introduction}
\label{sec:intro}
In the era of modern data science,  complex data structures are increasingly encountered.  
An  important  but largely unexplored setting is where a response variable takes values in a non-Euclidean metric space without vector space operations or inner product. 
Examples of such random objects \citep{mull:16} include distributional data in Wasserstein space~\citep{pete:16, mata:21}, 
symmetric positive definite matrix objects~\citep{dryd:09}, spherical data~\citep{di:14}, phylogenetic trees~\citep{bill:01} and data on  finite-dimensional Riemannian manifolds~\citep{bhat:03,bhat:05, afsa:11,eltz:19}, 
among other data types.  Data modeling and analysis for metric space valued data is challenging  due to the absence of any linear structure. For example, the  definition of a sample or population mean as an average or expected value is not applicable and is replaced by barycenters or Fr\'echet means~\citep{frec:48}. Similarly, regression approaches to quantify the dependence  between a random object response and Euclidean predictors require a notion of a  conditional Fr\'echet mean~\citep{pete:19} with several approaches for corresponding regression models \citep{hein:09,dong:22,scho:22,zhan:21:1,tuck:21}.

Technological advances have made it possible to record and efficiently store repeated measurements of images~\citep{peyr:09,gonz:18}, shapes~\citep{smal:12}, networks~\citep{tsoc:04} and other  random objects. There are only few methods available to analyze time courses of random objects and only for the case where time courses  are continuously recorded  and  fully observed  over time \cp{mull:20:10}.  But when such data are recorded in longitudinal studies with repeated observations of random objects, these  are often sparsely recorded over time, posing a substantial additional challenge for statistical analysis.  To our knowledge, there is currently no statistical method available  to handle longitudinal random objects. This paper presents the first approach for the statistical analysis of such data.  
For sparsely sampled trajectories as we consider here it is of interest to gain information about the  actual  individual time courses, i.e., the underlying metric-space valued curves that produce the observed measurements but are latent, due to the sparse measurement scheme.  

Flexible nonparametric recovery methods have been extensively studied for the case of scalar responses based on versions of functional principal component analysis   \cite[see, e.g.,][]{stan:98,rice:01,mull:05:4,sent:11,yao:15,    mull:21:2,li:22}. However, all of these approaches require that the data are in a linear space and thus cannot be extended to the case of object data, where one cannot make use of vector space operations. A second and more restrictive approach are 
classical Euclidean linear mixed effects models~\citep{lair:82,digg:02,verb:10}, where  
the individuals in the population are assumed to follow the same general linear model but with random intercepts and slopes that are
subject-specific, with various extensions  \cite[see, e.g.,][]{wu:09,schi:15a,alla:17,yue:20,pell:21}. 
Our goal in this paper is to address the challenges to extend random effects models to the case of object data.

Given a covariate vector   $Z_i\in \real^p$, $p\geq 1$,  for the $i^{\text{th}}$ subject, $1 \le i \le n$,  repeated  measurements   $Y_i = (Y_{i1}, Y_{i2},\dots, Y_{in_i})$ and  measurement times  $T_i = (T_{i1}, T_{i2},\dots, T_{in_i})$,  the mixed effects linear regression  for repeated measurements/longitudinal data is 
\begin{align}
	\label{eucl:re_summary}
	\expect\p{Y_i(t)| \nu_{i}, T_{ij} = t} = \nu_i t, \quad \expect\p{\nu_i|Z_i = z} = \beta^\intercal z, \end{align}
where the $\nu_i$ are subject-specific random slopes that determine  trajectories $\nu_i t$ and depend linearly on the  baseline covariate vector $Z$. Here  $\beta \in \real^p$ is a fixed parameter vector. A typical additional assumption is $Y_i(t)=\nu_i t + \varepsilon(t)$  for zero mean finite variance additive errors and also joint Gaussianity of all random components. 
As we  aim to generalize model~\eqref{eucl:re_summary} to the case of sparse random object observations $Y_{ij}$,  where an additive structure for the model is not available, the trajectories $\nu_i t$ are written without intercepts; in the real case, this form
can be obtained by centering predictors and responses for each subject. 

\bco

In classical mixed effects models  the functions $ \gamma_i$ in~\eqref{eucl:re_summary} and $\zeta$ are    linear. 
When $\gamma_i(t) = \nu_i t$ and $\zeta(z) = \beta^\intercal z$, the above model boils down to 
\begin{align}
	\label{eucl:re_summary}
	\expect\p{Y_i(t)| \gamma_{i}, T_{ij} = t} = \nu_i t,\quad \expect\p{\nu_i|Z = z} =  \beta^\intercal z,
\end{align}
where $\nu_i$ is the random effects for the $i^{\text{th}}$ individual and $\zeta(z) = \beta^\intercal z$ illustrates the effect curve of the external/baseline covariates $z \in \real^p$, $p\geq 1.$
The above model posits natural heterogeneity among individuals in terms of their initial level of response and changes in the response over time. The effects of baseline/external covariates (e.g., due to treatments, exposures, or background characteristics of the individuals) is incorporated into the model as fixed effects in the model, which provides a flexible way to explain the variability among repeated measures.

\bco

Extensions of the classical mixed effects models for Euclidean data have been well-studied to include more flexible non-parametric approaches via a spline/ basis representation of the data~\cite{rice:01, stan:98, lin:01, wang:05}, by local polynomial kernel regression~\citep{wu:02}, considering a linear additive model~\citep{buja:89},through a principal component (Karhunen-Lo\'eve) basis expansion~\citep{guo:02, jian:10}, among other methods.
In recent literature, various specialized tools and methods have been employed to explore different aspects of longitudinal data such as generative statistical models~\citep{schi:15a}, hierarchical models in univariate Riemannian manifold setting~\citep{alla:17}, tensor mixed effects model to analyze high-dimensional Raman mapping data~\citep{yue:20}, random forest methods on hierarchical data in exponential families~\citep{pell:21}, incorporating measurement error and missing observations~\citep{wu:09}, and so on.

To summarize, all the classical methods inherently use the linear structure of the space where the responses lie and use a suitable version of the model~\eqref{eucl:re_summary} to target the mean response trajectories. Thus, these methods rely in a fundamental way on the vector space structure of the space. When there is no global or local linear structure, a new methodology is needed and this work contributes to such development.

\fi

A key observation that makes it possible to generalize model \eqref{eucl:re_summary} to the case of object data is that the linearity assumption from a more general perspective corresponds to the assumption that responses are scattered around a geodesic, which in the case of real-valued data is a line.  
Accordingly we consider in the following geodesic metric spaces; we will model subject-specific random trajectories as geodesics in such spaces.  Noisy observations of random objects are sparse in time and  located around the geodesic, where noise is modeled through perturbation maps that are applied to the true random objects, as  in metric spaces there is no framework for additive noise.  To obtain asymptotic results, we consider the case of small errors and develop an approach that makes it possible to recover the subject-specific geodesic trajectories, using 
global Fr\'echet regression for random object responses~\citep{pete:19} as an auxiliary tool.

In Section~\ref{sec:model} we provide a brief review of metric geometry and geodesics and provide further motivation for the proposed model.  In Section~\ref{sec:theory}, we discuss the connection between the underlying subject-specific geodesic path and the path estimated by the global Fr\'echet regression method and proceed to establish theoretical guarantees for the asymptotic convergence of model components,  including rates,  based on M-estimation theory.  Our motivating application examples deal with samples of probability distributions, data lying on the unit sphere in $\real^3$ and correlation matrices, which are illustrated with simulations in Section~\ref{sec:simu}.  Real data applications for resting state fMRI longitudinal data from ADNI and demographic data are discussed in Section~\ref{sec:data}. 

\section{Preliminaries and Model}
\label{sec:model}
\subsection{Preliminaries on Metric Spaces}
\label{sec:geod}

In the following,  $(\M,d)$ denotes  a metric space that is complete, separable and totally bounded and we refer to the elements $Y \in \M$ as random objects. We consider sets  $\T = [0,1]$ and $\S \subset \real^p$ for $p\geq  1$ and  a random tuple $(Y, T, Z)$ with a joint distribution on the product space $\M\times\T\times \S,$ where in a regression setting 
$Y \in \M$ is a random object response,  $T \in \T$ is a random time point where the random object $Y$   is observed  and $Z$  a (baseline) covariate with $Z \in \S$. We focus on a longitudinal setting, where one observes  $n$  subjects and  $n_i>1$ observations are made  at random times $T_{ij} \in \T$ for the $i^{\text{th}}$ subject with  
corresponding observations $Y_{ij} = Y_i(T_{ij})\in \M$. 

A geodesic in a geodesic metric space connecting two distinct points is the shortest path connecting the two points. Geodesics in a metric space are analogous to straight lines in a Euclidean space. In a uniquely geodesic metric space $\M$ with metric $d$, a constant speed geodesic $\gamma_{\nu_0,\nu_1}(t) \in \M, \  t \in [0,1]$, connecting two points $\nu_0$ and $\nu_1$ is characterized by $\gamma_{\nu_0,\nu_1}(0) = \nu_0$, $\gamma_{\nu_0,\nu_1}(1) = \nu_1$ and $d(\gamma_{\nu_0,\nu_1}(t_1),\gamma_{\nu_0,\nu_1}(t_2)) = |t_1 -t_2|d(\nu_0,\nu_1)$. If for any two points in a metric space there exists a geodesic that connects them,  the space is a geodesic space  and it is uniquely geodesic if for every pair of points $x,y \in \M,$ there is a unique geodesic $\gamma_{\nu_0,\nu_1}:[0,1] \mapsto \M$ from $x$ to $y$. For further details and background we refer to \cite{bura:01} and
the review in Section 2 of  \cite{lin:21}. 
Given a geodesic $\gamma_{\nu_0,\nu_1}(t)$ defined on $t\in [0,1],$ if the geodesic property as defined above continues to hold for $\gamma_{\nu_0,\nu_1}(t)$ with $t\in[t_1,t_2]$ where $t_1<0<1<t_2$, we say that the geodesic can be extended from $[0,1]$ to $[t_1,t_2]$~\citep{ahid:20}. 
We assume throughout  that $(\M,d)$ is a uniquely extendable geodesic space, i.e., it is a uniquely geodesic space, where all  geodesics can be extended. 
It is obvious that the Euclidean space, where the geodesic path connecting two points $a, b\in \real$ is simply the line connecting the two points,  is a uniquely extendable geodesic space.  Other examples  of  uniquely extendable geodesic spaces are as follows.  

\noindent \textit{Example 1: Space of distributions with the Wasserstein metric.}
For a closed interval $Q\subset\real$, the Wasserstein space $\mathcal{W}_2(Q)$ of probability distributions on  $Q$  
with finite second moments is endowed with the $L_2$-Wasserstein distance
\begin{align*}
	d_W(\mu,\nu) = \p{\int_0^1 [F_{\mu}^{-1}(s) - F^{-1}_{\nu}(s)]^2 ds}^{1/2}, \text{  for } \mu,\nu \in \mathcal{W}_2(Q),
\end{align*}
where $F^{-1}_{\mu}$ and $F^{-1}_\nu$ denote the quantile functions of $\mu$ and $\nu$, respectively. We further require the distributions to be continuous, i.e., to possess densities. Then  $(\mathcal{W}_2(Q), d_W)$ is a uniquely geodesic space~\citep{ambr:08}. Given any $\mu, \nu \in \mathcal{W}_2(Q)$ where $\mu \ne \nu$, 
there is a unique  geodesic that connects  $\mu$ and $\nu$, 
given by $\gamma_{\mu,\nu}(t) = [t(F_\nu^{-1}\circ F_\mu - \text{id}) + \text{id}]\#\mu,\ t \in[0,1]$. For  a measurable function $h: \, Q \to Q$, $h\#\mu$ is a pushforward  measure such that $h\#\mu(A) = \mu(\{r \in Q : h(r) \in A\})$ for any set $A \in \mathcal{B}(Q)$, the Borel $\sigma$-algebra on $Q$. For the  extendibility of geodesics in the space of continuous probability measures we refer to 
\cite{ahid:20,mull:21:3}.\vs\vs

\noindent\textit{Example 2: Space of positive definite matrices.}  
The space  of positive definite symmetric $K \times K$ matrices  $\mathcal{S}_K$, equipped with the Frobenius inner product  $\langle A, B \rangle_F = \text{tr}(A^\intercal B)$ and the induced Frobenius metric $d_F(A, B) = \ltwoNorm{A-B}_F$, $A, B \in \mathcal{S}_K$, where $\ltwoNorm{A}_F$ is the usual Euclidean matrix norm, possesses unique geodesics, which are straight lines in the Euclidean vector space given by $\gamma_{A, B} :[0,1]\to \mathcal{S}_K$ with $\gamma_{A, B}(t) = tA +(1-t)B.$ 
Other metrics $d$  for which $\mathcal{S}_K$ is a uniquely geodesic space include the log-Euclidean metric~\citep{arsi:07}, the power  metric family~\citep{dryd:10}, the Log-Cholesky metric~\citep{lin:19} and the Bures-Wasserstein metric \citep{taka:11}; these geodesics are extendible as long as $A, B$ are strictly positive definite. 
A popular metric on $\mathcal{S}_K$ 
that has been successfully used  in various practical applications  for covariances is the square root power metric \citep{pigo:14,tava:19}, where 
 $d_{1/2}(A, B) = \ltwoNorm{A^{1/2}-B^{1/2}}_F$, $A, B \in \mathcal{S}_K$; we will use this metric in Section 5
 to illustrate the proposed random effects model for neuroimaging data. The geodesics in this metric are 
 $\gamma_{A,B}(t)=(tA^{1/2} +(1-t)B^{1/2})^2$.\vs\vs
 

\noindent\textit{Example 3: Spheres with geodesic metric}.
A $(p-1)$-dimensional sphere $S^{p-1} = \{ x \in \real^p: \,\, \|x\|=1\}$ embedded in $\real^p$ is a complete Riemannian manifold.  The geodesic metric $d_g$ between two points $x,y$ on the surface of the unit sphere $S^n$ is given by $d_g(x,y) = \arccos \langle x, y \rangle.$
Consider $M = S^2$ the 2-sphere with the spherical geodesic metric. Then the
great circles are geodesics. The great circle passing through two points $x,y \in S^2$ can be parametrized as $\gamma_{u,v}(t) = (\cos t) u + (\sin t)v.$ However, this space is not uniquely geodesic as two polar points can be  connected by arbitrarily many different geodesics. 
In order to make the space a uniquely geodesic space one can slice off the subset of the sphere with $x_1 \le -1 +\gamma$ for any 
small $0 <\gamma \le 1/2, $ which includes the half sphere, where $x_1$ is the first coordinate of $x$. Since the sphere with the slice removed is an open set, the great circle geodesics are extendable.\vs\vs

\noindent\textit{Example 4: The space of phylogenetic trees.}
Phylogenetic trees are  of interest in  evolutionary biology, where they are used to represent the evolutionary history of a set of organisms. In a seminal paper ~\citep{bill:01}, phylogenetic trees with $m$ leaves are modeled by metric $m$-trees endowed with a metric that turns the space of phylogenetic $m$-trees into a metric space, as follows: A leaf is a vertex that is connected by only one edge, and a metric $m$-tree is a tree with $m$ uniquely labeled leaves and positive lengths on all interior edges, where an edge is called an interior edge if it does not connect to a leaf. A collection of $m$-trees that have
the same tree structure (taking leaf labels into account) but different edge lengths can be identified with the orthant $(0,\infty)^r$, where $r$ is determined by the tree structure and corresponds to the number of interior edges of each tree in the collection. 
With this identification between points and metric $m$-trees, the BHV metric $d_T$ on the space $\mathcal{T}_m$ of all metric $m$-trees is defined as follows: For two trees in the same orthant, their distance is the Euclidean distance of their edge lengths, while for two trees from different orthants, their distance is the minimum length over all paths that connect them and consist of only connected segments, where a segment is a straight line within an orthant. The minimum length path is the geodesic, which is extendable within the orthants where it starts and ends. According to Lemma 4.1 of~\cite{bill:01}, $\mathcal{T}_m$ is a 
unique geodesic space. It is a CAT$(0)$ space. More generally, each geodesic 
CAT$(0)$   metric space  is a unique geodesic space \cite[for a brief review see, e.g.,][]{lin:21}.   
\subsection{Preliminaries on noisy trajectories}
\label{sec:err}
Since  the metric space where the random object responses reside  is devoid of any vector-space structure,  one  cannot use classical additive error models.   Noise in observations can instead be quantified by 
perturbation maps~\citep{chen:20} $\mathcal{P}: \M \to \M$, characterized by 
\begin{align}
\label{pert:map}
\mu' = \argminsp \expect{\b{d^2(\mathcal{P}(\mu'),\mu)}} \ \text{for all } \mu' \in \M.
\end{align} 
We assume that  for the $i^{\text{th}}$ individual, noise-contaminated random objects  $Y_{ij}$ recorded at $T_{ij}$ are centered around an underlying trajectory $\alpha_i$. With  perturbation maps~\eqref{pert:map}, the observed data are 
\begin{align}
\label{pert:noisy:resp1}
Y_{ij} = \mathcal{P}_{ij}\p{\alpha_i(T_{ij})}, \ j=1,\dots,n_i,\ i =1,\dots,n.
\end{align} 
In connection with the classical mixed effects model in~\eqref{eucl:re_summary}, the perturbation map replaces  additive errors and the underlying trajectory is $\alpha_i(t) =\nu_i t$. 
The size of the error is quantified as $\expect{\b{d^2(\mathcal{P}_{ij}(\alpha_i(t)), \alpha_i(t))}}$, which  is bounded owing to the total boundedness  of the metric space, and corresponds to  the error variance for  classical Euclidean responses. 

For the classical linear mixed model $\alpha_i(t) = \nu_it$ is a line in the Euclidean space and therefore a geodesic. Thus a defining  feature of the classical linear mixed effects model is to fit geodesics to the data. A natural extension to the case of a general geodesic space is then to replace linearity by geodesicity, where observed data are assumed to cluster around a true geodesic.
For the remainder of the paper, the  underlying trajectory $\alpha_i$ for the $i^{\text{th}}$ individual  is assumed to be a uniquely extendable geodesic $\alpha_i=\gamma_{\nu_{i0},\nu_{i1}}^{(i)}$  
in the metric space $(\M,d)$ connecting the points $\nu_{i0}$ and $\nu_{i1}$. This leads to the following general model for the observed data,
\begin{align}
\label{pert:noisy:resp}
Y_{ij} = \mathcal{P}_{ij}\b{\geodi{T_{ij}}}, \ j=1,\dots,n_i,\ i =1,\dots,n.
\end{align}

\subsection{Random effects model for $\M$-valued data}
\label{sec:MEModel}
In a uniquely geodesic space $\M$ the randomness of  the geodesic path $\gamma_{\nu_0,\nu_1}(\cdot)$ is incorporated through the two endpoints $\nu_0$ and $\nu_1$ that determine the geodesic. For the $i^{\text{th}}$ individual, the underlying true geodesic path that connects the end-points $\nu_{i0}$ and $\nu_{i1}$  
is $\geodi{t} :[0,1]\to (\M,d).$
 We assume throughout that with probability $1$ the random geodesic that generates the observations is unique, an assumption that is satisfied for unique geodesic spaces such as those discussed in Examples 1-4 in Section \ref{sec:geod}.  
We also require the following assumption for the data generation mechanism.
\ben[label = (A\arabic*), series = fregStrg, start = 1]
\item \label{ass:pert}  Observation times  $T_{ij}$,  random perturbation maps $\mathcal{P}_{ij}$ and the random mechanism that generates the underlying geodesic trajectory $\geodi{t}$ $t \in [0,1]$ (or alternatively generates the two endpoints $\geodpti{0}$ and $\geodpti{1}$) are all independent and i.i.d.
\een
The proposed random effects model at the subject level for  $\M-$valued responses is  
\begin{align}
\label{model:pert:noisy:resp}
\geodi{T_{ij}} = \argminsp \expect{\b{d^2(Y_{ij},\mu)|\gamma_{\nu_{i0},\nu_{i1}}^{(i)}, T_{ij}}}, \quad  Y_{ij} = \mathcal{P}_{ij}\b{\geodi{T_{ij}}}, \ j =1,\dots,n_i.
\end{align} 
Once the random effects inherent in the subject-specific geodesics are recovered from the noisy observations, we regress  the entire geodesic paths $\br{\geodi{t}: t \in [0,1]}$ that constitute the responses on the  predictors $Z_i\in \S\subset\real^p,$ $p\geq 1$. This is implemented  through modeling the conditional Fr\'echet mean $\expect_{\oplus}\b{\{\geodi{t}: t\in[0,1]\}|Z_i = z}$
through applying a global Fr\'echet regression step \citep{pete:19}.

Since a  geodesic is determined by the two endpoints, the geodesic path $\br{\geodi{t}: t \in [0,1]}$ can be represented as a 
$\M-$valued pair $\p{\geodpti{0},\geodpti{1}} \in \p{\mathcal{D}_{\M},d_{\M}}$, where  the space $\p{\mathcal{D}_{\M},d_{\M}}$ is the  product metric space $(\M,d) \times (\M,d)$ with the  metric
\begin{align}
\label{prod:metric}
d_{\M}\p{(a_1,b_1),(a_2,b_2)} := \sqrt{d^2(a_1,a_2) + d^2(b_1,b_2)}, \text{ for all } a_1,a_2,b_1,b_2 \in (\M,d). 
\end{align}
In the context of metric geometry such product metric spaces with a $l_2$-type metric that combines the  metrics of the original spaces  have been extensively studied. In particular, it is well known that $\mathcal{D}_{\M}$ is a geodesic space if and only if $\M$ is geodesic ~\citep{bura:01}. This decomposition enables us to model the effective object response pair separately as
\begin{align}
\label{model:accross:subj:final}
\zeta_\oplus(z)  = \expect_{\oplus}\b{\p{\geodpti{0},\geodpti{1}}|Z_i = z} &=  \underset{(\mu_1,\mu_2) \in (\mathcal{D}_{\M},d_{\M})}{\argmin \ } \expect\b{d_{\M}^2\p{(\mu_1,\mu_2),(\geodpti{0},\geodpti{1})} |Z_i =z}\nonumber\\
&=  \underset{(\mu_1,\mu_2) \in (\mathcal{D}_{\M},d_{\M})}{\argmin \ } \expect\b{d^2\p{\mu_1,\geodpti{0}} + d^2\p{\mu_2,\geodpti{1}} |Z_i =z}.
\end{align}
This optimization problem is separable with optimal solution $\zeta_\oplus(z) =$ $(\gamz,\ggamz)^\intercal$ where 
\begin{align}
\label{model:accross:subj:final2}
& \gamz =   \underset{\mu_1 \in \M}{\argmin \ } \expect\b{d^2\p{\mu_1,\geodpti{0}}|Z_i =z},\quad 
\ggamz =  \underset{\mu_2 \in \M}{\argmin \ } \expect\b{d^2\p{\mu_2,\geodpti{1}}|Z_i =z}.
\end{align}

To implement the second step regression for higher dimensional predictors $Z\in \S \subset \real^p,$ $p\geq 2$, we use the global Fr\'echet regression (GFR) ~\citep{pete:19} method, which is a generalization of multiple linear regression for random object responses, and thus provides a direct extension of the multiple linear regression step for the baseline covariate effect that is implemented in classical random effects modeling for Euclidean responses. 
For Euclidean  data, the GFR approach is equivalent to fitting  a multiple linear regression model by least squares.     

Employing the GFR approach, defining a weight function  $s(Z,z) = 1 + (Z - \mu_Z)^\intercal \Sigma_Z^{-1} (z - \mu_Z)$ 
with $\mu_Z = \expect{(Z)}$ and $\Sigma_Z = \var(Z),$
the regression step in model~\eqref{model:accross:subj:final2} can be written as   $\zeta_\oplus(z) =$ $(\gamz,\ggamz)^\intercal$,  where 
\begin{align}
\label{model:accross:subj:final:gfr}
&\zeta_{k}(z) = \argminsp \expect\b{s(Z,z)d^2\p{\mu, \geodpti{k}}},\ k =0,1.
\end{align}
Combining a subject-specific approach  in model~\eqref{model:pert:noisy:resp}  with   model~\eqref{model:accross:subj:final2} for the impact of the covariate $Z$ thus provides a direct generalization of the standard random effects  model~\eqref{eucl:re_summary}. 

\section{Estimation and theory}
\label{sec:theory}
Consider the  global Fr\'echet regression (GFR) model with a response $Y \in (\M,d)$ and a predictor $T \in \T\subset [0,1]$ given by 
\begin{align}
	\label{gfr:smoothing:first_step}
	m_\oplus(t) = \argminsp \expect\b{w(T,t)d^2\p{\mu, Y}},
\end{align}
where $w(T,t) = 1 + (T - \mu_T)^\intercal \Sigma_T^{-1} (t - \mu_T)$ are  weight functions that are linear in $t$, with $\mu_T = \expect{(T)}$ and $\Sigma_T = \var(T).$ Based on the observations $(Y_{ij},T_{ij})$ $j = 1,\dots,n_i,$ for any given subject $i$, $i=1,\dots,n,$ following~\eqref{gfr:smoothing:first_step}, a subject-specific version of the GFR model is
\begin{align}
	\label{gfr:smoothing:first_step_subj_i}
	\mopi{t} = \argminsp \expect\b{w(T_{ij},t)d^2\p{\mu, Y_{ij}}},
\end{align}
where the weight function $w$ is defined as before.
This model will be implemented to recover individual trajectories from the data available separately for each subject, where we first assume the data lie exactly on the underlying geodesic and subsequently consider the small error case, dealing with additional perturbations of the responses. 

Using a similar idea as Theorem 1 of~\cite{fan:21} the following result shows that in the noise-free case the geodesic paths coincide with the GFR path.
\bthm
\label{thm:jianing:modified}
Consider the sample $\p{T_{ij}, Y_{ij}}$, $T_{ij} \in [0,1]$ $j = 1,\dots,n_i.$ For each subject $i$ assume that there exists a geodesic $\geodi{t} \in (\M,d),\ t \in [0, 1]$ that uniquely connects the endpoints $\geodpti{0}= \geodi{0}$ and $\geodpti{1}=\geodi{1}$ such that the responses $Y_{ij} = Y_i(T_{ij})$ are located exactly on this geodesic, that is, for each $Y_{ij} \in (\M,d)$  there exists a $u_{ij}\in (0,1)$ with  $Y_{ij} = \geodi{u_{ij}}.$ If the predictors $T_{ij}$ for any given subject $i$ are such that $T_{ij} = au_{ij} +b, \,  j  = 1,\dots,n_i,$ for some constants $a,b \in \real,$ implementing the 
global Fr\'echet regression  in~\eqref{gfr:smoothing:first_step_subj_i} exactly recovers the geodesic $\geodi{t} : t\in [0,1].$ If the geodesic is extendable from $[0,1]$ to $[s_1, s_2]$ and the extension is unique   in the 
sense that it is the only geodesic connecting $\geodi{s_1}$ and $\geodi{s_2},$ then the global Fr\'echet regression recovers the extended geodesic.
\ethm

Under the assumptions of Theorem~\ref{thm:jianing:modified} 
the GFR path $\br{\mopi{t}: t \in [0,1]}$ coincides with the underlying geodesic path $\br{\geodi{t}: t \in [0,1]}$ and the  latter can be represented by the two endpoints $(\geodpti{0},\geodpti{1} )$ with  $\mopi{t} = \geodi{t}$ for $t=0,1$. If  the geodesics are uniquely extendable, the pair $(\mopi{0},\mopi{1})$ effectively represents the $\M-$valued random effect for the $i-$th subject and therefore serves as response  for a second  Fr\'echet regression as per  model~\eqref{model:accross:subj:final}, \eqref{model:accross:subj:final2},  with the covariate $Z$ as predictor. 

In practical implementation, we replace   $(\mopi{0},\mopi{1})$ by the empirical version of GFR
\begin{align}
	\label{gfr:estd:first_step}
	\hmopi{t} = \argminsp \frac{1}{n_i} \sum_{j=1}^{n_i} w(T_{ij},t)d^2\p{\mu, Y_{ij}},   \quad t=0,1,
\end{align}
where the  empirical weights are  $w = 1 + (T_{ij} - \bar{T}_i)^\intercal \hat{\Sigma}_{T_i}^{-1} (t - \bar{T}_i),$ with $\bar{T}_i$ and $\hat{\Sigma}_{T_i}$ being the sample mean and covariance matrix for the predictor $T_{ij},$  $j = 1,\dots,n_i$ for the $i^{\text{th}}$ subject. 
With estimated object responses $(\hmopi{0},\hmopi{1})$ in hand, 
we proceed with the GFR implementation to recover the effect of covariates $Z$, where 
$\hat{\zeta}_\oplus(z) = (\hgamz,\hhgamz)^\intercal$ and 
\begin{align}
	\label{model:accross:subj:est:final2}
	\hat{\zeta}_k(z) = &  \underset{\mu_1 \in \M}{\argmin \ } \frac{1}{n}\sum_{i=1}^n s_{in}(Z_i,z) d^2\p{\mu_1,\hmopi{k}}, \ k = 0,1,
\end{align}
where  the  empirical GFR weights  are given by  
\begin{align}
\label{model:accross:subj:final2:wt}
  s_{in}(Z_i,z) = 1 + (Z_i - \bar{Z})^\intercal \hat{\Sigma}_Z^{-1} (z - \bar{Z}),
\end{align} 
$\bar{Z}$ and $\hat{\Sigma}_Z$ being the sample mean and covariance matrix for the predictor $Z_i,$ $i=1,\dots,n.$ 
\bthm
\label{thm:no_error:gfr:second_step}
Under assumptions~\ref{ass:gfr:second_step:P0}-\ref{ass:gfr:second_step:P2} in the Appendix 
it holds that
\[
d_{\M}\p{\hat{\zeta}_\oplus(z), \zeta_\oplus(z)} = \Op(n^{-1/2}).
\]
\ethm
Next we  discuss the  more realistic case where  responses do not lie exactly on the underlying geodesic paths but instead are perturbed from those on the path as per~ \eqref{pert:map}, \eqref{pert:noisy:resp}.  
To this end, let $\p{\Omega^\ast, \mathcal{F}^\ast,P^\ast}$ be the underlying probability space on which the observed data $(T_{ij},Y_{ij})$ are defined for the $i^{\text{th}}$ subject,  $i=1,\dots,n, \, j=1,\dots, n_i$. 
Since the mechanism that generates the data are independent as per~\ref{ass:pert}, $\p{\Omega^\ast, \mathcal{F}^\ast, P^\ast}$ can be perceived as a product space of two probability spaces: $\p{\Omega_1, \mathcal{F}_1, P_1}$, on which the $\M$-valued geodesic $\{\geodi{t}: t\in [0,1]\}$ connecting the two points $\nu_{i0}$ and $\nu_{i1}$, is defined;  and $\p{\Omega_2, \mathcal{F}_2, P_2},$ on which the observed time points $T_{ij}$  and the random perturbation maps $\mathcal{P}_{ij}$ associated with the noisy observation $Y_{ij}$ are defined. 
Thus, one can attribute the randomness of the noisy observations to three sources,   $Y_{ij} = \mathcal{P}_{ij}\b{\geodi{T_{ij}}} = f(\omega_1,\omega_2,\omega_3)$, where $\omega_1$ is a random element in $\p{\Omega_1, \mathcal{F}_1, P_1}$ that generates the endpoints of the true geodesic trajectory, thus generating the underlying geodesic; $(\omega_2,\omega_3) \in \p{\Omega_2, \mathcal{F}_2,P_2},$ where $\omega_2$ generates the $T_{ij}$ and $\omega_3$ generates  $\mathcal{P}_{ij}$ for $j=1,\dots,n_i$; $i=1\dots,n.$ For the special case of random effects models in Euclidean space,  $\p{\Omega_1, \mathcal{F}_1, P_1}$ is the underlying probability space for random slope and intercept. 

Note that fixing some element $\omega_1\in \Omega_1$ corresponds to 
a realization of the $\M$-valued underlying geodesic process. Also, as per assumption~\ref{ass:pert},  given a $\omega_1 \in \p{\Omega_1, \mathcal{F}_1,P_1},$ $(T_{\cdot j}, \mathcal{P}_{\cdot j})$ are independent in $\p{\Omega_2, \mathcal{F}_2,P_2}$ for all $j = 1,\dots, n_i$ and do not depend on $\omega_1.$ Suppose that for a given $\omega_1 \in \p{\Omega_1, \mathcal{F}_1,P_1}$, the geodesic $\nuom$ is observed at $m$ random time points.

We use notations $\smash{\nuom(\cdot)}, \smash{\mathcal{P}(\nuom(\cdot))}$ and $T$ to represent the corresponding quantities for the underlying geodesic, noisy observation and the random time point, respectively, for any given $\omega_1 \in \p{\Omega_1, \mathcal{F}_1, P_1}$. Denote by $\expect_{\Omega_2}$ the expectation with respect to the probability measure $P_2.$ 
For any $t \in [0,1]$, define
$
\nuom(t) = \argminsp \expect_{\Omega_2}{\b{d^2(\mathcal{P}(\nuom(\cdot)),\mu)| T = t}}.
$
We make the following small errors assumption, which mean  that errors implemented in  the form of perturbations are asymptotically negligible, uniformly across all realizations of the geodesic paths,  
\ben[label = (A\arabic*), series = fregStrg, start = 2]
\item \label{ass:pert:error} 
$\expect_{\Omega_2}{\b{d^2\p{\pert\p{\nuom(T)}, \nuom(T)}}}  = O\p{\alpha_n^2},
$
with $\alpha_n \to 0$ and $n\alpha_n^2 \to \infty.$ 
\een

For classical Euclidean linear random effects models with an additive error structure, this small errors assumption is not required due to the availability of  additive operations,  permitting the application of the law of large numbers and central limit theorem.  None of these is available in general geodesic spaces. 
A small error assumption is   commonly required  in nonlinear models   with  measurement errors and instrumental variable models~\citep{amem:85, chen:11, carr:04, carr:06, sche:16}. 
Observing  that the underlying true geodesic trajectory for the $i^{\text{th}}$ individual given by $\geodi{\cdot}$ is a random realization  corresponding to some $\omega_1$ in the probability space $\p{\Omega_1,\mathcal{F}_1, P_1}$, define the GFR model at the population level for any $\omega_1 \in \p{\Omega_1, \mathcal{F}_1, P_1}$ as
\begin{align}
	\label{true_obj:om}
	\mustarom(t) &= \argminsp \Hom\p{\mu,t},\ \Hom\p{\mu,t} = \expect_{\Omega_2}{\b{w(T,t)d^2(\nuom\p{T},\mu)}},
\end{align}
where $w(T,t) = 1 + (T - \mu_T)^\intercal \Sigma_T^{-1} (t - \mu_T)$ is the GFR weight function  with $\mu_T = \expect{(T)}$ and $\Sigma_T = \var(T)$, as before, and  $\Hom$ would be the objective function to minimize using global Fr\'echet regression with a fixed target response on the geodesic for a given $\omega_1 \in \p{\Omega_1, \mathcal{F}_1,P_1},$ if there was no error in the observations. 
Since in the error-free case  the GFR path recovers the geodesic entirely, $\geodi{t}$ equals $\mustarom(t), t \in \T$ for some $\omega_1 \in \p{\Omega_1,\mathcal{F}_1, P_1}, \, i=1,\dots,n.$

On the other hand, a GFR model based on the observed noisy responses, for any given $\omega_1 \in \p{\Omega_1, \mathcal{F}_1, P_1}$, can be defined as
\begin{align}
	\label{noisy_obj:om}
	\mutilom(t) &= \argminsp \Gom\p{\mu,t},\ \Gom\p{\mu,t} = \expect_{\Omega_2}{\b{w(T,t)d^2(\mathcal{P}\p{\nuom(\cdot)}\p{T},\mu)}},
\end{align}
where the weight function $w(T,t)$ for the global Fr\'echet regression is defined as before. In our notation the GFR path $\mopi{t}\in\M$  for the $i^{\text{th}}$ subject corresponds to $\mutilom(t), t\in \T$ for some $\omega_1 \in \p{\Omega_1,\mathcal{F}_1,P_1}$. 
In other words, the quantities $\geodi{\cdot}$ and $\mopi{\cdot}$, for $i = 1,\dots,n,$ are the subject-level realizations of $\mustarom(\cdot)$ and $\mutilom(\cdot)$, respectively for some random element $\omega_1 \in \p{\Omega_1,\mathcal{F}_1,P_1}$.
We require the following assumptions for all $\omega_1 \in \p{\Omega_1,\mathcal{F}_1,P_1}$. 
\ben[label = (K\arabic*), series = fregStrg, start = 1]
\item \label{ass:noisy:true:obj:min}  For any given $t \in \T$, the Fr\'echet means $\mustarom(t)$ and $\mutilom(t)$ exist and are unique, and for any $\eps>0$ it holds that 
$\underset{d\p{\mustarom(t),\mu}>\eps}{\inf\ } \b{\Hom\p{\mu,t} -\Hom\p{\mustarom(t) ,t} } >0$\\ 
and $\underset{d\p{ \mutilom(t),\mu}>\eps}{\inf\ } \b{\Gom\p{\mu,t} -\Gom\p{\mutilom(t) ,t} } >0.$
\item \label{ass:noisy:true:obj:curv} There exist constants $C_1>0,$ $\beta_1>1,$ such that for all $\eta>0,$
\begin{align*}
	\underset{\omega_1 \in \p{\Omega_1,\mathcal{F}_1,P_1}}{\inf\ } \underset{d\p{\mustarom(t),\mu}<\eta}{\inf\ } \b{\Hom(\mu,t) - \Hom\p{\mustarom(t),t} - C_1 d\p{\mustarom(t),\mu}^{\beta_1}} \geq 0.
\end{align*}
\een
Assumption~\ref{ass:noisy:true:obj:min} is commonly used  to establish  consistency of an M -estimator (see Chapter 3.2 in~\cite{vand:00}). It ensures  weak convergence of the empirical process $\Hom-\Gom$, which in turn implies convergence of the minimizers \citep{chen:20}. 
Assumption~\ref{ass:noisy:true:obj:curv} relates to the curvature of the objective function and is needed to control the behavior of the true and perturbed objective functions $\Hom$ and $\Gom$,   respectively, near their minimizers. These assumptions are satisfied for many random objects of interest 
\citep{pete:19}).

The following lemma establishes a connection between the population level Fr\'echet means of the responses lying exactly on a geodesic (given in~\eqref{true_obj:om}) and the perturbed responses situated near but not on  the geodesic (given in~\eqref{noisy_obj:om}).
\blem
\label{lem:noisy:true:min:convP}
Under assumptions~\ref{ass:pert},\ref{ass:pert:error} and \ref{ass:noisy:true:obj:min},~\ref{ass:noisy:true:obj:curv}, for any given $t \in \T,$
\[\supom d\p{\mustarom(t), \mutilom(t)} = \O(\alpha_n),  
\]
where $\alpha_n$ is as defined in Assumption~\ref{ass:pert:error}.
Further, for any $i = 1, \dots, n$ and  any $t \in \T,$
\[
d\p{\mopi{t},\geodi{t}} = \O(\alpha_n). 
\]
\elem
The above lemma implies that for any individual $i$, the underlying geodesic trajectory $\geodi{\cdot}$ can be recovered pointwise with asymptotically negligible error by the GFR path for the $i^{\text{th}}$ individual arbitrarily closely; $i=1,\dots,n.$
This suggests to estimate the underlying subject-specific geodesic from the noisy observations for each subject  
by the same method as before, obtaining the GFR path as in   \eqref{gfr:estd:first_step}.
Pointwise consistency of estimates is sufficient  as one only needs to  recover the endpoints $(\geodpti{0},\geodpti{1})$ of the geodesic.  
We  follow the same approach as before to infer the effect of the covariate $Z$ by implementing 
\eqref{model:accross:subj:est:final2}. This  is justified by the following result, which provides the rate of convergence of the regression of the metric space-valued random effects on the covariate $Z$. 
\bco
global Fr\'echet regression model as per~\eqref{model:accross:subj:final:gfr}, now with the response tuple $\p{\hmopi{0}, \hmopi{1}}$ and the predictors $Z_i \in \S \subset \real^p$; $p\geq 1$, for $i=1,\dots,n$. The second step GFR estimate is given by $\hat{\zeta}_\oplus(z) = (\hgamz,\hhgamz)^\intercal$ where
\begin{align}
	\label{step2:est:gfr}
	\hat{\zeta}_k(z) &= \argminsp M_{n}^{(k)}(\mu),\text{ where } M_{n}^{(k)}(\mu) = \frac{1}{n}\sum_{i=1}^n s_{in}(Z_i,z)d^2\p{\mu, \hmopi{k}},\ k = 0, 1,
\end{align}
where the empirical weights for the GFR estimator are given in~\eqref{model:accross:subj:final2:wt}. Combining the two-step models we show the asymptotic consistency results with a suitable rate of convergence for the overall estimate $\hat{\zeta}_\oplus(z)$, $z \in \S$ in the following theorem. The key assumptions for establishing the rate of convergence of any M-estimator, namely, the assumption of well-separateness of the minimizer, an upper bound on the entropy integral of the underlying metric space, and a local lower bound on the curvature of the objective functions, are employed in this context (see Appendix~\ref{sec:appen2}) in conjunction with the definition of the product metric in~\eqref{prod:metric}
\fi 
\bthm
\label{thm:step2:conv:rate}
Under assumptions~\ref{ass:pert},\ref{ass:pert:error},\ref{ass:noisy:true:obj:min},\ref{ass:noisy:true:obj:curv} and \ref{ass:gfr:second_step:P0},\ref{ass:gfr:second_step:P2} in the Appendix,  
for any $z \in \S\subset \real^p$ with $p\geq 1,$
$$d_{\M}\p{\hat{\zeta}_\oplus(z), \zeta_\oplus(z)} = \Op(\alpha_n^{1/2}).$$
\ethm
From the definition of $\alpha_n$ in assumption~\ref{ass:pert:error}, the rate is slightly slower than $n^{-1/4}$. 
\section{Simulation studies}
\label{sec:simu}
We report here only a subset of our simulation results for the important case of responses in the space of univariate distributions endowed with the Wasserstein metric, while additional simulation results  for spherical data can be found in Section~\ref{suppl:sec:simu:sphere} in the Supplement.

The Wasserstein space  of  probability distributions that we consider here  is as in Example 1, with 
\bco
on $(M,\mathcal{B}(M))$ with finite second moments, where $M$ is a closed interval in $\real$, is denoted by $\mathcal{W}_2(M)$ and is endowed with the 2-Wasserstein distance
\begin{align*}
	d_W(F_1,F_2) = \p{\int_0^1 [F^{-1}_1(s) - F^{-1}_2(s)]^2 ds}^{1/2},
\end{align*}
for any two distributions $F_1, F_2 \in \mathcal{W}_2(M)$, where $F^{-1}_1$ and $F^{-1}_2$ denote the quantile functions of $F_1$ and $F_2$, respectively; specifically, for any distribution $\mu = \mu(F) \in \mathcal{W}_2(M)$ with cumulative distribution function (cdf) $F$, we consider the quantile function $F^{-1}$ to be the left continuous inverse of F, that is, $F^{-1}(p):= \inf \{q \in M: F(q) \geq p\},$ for any $p \in (0, 1)$ (see eg. ~\cite{pana:19}). 
\fi
time-varying distributions as responses $Y$ which can be represented by their  quantile functions $Q_Y(\cdot)$. For each subject $i$, the random responses are constructed as repeated measurements around some underlying geodesic path $\geodi{\cdot}$ in the space $(\mathcal{W}_2(M),d_W)$. These underlying geodesic paths were 
generated conditional on a  covariate $Z_i \in \S \subset \real$, while the observed responses were sampled on these geodesics and then perturbed, implementing  
the following steps.
For each subject 

\noindent \underline{\textbf{Step 1.}} Generate $Z_i \overset{i.i.d.}{\sim} \text{Unif}(-1,1)$.

\noindent \underline{\textbf{Step 2.}} Generate $n_i$ random time points  $T_{ij}\overset{i.i.d.}{\sim} \text{Unif}(0,1)$. 
We consider a dense design, where  $n_i = 50$, as well as a sparse design, where  $n_i \in \br{2,\dots,5}$  with equal probability.  

\noindent \underline{\textbf{Step 3.}} Generate end points of the geodesics,  $\geodpti{0}$ and $\geodpti{1}$, that depend  on the external covariate $Z_i$ in the following way. 
Representing $\geodpti{0}$ and $\geodpti{1}$ as quantile functions $Q_{\smash{\geodpti{0}}}(\cdot)$ and $Q_{\smash{\geodpti{1}}}(\cdot)$, the  conditional expectation of $\geodpti{k}$ given $Z$ is modeled as
\begin{align}
	\label{sim:model}
	\expect{\b{Q_{\smash{\geodpti{k}}}(\cdot)|Z_i = z, T_{ij} = u}} = 
	\xi_{u,z} + \sigma_{u,z} \Phi_{[0,1]}^{-1}(\cdot),\ k = 0,1, \ j =1,\dots, n_i,
\end{align}
where $\Phi_{[0,1]}(\cdot)$ is the cdf of a standard normal distribution truncated on $[0,1].$ Specifically, the corresponding distribution function is given by
$$F(x) = \frac{\Phi((x - \xi_{u,z})/\sigma_{u,z}) - \Phi(-\xi_{u,z}/\sigma_{u,z})}{\Phi((1 - \xi_{u,z})/\sigma_{u,z}) - \Phi(-\xi_{u,z}/\sigma_{u,z})}\mathbf{1}_{[0,1]}(x) + \mathbf{1}_{(1,\infty)}(x), \quad x \in \real. 
$$

The distributional responses $\geodpti{0}$ and $\geodpti{1}$ are perturbed versions from model~\eqref{sim:model}. 
We consider four different simulation scenarios for  location-scale families with  varying sample sizes and perturbation levels, for both sparse and dense sample designs. The global parameters considered in the following data generation mechanisms are $\mu_0 = 0,\ \sigma_0 = 0.1,\ \beta_1 = 0.3,\ \beta_2 = 0.25,\ \gamma = 0.3,\ \nu_1 = 0.25,\ \nu_2 = 1.$

\textbf{Setting I.} The mean changes with the predictor values while the variance is constant. We generate the the auxiliary distribution parameters independently as $\mu_Y | (Z= z, T = u) \sim N_{[0,1]}(\xi_{u,z},\nu_1)$ and $ \sigma_Y| (Z= z, T = u) = \sigma_{u,z}$, where $\xi_{u,z} = \mu_0 + \beta_1 z + \beta_2 u$  and $\sigma_{u,z} = 0.1$. The corresponding distribution is given by $Q_{\smash{\geodpti{k}}} = \mu_Y + \sigma_Y \Phi^{-1}$, $k=0,1,$ where $\Phi$ is the standard normal cdf. 

\textbf{Setting II.} The mean remains constant, while the variance changes w.r.t to the predictor values. Again, the distribution parameters are generated independently as $\mu_Y | (Z= z, T = u) \sim N_{[0,1]}(\xi_{u,z},\nu_1)$ and $ \sigma_Y| (Z= z, T = u) \sim \text{Gamma}\left(\smash{\frac{(\sigma_0 +\gamma z)^2}{\nu_2}}, \smash{\frac{\nu_2}{(\sigma_0 +\gamma z)^2}}\right),$ such that $\xi_{u,z} = \mu_0 + \beta_2 u$ and $ \sigma_{u,z} = \sigma_0 +\gamma z.$

\textbf{Setting III.} The mean and variance both vary w.r.t to the predictor values. To this end, $\mu_Y | (Z= z, T = u) \sim N_{[0,1]}(\xi_{u,z},\nu_1)$ and $ \sigma_Y| (Z= z, T = u) \sim \text{Gamma}\left(\smash{\frac{(\sigma_0 +\gamma z)^2}{\nu_2}}, \smash{\frac{\nu_2}{(\sigma_0 +\gamma z)^2}}\right)$, independently sampled such that $\xi_{u,z} = \mu_0 + \beta_1 z + \beta_2 u$ and $ \sigma_{u,z} = \sigma_0 +\gamma z.$

\textbf{Setting IV.} After sampling the distribution parameters as in the previous setting, the resulting distribution is then ``transported'' in Wasserstein space via a random transport map $T$, that is uniformly sampled from a family of perturbation/ distortion functions $\left\{ T_k: k \in \pm 1,\pm 2,\pm 3 \right\}$, where $T_k(a) = a -  \smash{\frac{\sin(\pi k a)}{|k\pi|}}.$ 
The transported distribution is given by $T\#(\xi_{u,z} + \sigma_{u,z} \Phi_{\smash{[0,1]}}^{-1}(\cdot))$,
where $T\#p$ is a push-forward measure such that 
$T\#p(A) = p(\{x : T(x) \in A\})$,  for any measurable function $T : \real \to \real,$ distribution $p \in \mathcal{W}$, and set $A \subset \real.$ We sample the random transport map $T$ 
uniformly from the collection of maps described above; $p$ denotes a truncated Gaussian distribution with parameters $\xi_{u,z}$ and $\sigma_{u,z}$, and $\mathcal{W}$ is the metric space of distributions equipped with the Wasserstein metric.
The distributions thus generated are not Gaussian anymore due to the transportation. The Fr\'echet mean can be shown to remain at  $ \xi_{u,z} + \sigma_{u,z} \Phi^{-1}(\cdot)$ as before.   
Then the  geodesic in the quantile space connecting $Q_{\smash{\geodpti{0}}}(\cdot)$ and $Q_{\smash{\geodpti{1}}}(\cdot)$ is given by $Q_{\smash{\geodi{\cdot}}} : t \mapsto (1-t)\ Q_{\smash{\geodpti{0}}} + t\ Q_{\smash{\geodpti{1}}}.$ For the $i^{\text{th}}$ subject, $n_i$ points are generated on the true underlying geodesic $Q_{\smash{\geodi{t}}},$ $t \in [0,1]$.

\noindent \underline{\textbf{Step 4.}} Perturb the true quantile functions $Q_{\smash{\geodi{T_{ij}}}}$ situated on a geodesic  
such that the observed responses remain  valid quantile functions. The perturbed/noisy distributional responses, represented as quantile functions,  are constructed as $\tilde{Q}: [0,1] \to [0,1]$ such that
\begin{align}
	\label{sim:pert:quant}
	\tilde{Q}(u) &= Q(s) + \eps \Delta(s),\ s\in [0,1], 
\end{align}
where $\Delta(s) = \alpha_n Q(s)(1-Q(s)),$ $0<\alpha_n<1$  and $\eps = \pm \alpha_n$ with equal probability $1/2.$
With a sufficiently small choice of $\alpha_n \in (0,1)$, $\tilde{Q}$ is an increasing quantile function in $[0,1]$. From the construction we have $\Delta(s) \leq \min\br{Q(s), 1-Q(s)}$ for all $s\in [0,1].$ Note that for $0<\alpha_n<1$,
$\tilde{Q}' = (Q \pm \alpha_n \Delta)' >0$,
as long as the true quantile functions $Q$ are strictly increasing  
and $\expect{\p{\tilde{Q}}} = Q$.
The observed responses are thus  per~\eqref{sim:pert:quant} 
	$\tilde{Q}_{\smash{Y_{ij}(T_{ij})}} = Q_{\smash{\geodi{T_{ij}}}} \pm \alpha_n^2 Q_{\smash{\geodi{T_{ij}}}}\p{1 - Q_{\smash{\geodi{T_{ij}}}}}.$  We implemented the proposed model as per  \eqref{gfr:estd:first_step} and  
	\eqref{model:accross:subj:est:final2}.
	
	\bco
	
	$ 
	
	The first-step global Fr\'echet regression (GFR) model was fitted at each subject level at the output points $0$ and $1$, with the distributional responses $\tilde{Q}_{Y_{ij}(T_{ij})}$  and Euclidean predictors $T_{ij}$,  $j = 1,\dots,n_i$. The estimates $\hmopi{0}$ and $\hmopi{1}$ are obtained for all $i =  1,\dots,n.$ Treating $(\hmopi{0},\hmopi{1})$ as distribution-valued responses and the baseline covariates $Z_i \in \S$ as Euclidean predictors, we then fit the second-step GFR as per~\eqref{model:accross:subj:est:final2} to obtain the final estimates $\hat{\zeta}_\oplus(z) = (\hgamz,\hhgamz)^\intercal.$
	The estimated distributions are computed on the geodesic connecting the points $\hgamz$ and $\hhgamz$, which is given by $t \mapsto t Q_{\hgamz} + (1-t) Q_{\hhgamz},$
	$t \in [0,1]$.
	
	\fi
	
\begin{figure}[!htb]
		\centering
		\includegraphics[width =.5 \textwidth]{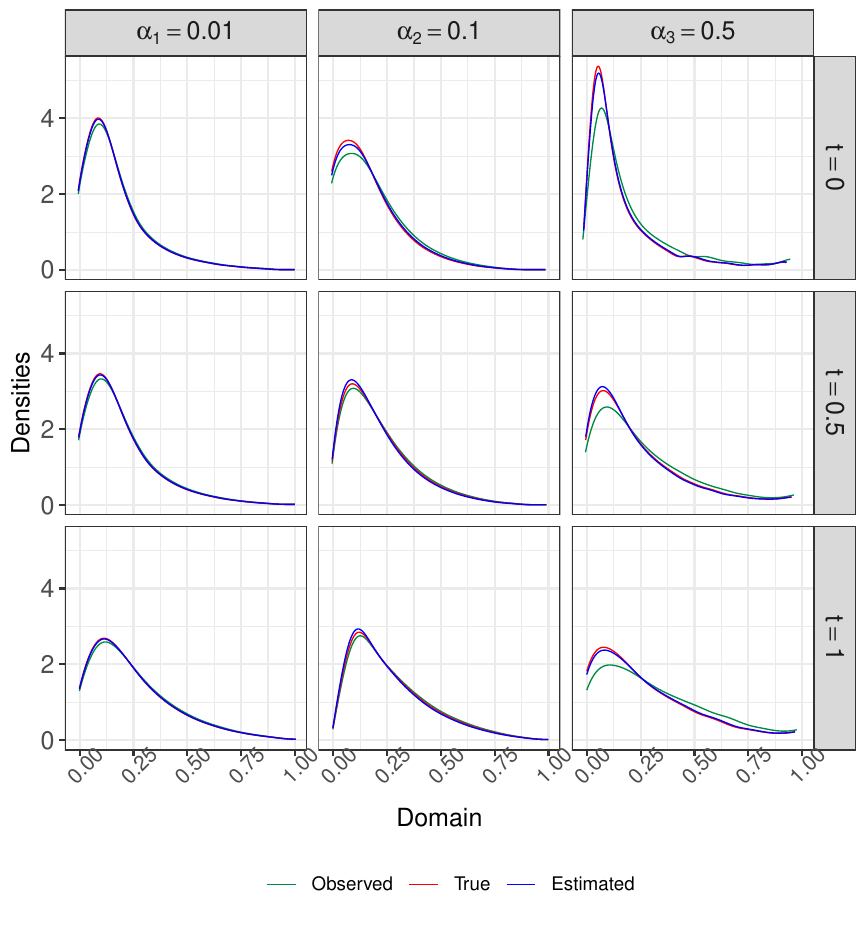}
		\centering
		\caption{Visualization of the true (red), observed (perturbed, green), and estimated (blue) distributional object responses as densities  for a randomly selected simulation sample generated under setting IV with a sparse design where each subject has $2$ to $5$ repeated measurements, comparing  varying perturbation levels $\alpha_n = 0.01,0.1,0.3$ (left, middle and right). The densities lying on a geodesic in the Wasserstein space of distributions are displayed at three different time points,  $t =0, 0.5$, and $1$ (top, middle, and bottom  rows, respectively).}
		\label{Fig:Sim:dens_example}
	\end{figure}

	The effect of the perturbation parameter $\alpha_n$ is demonstrated in Figure~\ref{Fig:Sim:dens_example} for one simulation run in Setting IV. True, observed, and predicted distributions 
	are shown for the sparse design case.  The predicted distributions are obtained 
	for the observed values of the covariate/predictor $Z_i$  for all $t \in [0,1]$, represented as densities. 
	For small perturbations, the observed distributions are seen to be close to the underlying true geodesic path of distributions, while for  larger levels of perturbation  deviations are larger. However, estimated/predicted  distributions throughout remain  close to the true distributions, providing  evidence for the efficiency of the proposed random effects model. 

\begin{figure}[!htb]
		\centering
		\hspace{-.2cm}\includegraphics[width = .6\textwidth]{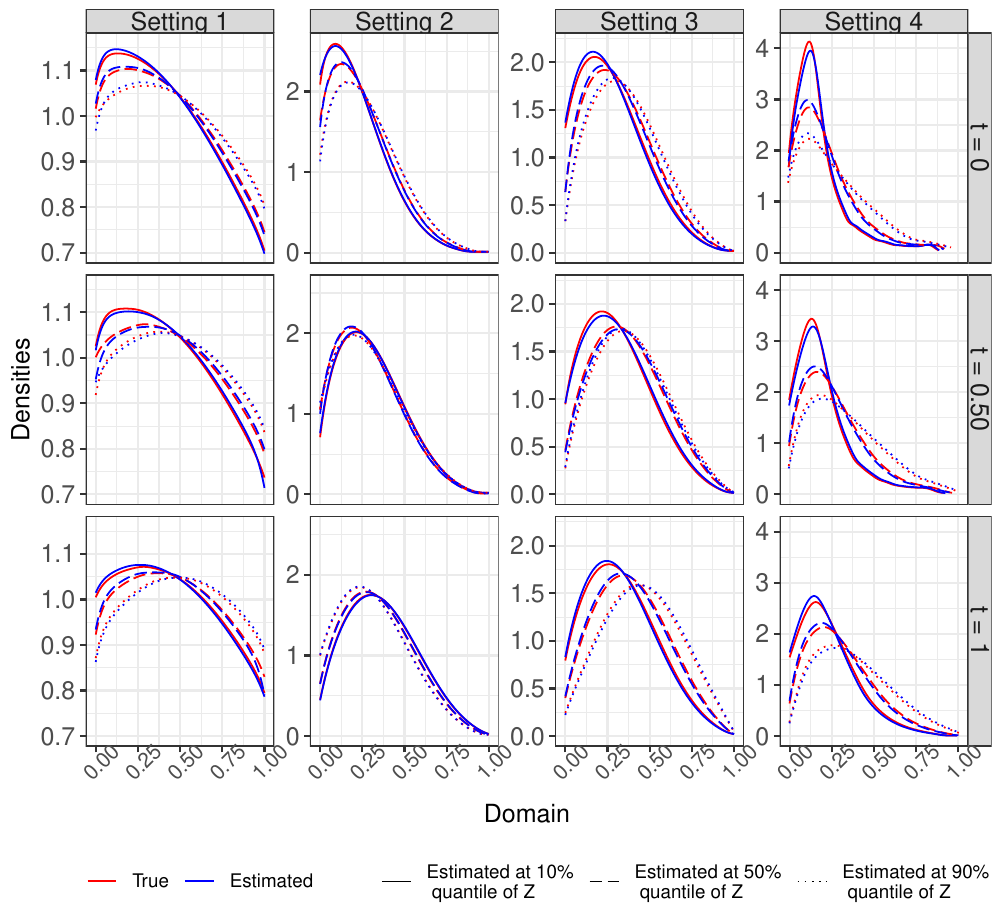}
		\centering
		\caption{The  time-dynamic effect of the baseline covariate for distributional objects represented as densities for a randomly selected simulation sample, displaying   true (red) and estimated (blue) densities for simulation settings I-IV (columns from left to right). Data were generated under a
			sparse design,  where each subject has $2$ to $5$ repeated measurements and where response distributions  were perturbed  with a fixed small perturbation level $\alpha = 0.1$. Estimated/predicted densities are shown for   
			the $10\%$ (solid), $50\%$ (long-dashed) and $90\%$ (dotted) quantile levels of the covariate. The top, middle, and bottom panels correspond to the prediction/estimation at times $t =0, 0.5$, and $1$, respectively.}
		\label{Fig:Sim:dens_example2}
	\end{figure}

	We illustrate  the effects of the  covariate $Z$ on the model fits across different simulation settings for one simulation run in  Figure \ref{Fig:Sim:dens_example2}. Again data are generated for a sparse design for each of the  settings mentioned above with sample size $n=500$, where the observed distributions are generated around the true underlying geodesics in the Wasserstein space 
	and observations are perturbed at  perturbation  level $\alpha_n = 0.1.$ To assess the covariate effects,  we fitted the model at covariate levels that correspond to the  $10\%,$ $50\%,$ and $90\%$ quantiles of the  covariate $Z$. 
	One finds that at all time points and across all settings the predicted densities closely approximate the truth. This demonstrates that in the small error case  the proposed random effects model and its implementation is well suited to recover the true trajectories when given the covariate information.     
	For setting I with  a location shift in the data generation mechanism, the modes of the densities shift towards the right, i.e., a higher value of the covariate is associated with a right shift in the estimated densities. For setting II, a higher value of the covariate is associated with an increase in the spread of the distribution. Settings III and IV capture the combined effect of location and scale shifts. The location, spread, skewness, and overall shape of the predicted densities change as expected with increasing levels of the predictor values. One also observes a rightwards  shift of the distributions over time,  
	an expected consequence of the generation of the geodesics in distribution space.
	
\begin{figure}[!htb]	
		\centering
		\includegraphics[width=.5\textwidth]{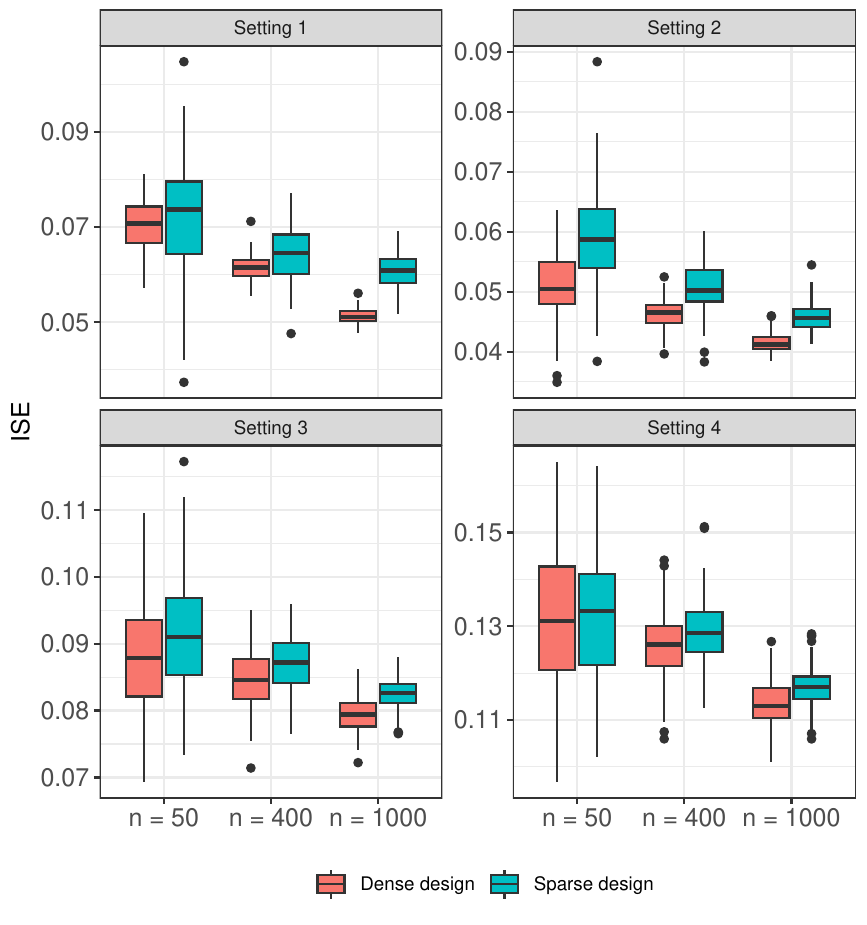}
		\centering
		\caption{Boxplots of Integrated Squared Errors (ISE) calculated as per~\eqref{sim:mise:dens}, over $500$ simulation runs for the four simulation settings  (displayed in the panels clockwise from the top left corner). Results are shown for sample sizes $n= 50,400,1000$ for both sparse (blue) and dense (red) designs.}
		\label{Fig:Sim:Boxplot1}
	\end{figure}

	We further studied the effect of sample size and sample design (sparse or dense) for the four simulation settings on the performance of the proposed method while keeping the perturbation level fixed at $\alpha= 0.1$. The results of $500$ Monte Carlo simulation runs are shown in Figure \ref{Fig:Sim:Boxplot1}, where we display boxplots of 
	Integrated Square Error (ISE) as a measure of discrepancy between the true and the estimated distributions. Specifically, 
	\begin{align}
		\label{sim:mise:dens}
		\text{ISE}_r = \int_{z \in \S} \int_{t \in [0,1} d_W (Y^r(t,z), \hat{Y}^r(t,z)) dt dz,
	\end{align}
	where $Y^r(t,z)$ and $\hat{Y}^r(t,z)$ denote, respectively, the true distributional object lying on a geodesic (without perturbation) in the Wasserstein-2 space and the estimated object at time point $t$ and covariate value $z$ for the $r^{\text{th}}$ simulation run, where $r = 1,\dots,500.$ 
	We observe a decrease in ISE for increasing sample size and deviations  are generally higher if both the location and scale parameters are varied as a function of the covariate. 
	
	\subsection{Simulation study: Responses lying on the surface of a sphere}
	\label{sec:simu:sphere}
	We applied the proposed approach targeting general random objects
	as responses lying on the surface of a sphere. The numerical results describing the data generation mechanism and evaluating the performance of the proposed method are discussed in details in subsection~\ref{sec:simu:sphere} of the Supplement.

	\section{Data analysis}
	\label{sec:data}
	\subsection{Longitudinal  fMRI data}
	\label{sec:data2}
	Resting-state functional Magnetic Resonance Imaging (fMRI) methodology makes it possible to study brain activation and to identify brain regions or cortical hubs that  exhibit similar activity  when subjects are in the resting state \citep{alle:14}. FMRI measures brain activity by detecting changes in blood-oxygen-level-dependent (BOLD) signals in the brain across time. 
	The analysis of brain functional connectivity at the subject level typically relies on a specific spatial parcellation of the brain into a set of regions of interest (ROIs). Temporal coherence between pairwise ROIs is usually measured by the so-called  Pearson correlation coefficient matrix (PCC) of functional connectivity obtained from the fMRI time series, which is an $m\times m$ correlation matrix if one has  $m$ distinct ROIs. In this analysis, we will use PCC matrices derived from fMRI as responses. 
	Alzheimer's Disease has been found to be associated  with anomalies in the functional integration of ROIs \citep{damo:12,zhan:10c} that may be time-varying, along with changes in the brain due to aging for cognitively normal subjects. 
	This provides the motivation to explore the time-varying regression relationship between the connectivity correlation matrix objects and  relevant external covariates.
	
	Available data are from the  Alzheimer's Disease Neuroimaging Initiative (ADNI) database (adni.loni.usc.edu), where PCC matrices 
	derived from fMRI signals are observed sparsely over time for each subject in a sample of 
	$n = 340$ subjects composed of  $155$ Cognitive Normal (CN) subjects and $185$ Alzheimer's patients with mild cognitive impairment (MCI) with ages ranging from $55.7$ to $94.8$ years. At least 2 scans are available for each subject but not more than 
	9 scans, with a median of 4 scans, so these are  sparsely sampled longitudinal data. We normalized the time scale of the  measurements to the interval 
	$\T = [0,1]$, where for each subject the time at which the first scan is recorded is defined as the origin of the time scale $t=0$ 
	and $t=1$ is  7 (9) years after the first scan for the CN (MCI)  subjects.
	The pre-processing of the BOLD signals adopted  standard procedures of slice-timing correction, head motion correction, and other standard  steps. 
	Then $m= 90$ brain seed voxels for each subject were extracted for the ROIs  of the automated anatomical labeling (AAL) atlas~\citep{tzou:02} to parcellate the whole brain into $90$ ROIs, with $45$ ROIs in each hemisphere, and the signals were converted   
	to a $90 \times 90$ PCC matrix, which corresponds to the available observation for each time point and subject. 
	
	The structure of the space of random objects always  depends on the choice of the metric, which is often chosen for convenience and  interpretability in the context of  specific data applications. Here we endow the space of symmetric positive definite correlation matrices $\mathcal{M}$ with the power-Euclidean metric $d_P$ with the power $\alpha = 1/2$~\citep{dryd:10}, 
	\begin{align}
		\label{power:metric}
		d_P(A,B) = \frac{1}{\alpha}\ltwoNorm{A^\alpha - B^\alpha}_F \quad \text{for any} \quad A,B \in \mathcal{M}.\end{align}
	Here  $S^\alpha = U\Lambda^\alpha U^\intercal $, for the usual spectral decomposition of $S = U\Lambda U^\intercal $ with  an orthogonal matrix $U$ and a diagonal matrix $\Lambda$ with strictly positive entries and  $\ltwoNorm{\cdot}_F$ denotes the Frobenius norm. 
	The space $\M$ is a  uniquely extendable geodesic space. 
	To implement the proposed random effects model, in a first step we recovered the underlying subject-specific trajectories by estimating the matrices at  the endpoints $0$ and $1$, and then regressed these on the covariate $Z$, which was chosen as  
	the two-dimensional vector (Age, ADAS-Cog-13 score) for each subject, obtained at the time of the first scan $t=0$. 
	For Alzheimer's studies, the ADAS-Cog-13 score (henceforth referred to as the C score)  is a widely-used measure of cognitive performance. It quantifies  impairments across cognitive domains 
	\citep{kuep:18}; higher scores indicate more serious cognitive deficiency.  
	
\begin{figure}[htb]
		\centering 
		\begin{subfigure}{0.25\textwidth}
			\includegraphics[width=\textwidth]{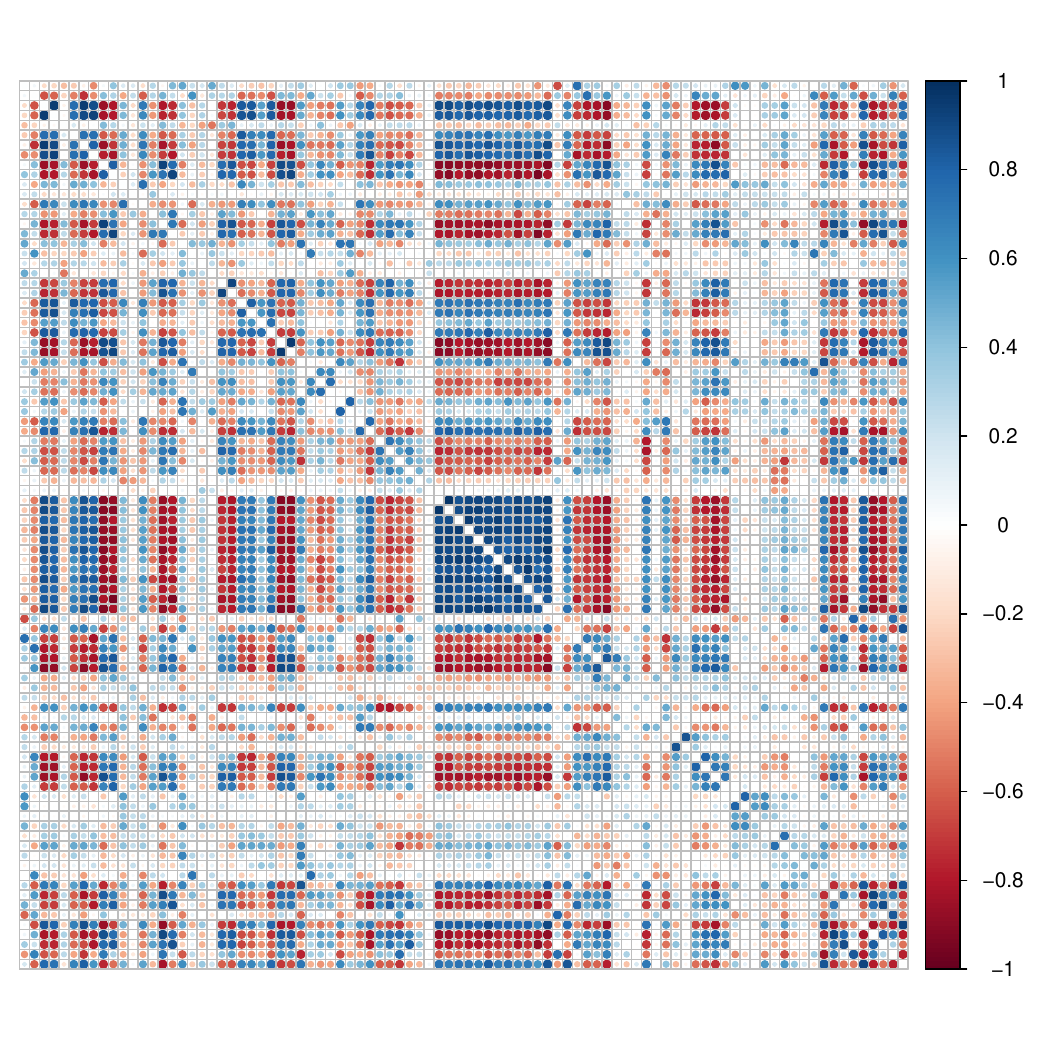}
		\end{subfigure}\hfil 
		\begin{subfigure}{0.25\textwidth}
			\includegraphics[width=\textwidth]{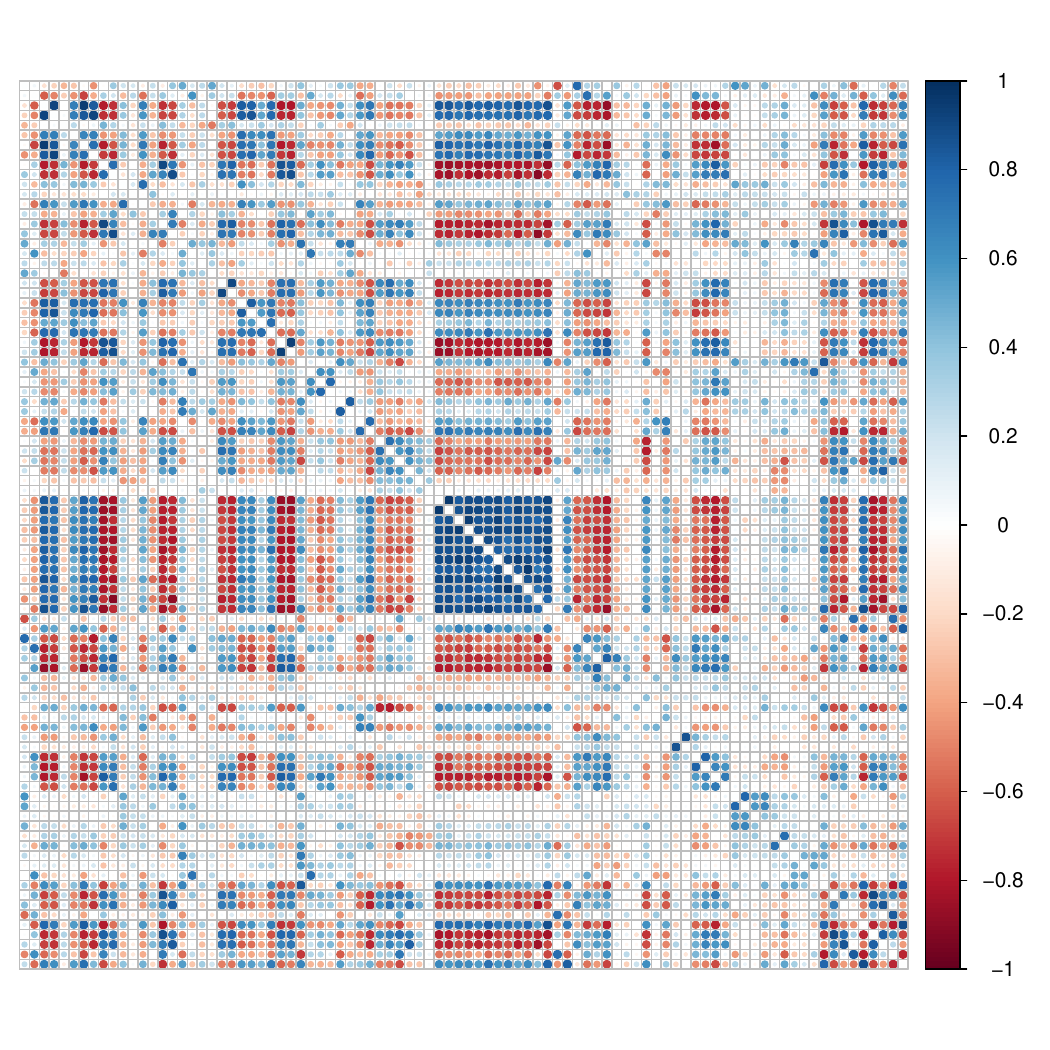}
		\end{subfigure}\hfil 
		\begin{subfigure}{0.25\textwidth}
			\includegraphics[width=\textwidth]{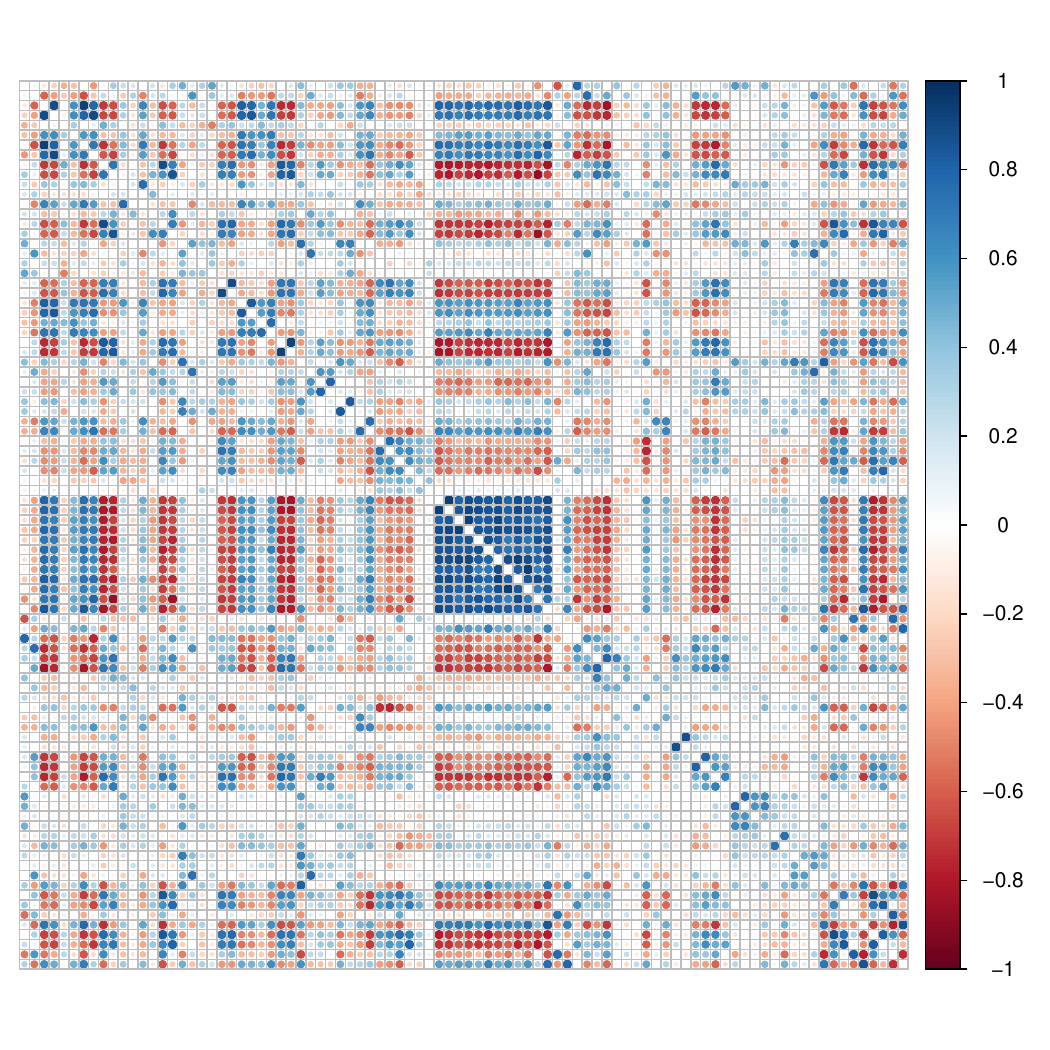}
		\end{subfigure}
		
		\medskip
		\begin{subfigure}{0.25\textwidth}
			\includegraphics[width=\textwidth]{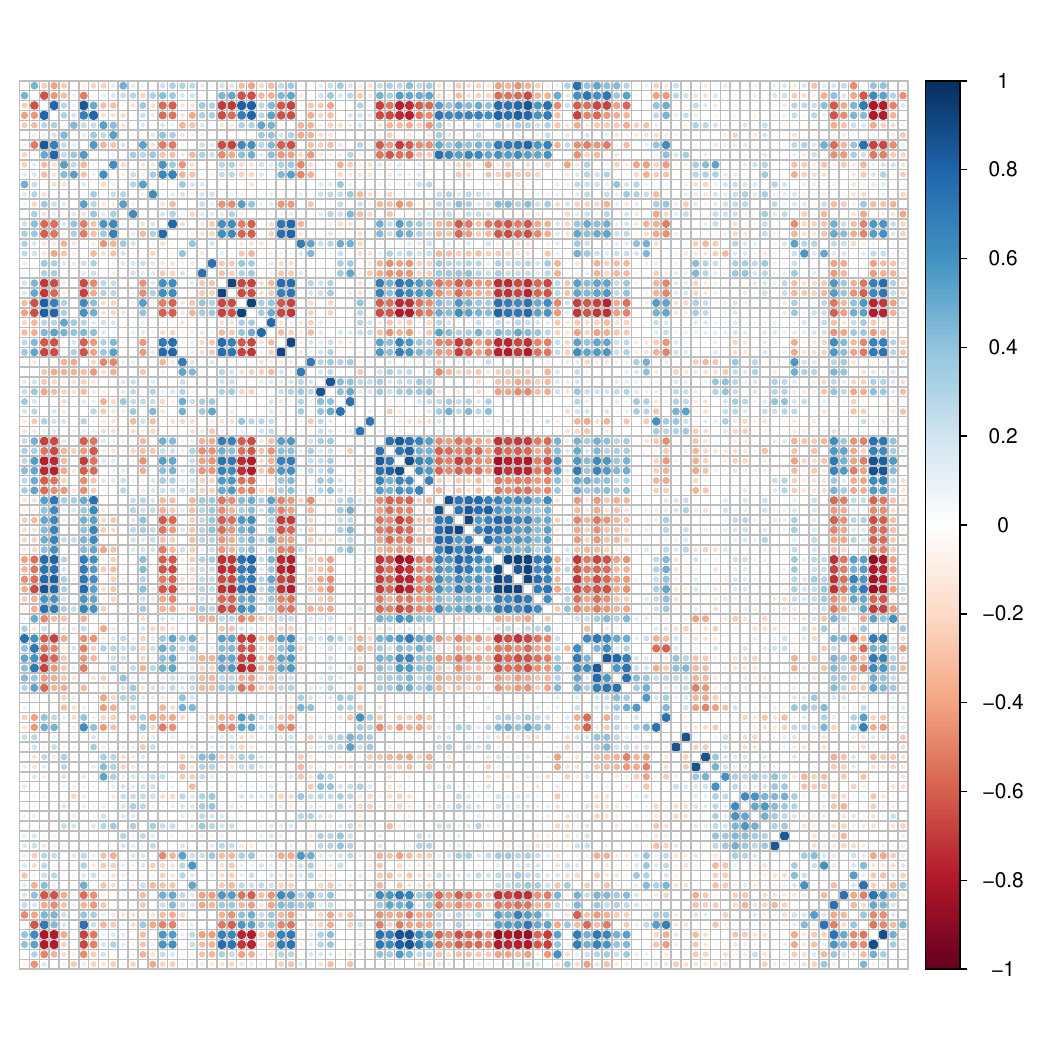}
		\end{subfigure}\hfil 
		\begin{subfigure}{0.25\textwidth}
			\includegraphics[width=\textwidth]{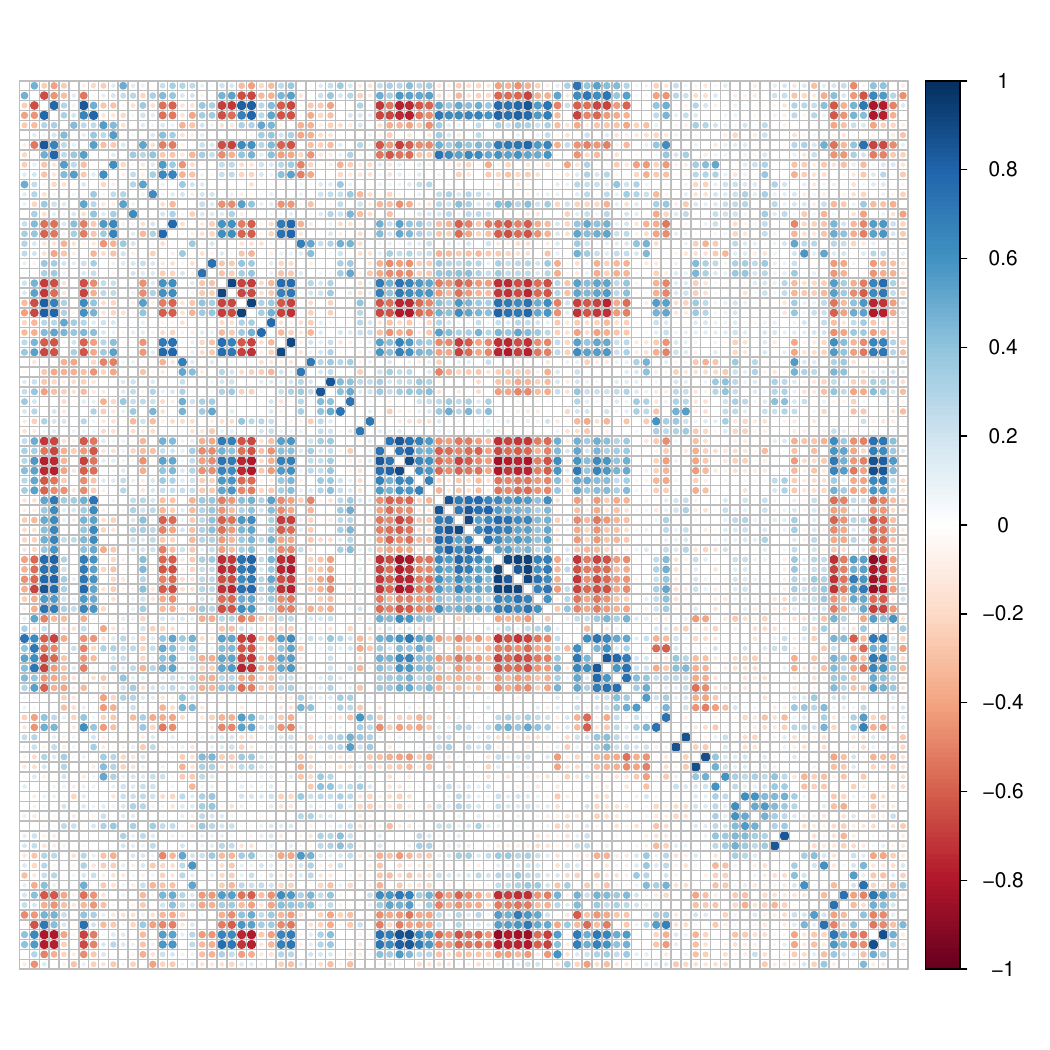}
		\end{subfigure}\hfil 
		\begin{subfigure}{0.25\textwidth}
			\includegraphics[width=\textwidth]{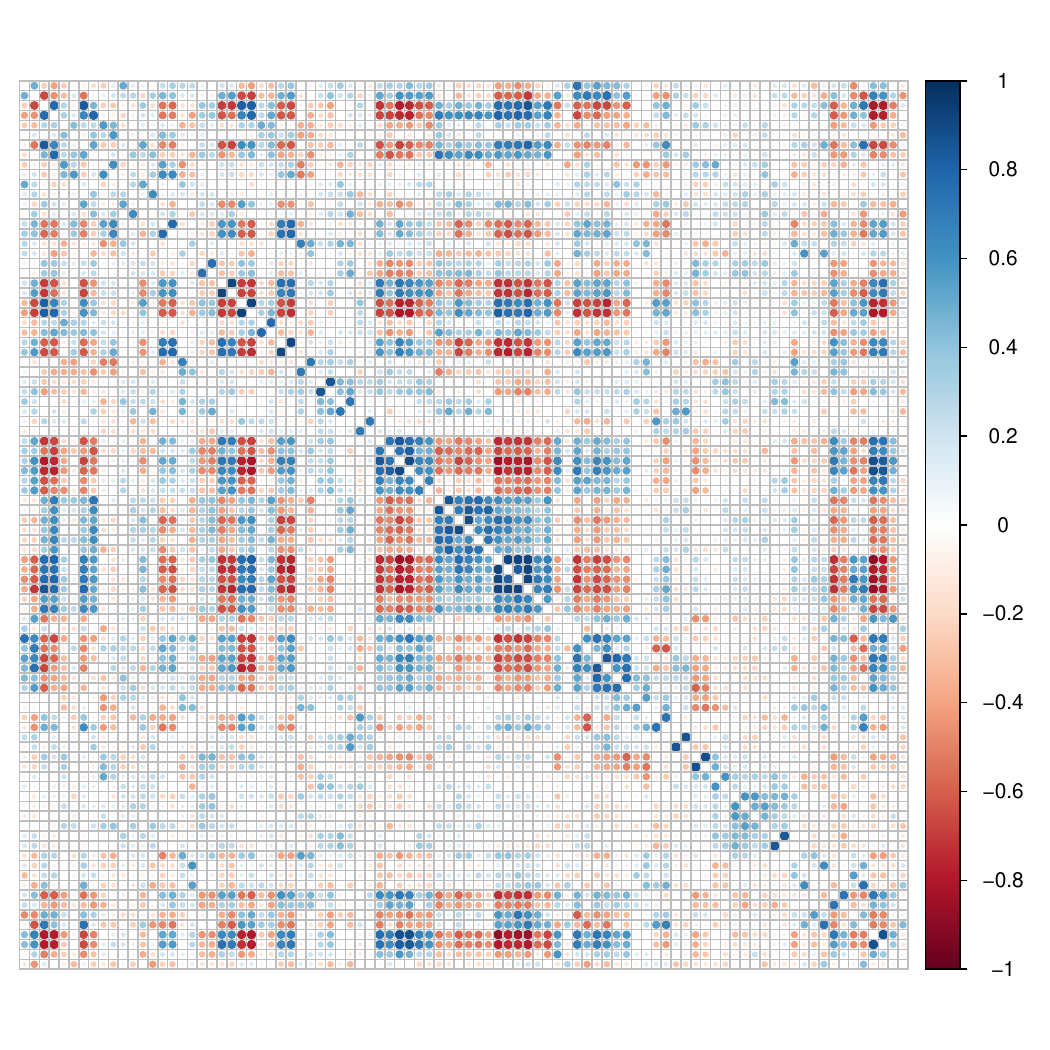}
		\end{subfigure}
		\centering
		\caption{Estimated/predicted PCC matrices obtained from the proposed random effects model  for CN subjects. The top and bottom rows correspond to the predicted correlation matrices at times $t=0$ and $t=1$ respectively, while within each row the left, middle, and right panels depict the fits at the $10\%, 50\%$, and $90\%$  quantiles of the C-score with  age fixed at its mean level.  Positive (negative) values for correlations are drawn in red (blue).} 
	\label{Fig:Data:adni:corrplot_over_score_CN}
\end{figure}

	
	To illustrate  the effect of the C-score, we fix the age of the subjects at its mean level (74 years) and provide the fitted model at the $10\%, 50\%$, and $90\%$  quantiles of the C-score. Figure~\ref{Fig:Data:adni:corrplot_over_score_CN} demonstrates the trend for the temporal correlations for varying C-score levels at different times of the study for the CN subjects. 
	One finds that  the overall correlation strengths diminish with higher C-scores. Further, comparing the rows for each panel, we find that correlations are overall weaker  at time $1$ than at time $0$. A similar pattern with overall weaker connections emerges  for the MCI subjects (see Figure~\ref{Fig:Data:adni:corrplot_over_score_MCI} in the Supplement).	

\begin{figure}[!htb]
	\centering
	\includegraphics[width=.5\textwidth]{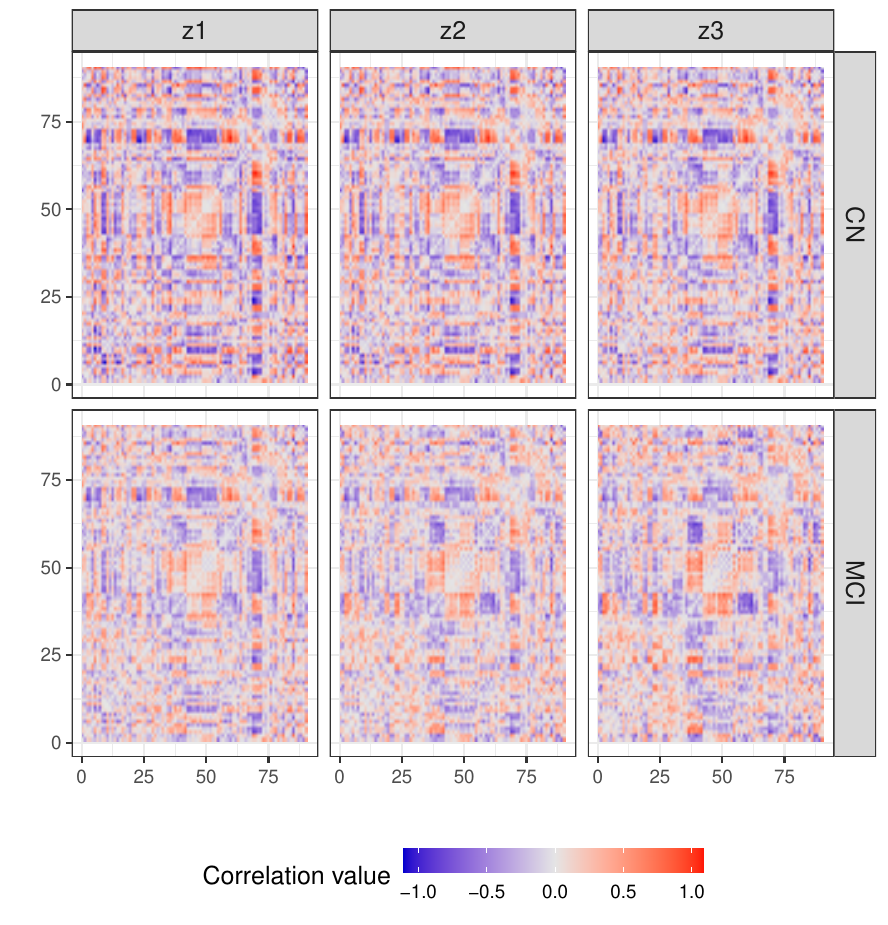}
	\caption{Differences of the predicted PCC matrices at time $1$ and at time $0$ for both CN (top) and MCI (bottom) subjects. The columns, from left to right, correspond to the differences of the  predicted PCC matrices, fitted at the $10\%, 50\%$, and $90\%$  quantiles of the C-score, respectively, while the  second covariate age is fixed at  its mean level. Higher (lower) values of the difference  are shown in red (blue).}
	\label{Fig:Data:adni:diff:corrplot:over_score}
\end{figure}

To further elicit the time-varying effects of the  C-score on the PCC matrix geodesics,  we subtract the predicted matrices at time $0$ from the predicted matrices at time $1$, for each of the three covariate quantiles, separately for CN and MCI subjects. 
In Figure~\ref{Fig:Data:adni:diff:corrplot:over_score}, the columns (from left to right) display the difference of the fitted PCC matrices at time $1$ and time $0$, at the $z_1 = 10\%, z_2 = 50\%$, and $z_3 = 90\%$  quantiles of the C-score, respectively, while the other covariate age is fixed at  its mean level. The top (bottom) row corresponds to the CN (MCI) subjects.
For higher score levels, 
the inter-hub connections are found to become weaker. The effect is clearly more pronounced for the MCI subjects as compared to the CN subjects, MCI subjects losing connectivity at a faster rate.

We also converted the PCC matrices 
into simple, undirected, weighted networks to facilitate interpretation by setting diagonal entries to $0$ and hard thresholding the absolute values of the remaining correlations. We kept the  $15\%$ strongest connections and discarded the others~\citep{schw:11}, converting the PCC into weighted adjacency matrices. 
The adjacency matrix computed from a PCC matrix is given by $A = (a_{ij})_{i,j =1.\dots,m}$, indicating the $i$-th and $j$-th hubs in the brain are either connected by an edge of weight $a_{ij} > 0$, or else unconnected if $a_{ij} = 0$.   
To represent the resulting estimated brain networks for changing covariate levels we use network summaries such as modularity, a summary measure of network segregation~\citep{newm:06a} and global efficiency~\citep{alex:13}, a measure of network integration. 
With $a_{ij}$ representing the edge weight between nodes $i$ and $j$, 
modularity is  defined as
$  Q = \frac{1}{2L} \sum_{i,j} \left[a_{ij} - \frac{k_ik_j}{2L}\right]\delta(c_i,c_j),$
where  $L$ is the sum of all of the edge weights in the graph, $k_i$ is the sum of the weights of the edges attached to node $i$, $c_i,c_j$ are the communities of the nodes; and $\delta(x,y) = 1 \text{ if } x \neq 1$ and $0$ otherwise.  
Table~\ref{tab:Data:adni:mod:CN_MCI} shows 
modularity and global efficiency of the brain networks for CN and MCI subjects at times $0$ and $1$ 
estimated at the $10\%, 50\%$, and $90\%$  quantiles of the C-score, respectively, while covariate age is fixed at its mean level. Both indices decrease for higher  C-scores and over  time where the decrease over time is much more pronounced for MCI subjects, in line with the previous findings for PCC matrices. 
\begin{table}[!htb]
		\centering
	\caption{\small{Modularity and global efficiency of the estimated brain networks obtained for CN and MCI subjects by hard thresholding at times $t=0$ and $t=1$,  
		for the $10\%, 50\%$, and $90\%$  quantiles of the C-score, while covariate age is fixed at its mean level.}}
	\label{tab:Data:adni:mod:CN_MCI}
	\centering
\small{	\begin{tabular}{c|cccc|cccc}
		\hline
		&
		\multicolumn{4}{c|}{CN} &
		\multicolumn{4}{c}{MCI} \\ \hline
		&
		\multicolumn{2}{c|}{Modularity} &
		\multicolumn{2}{c|}{Global Efficiency} &
		\multicolumn{2}{c|}{Modularity} &
		\multicolumn{2}{c}{Global Efficiency} \\ \hline
		&
		\multicolumn{1}{c|}{$t = 0$} &
		\multicolumn{1}{c|}{$t = 1$} &
		\multicolumn{1}{c|}{$t = 0$} &
		$t = 1$ &
		\multicolumn{1}{c|}{$t = 0$} &
		\multicolumn{1}{c|}{$t = 1$} &
		\multicolumn{1}{c|}{$t = 0$} &
		$t = 1$ \\ \hline
		\begin{tabular}[c]{@{}c@{}}10\% Quantile\\ of Total Score\end{tabular} &
		\multicolumn{1}{c|}{0.534} &
		\multicolumn{1}{c|}{0.479} &
		\multicolumn{1}{c|}{0.499} &
		0.387 &
		\multicolumn{1}{c|}{0.536} &
		\multicolumn{1}{c|}{0.593} &
		\multicolumn{1}{c|}{0.520} &
		0.368 \\ \hline
		\begin{tabular}[c]{@{}c@{}}50\% Quantile\\ of Total Score\end{tabular} &
		\multicolumn{1}{c|}{0.528} &
		\multicolumn{1}{c|}{0.474} &
		\multicolumn{1}{c|}{0.485} &
		0.371 &
		\multicolumn{1}{c|}{0.535} &
		\multicolumn{1}{c|}{0.541} &
		\multicolumn{1}{c|}{0.486} &
		0.365 \\ \hline
		\begin{tabular}[c]{@{}c@{}}90\% Quantile\\ of Total Score\end{tabular} &
		\multicolumn{1}{c|}{0.505} &
		\multicolumn{1}{c|}{0.462} &
		\multicolumn{1}{c|}{0.472} &
		0.355 &
		\multicolumn{1}{c|}{0.531} &
		\multicolumn{1}{c|}{0.465} &
		\multicolumn{1}{c|}{0.387} &
		0.322 \\ \hline
	\end{tabular}}
\end{table}

\begin{figure}[!htb]
	\centering
	\includegraphics[width = .5\textwidth]{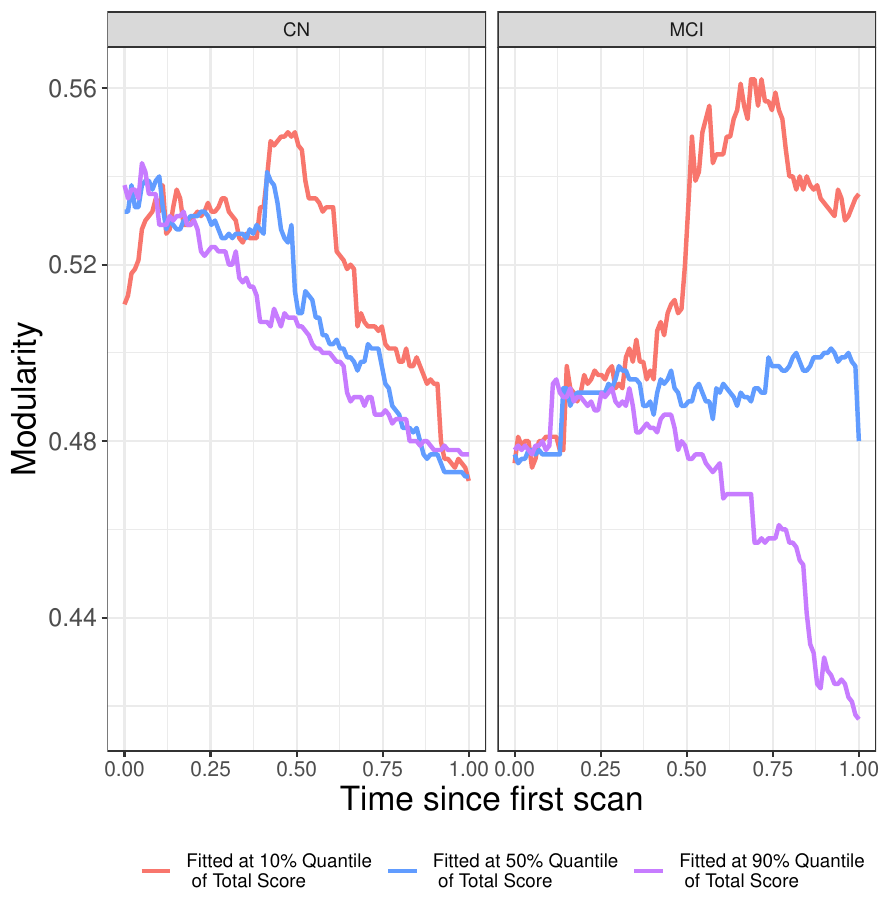}
	\caption{Modularity of the estimated brain networks over time for the CN and MCI subjects in the left and right panels, respectively). The covariate levels at which the networks are estimated are 
		the $10\%$ (red),  $50\%$ (blue)  and $90\%$ (purple) quantiles of the C-score, with the other covariate age fixed at its mean level.}
	\label{Fig:Data:adni:mod:CN_MCI}
\end{figure}

We  also evaluated the modularity of the predicted brain networks continuously over time between time $0$ and $1$ from the estimated PCC matrices on geodesics in the space of correlation matrices, 
see  Figure~\ref{Fig:Data:adni:mod:CN_MCI}. 
The modularity for the highest C-scores  generally is lowest and declines throughout time,   which suggests 
less and more rapidly declining connectivity.  In contrast, modularity 
for low and median C-scores stays stable for a longer period, where the contrast is even higher for MCI subjects. This  indicates that connectivity decline is higher for those starting with higher C-scores and lower connectivity.  

The validity of the fits obtained with the proposed random effects model can be assessed  by its out-of-sample prediction
performance.  We  randomly split the dataset into a training set with sample size $n_{\text{train}}$ and a test set with the remaining $n_{\text{test}}$ subjects.  We then take the fitted objects $\hat{\zeta}_\oplus(\cdot)$ obtained  from the training set and predict the responses in the test set using the covariates present in the test set. As a measure of the efficacy of the fitted model, we  compute the root mean squared prediction error 
\begin{center} $\text{RMPE} = \left[\frac{1}{n_{\text{test}}}\sum_{i=1}^{n_{\text{test}}} \frac{1}{n_i} \sum_{j = 1}^{n_i} d_{P}^2\p{Y^{\text{test}}_{ij}, \hat{Y}^{\text{test}}_{ij}}\right]^{-1/2}$, \end{center} 
where $Y^{\text{test}}_{ij} = Y^{\text{test}}_{ij}(T_{ij})$ denotes the $i^{\text{th}}$ observed response at time $T_{ij}$, $j = 1,\dots,n_i$, for the $i^{\text{th}}$ subject in the test set, $\hat{Y}^{\text{test}}_{ij} = \hat{Y}^{\text{test}}_{ij}(T_{ij})$ denotes the predicted 
object at the covariate level $Z_i$ for the  predictors in the test set, and  $d_P$ the power metric in $\M$, with power $\alpha = 1/2$. We repeat this process $100$ times and compute the  RMPE for each split for $n = 155$ ($n =185$) for the CN (MCI) subjects, separately, with results  in Table~\ref{Tab:rmpe:data:adni}.
\begin{table}[!htb]
		\centering
	\caption{Average Root Mean Prediction Error (RMPE) over $100$ repetitions, as obtained from predicted responses from the proposed two-step method.  Here, $n_{\text{train}}$ and $n_{\text{test}}$ denote the sample sizes for the  split training and testing data for CN and MCI subjects.}
	\label{Tab:rmpe:data:adni}
	
	\centering
	\begin{tabular}{  ccccccc } 
		\hline
		&$n_{\text{train}}$ &  $n_{\text{test}}$ &  First Quartile & Mean & Median & Third Quartile.\\ 
		\hline
		CN&	$100$ & $55$ &  $0.134$ & $0.204$  & $0.194$ & $0.266$\\
		\hline
		MCI& $120$ & $65$ &  $0.139$ & $0.199$  & $0.202$ & $0.271$\\
		\hline
	\end{tabular}
\end{table}
\subsection{Human mortality data: Remaining life distributions as object responses}
\label{sec:data1}
We also analyzed lifetables reflecting remaining life distributions human mortality across $28$ countries correspond to distributional responses, coupled with various country-specific covariates with the proposed random effects model. 
Details can be found in the Supplement. 
\section{Concluding remarks}
We present a novel random/mixed effects modeling framework for longitudinal/repeated measurements data when  data are random objects that  reside  in a geodesic metric space. The model is an extension of classical random effects models. The basic  linearity assumptions in the Euclidean setting become geodesic assumptions for object data; linearity emerges as a special case. 

The  proposed model and approach has two components. The first component  is concerned with   modeling  and implementing  the relation between sparse observations and  the underlying geodesics and  reflects  the subject-specific random effects, in analogy to the classical model, where  the intercept and slope of random regression lines constitute the random effects.  The second component characterizes  the fixed effects that are common to all subjects by incorporating information from external covariates. The connection between the intrinsic geometry of the underlying metric space and conditional Fr\'echet means implemented through Fr\'echet regression is the key to achieve  interpretable estimation with asymptotic convergence guarantees.

\section*{A. Technical assumptions on second-step Fr\'echet regression}
\label{sec:appen2}

In Section~\ref{sec:theory} the final estimates obtained from regressing  object responses $(\hmopi{0},\hmopi{1}) \in (\M \times \M, d_{\M})$ on the  Euclidean predictor $Z_{i} \in \S \subset \real^p$, $p\geq 1$, $i=1,\dots,n$, using model~\eqref{model:accross:subj:final:gfr},   are  $\hat{\zeta}_\oplus(z) = (\hgamz,\hhgamz)^\intercal$,  where
\begin{align}
	\label{step2:est:gfr2}
	\hat{\zeta}_k(z) &= \argminsp M_{n}^{(k)}(\mu),\text{ where } M_{n}^{(k)}(\mu) = \frac{1}{n}\sum_{i=1}^n s_{in}(Z_i,z)d^2\p{\mu, \hmopi{k}}, \ k =0,1, 
\end{align}
with  empirical weights for the GFR estimator as  in~\eqref{model:accross:subj:final2:wt}.
Define the intermediate targets
\begin{align}
	\label{step2:inter:gfr}
	\tilde{\zeta}_k(z) &= \argminsp \tilde{M}_{n}^{(k)}(\mu),\text{ where } \tilde{M}_{n}^{(k)}(\mu) = \frac{1}{n}\sum_{i=1}^n s_{in}(Z_i,z)d^2\p{\mu, \mopi{k}},\ k =0,1, 
\end{align}
where the empirical GFR weights are defined as before.
When  object responses lie on geodesics without error, the GFR paths  recover the underlying geodesic paths 
and 
estimates $\hat{\zeta}_k$ coincide with the  $\tilde{\zeta}_k$ in~\eqref{step2:inter:gfr}
for $k = 0,1$. 
Next  we list  the assumptions required for the theory of GFR~\citep{pete:19} that we adopt for this estimation step.  

\ben[label = (R\arabic*), series = fregStrg, start = 0]
\item \label{ass:gfr:second_step:P0} The objects $\zeta_k(z),\tilde{\zeta}_k(z)$, and $\hat{\zeta}_k(z)$, $k =0,1$, exist and are unique, the latter two almost surely and for any $\eps>0,$ 
\[
\underset{d(\mu,\zeta_k(z))>\eps}{\inf \ } M^{(k)}(\mu,z) - M^{(k)}(\zeta_k(z),z) >0, \ k =0,1.
\]
\item \label{ass:gfr:second_step:P1} For $k = 0,1,$ let $B_{\delta}(\zeta_k(z))$ be the ball of radius $\delta$ centered at $\zeta_k(z)$ and $N(\eps, B_{\delta}(\zeta_k(z)), d)$ be its covering number using balls of size $\eps$. Then
\[
\sqrt{1 + \log N(\eps, B_{\delta}(\zeta_k(z)), d)} d\eps = O(1) \text{ as } \delta \to 0.
\]
\item \label{ass:gfr:second_step:P2} There exist $\tilde{\eta}_k>0, \tilde{C}_k>0,$ possibly depending on $z$, such that $d(\mu,\zeta_k(z))<\tilde{\eta}_k$ implies
\[
M^{(k)}(\mu,z) - M^{(k)}(\zeta_k(z),z) \geq \tilde{C}_k d^2(\mu,\zeta_k(z)), \ k = 0,1.
\]
\een
Assumption~\ref{ass:gfr:second_step:P0} is commonly used  to establish the consistency of an M-estimator such as $\mopi{t}$; see Chapter 3.2 in~\cite{vand:00}. In particular, it ensures that weak convergence of the empirical process $\tilde{M}_{n}$ to the population process $M$ implies convergence of their minimizers. Furthermore, existence follows immediately if $\M$ is compact. The conditions on the covering number in Assumption~\ref{ass:gfr:second_step:P1} and curvature in Assumption~\ref{ass:gfr:second_step:P2} arise from empirical process theory and control the behavior of $\tilde{M}_{n} - M$ near the minimum, which is necessary  to obtain rates of convergence.

\section*{Acknowledgements}
\begin{singlespace}
	Data used in preparation of this article were
	obtained from the Alzheimer;s Disease Neuroimaging Initiative (ADNI) database (\url{adni.l
	oni.usc.edu}). As such, the investigators within the ADNI contributed to the design and
	implementation of ADNI and/or provided data but did not participate in analysis or writing
	of this report. A complete listing of ADNI investigators can be found at: \url{http://adni.lon
	i.usc.edu/wp-content/uploads/how to apply/ADNI Acknowledgement List.pdf}. Data
	collection and sharing for this project was funded by the Alzheimer's Disease Neuroimaging nitiative (ADNI) (National Institutes of Health Grant U01 AG024904) and DOD ADNI
	(Department of Defense award number W81XWH-12-2-0012).
	
\end{singlespace}
\begin{singlespace}
\bibliography{ore0}
\end{singlespace}


\newpage 
\section*{Supplementary Materials}

\section*{S.1. Additional data illustration and simulation results}
\label{suppl:sec:data:simu}
This section provides further illustrations of data applications and simulations. Random objects considered in the additional data demonstrations discussed in this section are univariate probability distributions with compact support endowed with the Wasserstein-2 metric
(applied to human mortality data) and data that reside on the surface of a sphere, endowed with the geodesic distance. Further illustrations of the proposed method include additional plots for the
ADNI study, continuing from Section~\ref{sec:data2} of the main manuscript.

\subsection*{S.1.1. Simulation study: Responses lying on the surface of a sphere}
\label{suppl:sec:simu:sphere}
We next implement our methodology when the responses lie on a Riemannian manifold. In particular, we consider responses lying on the surface of a unit sphere $S^2 \subset \real^3$ with the center being the origin. The geodesic distance between any two points $\omega_1$ and $\omega_2$ lying on the surface of the unit sphere $S^2$ is given by $d_g(\omega_1,\omega_2) = \arccos (\omega_1^\intercal\omega_2).$ We first model the conditional expectation of the end points of the underlying subject-specific geodesic, conditional on the covariates $Z$, as 
\begin{align}
	\label{sim:model:sphe}
	&\expect{\p{\nu_{ik}|Z_i =z, T_{ij}= u}} =\xi_{u, z}\nonumber\\
	=& (\sqrt{(1-z^2)}\cos(\pi u),\sqrt{(1- z^2)}\sin(\pi u), z),\  z\in (0,1),\ k = 0,1,\ j = 1,n_i.
\end{align}
The above quantifies the true time-varying regression function conditional on the baseline covariates.
In order to generate random realizations of the end-points according to model~\eqref{sim:model:sphe}, we first sample the time points at which the repeated measurements are made for each subject, denoted by $T_{ij}$, according to a sparse or a dense design as before (see Section~\ref{sec:simu} in the main manuscript) such that $T_{ij} \in [0,1].$ Further, the baseline covariates $Z_i$ are generated i.i.d. from  $Unif(0,1)$ for $j=1,\dots,n_i,$ $i=1\dots,n.$ 

The true responses on the surface of the sphere $S^2$ at the two end points of the underlying geodesic corresponding to the $i^{\text{th}}$ individual, for $i = 1,\dots,n$, are then constructed as follows. A bivariate noise random vector is generated  on the tangent space $T_{\geodi{u,z}}(\Omega).$ 
To this end, we define, for $j = 1, n_i$, $\psi_{ij} = \arcsin (T_{ij})$ and $\theta_{ij} = \pi T_{ij}.$ An
orthonormal basis for the tangent space is denoted by $(b_{ij}^{(1)},b_{ij}^{(2)}),$ where
$b_{ij}^{(1)} = (\cos(\psi_{ij}) \cos(\theta_{ij}), \cos(\psi_{ij}) \sin(\theta_{ij}), -\sin(\psi_{ij}))^\intercal$ and $b_{ij}^{(2)} = (\sin(\theta_{ij}), -\cos(\theta_{ij}),0)^\intercal.$
Adding a noise level $\sigma^2 = 0.2$,  bivariate random vectors $A_{ij} = c_{i1}b_{ij}^{(1)} + c_{i2}b_{ij}^{(2)}$ are computed, where $C_i = (c_{i1},c_{i2})^\intercal \overset{i.i.d.}{\sim} N_2(0, \sigma^2I_2)$. Finally, the responses are generated as
\[
\nu_{ik} = \cos\left( \lVert A_{ij}\rVert_E\right) \zeta_{T_{ij},Z_i} + \sin \left( \lVert A_{ij}\rVert_E\right) \frac{A_{ij}}{\lVert A_{ij}\rVert_E},\ j =1,n_i, \ k =0,1,
\]
with $\lVert \cdot \rVert_E$ being the Euclidean norm. The simulation steps above produce a point $\nu_{ik}$ on the surface of the two-dimensional sphere at the endpoints $k = 0,1$ of some underlying  geodesic paths on the surface of the sphere.
To complete this step, the geodesic path connecting $\nu_{i0}$ and $\nu_{i1}$ is given by 
$
t \mapsto  \smash{\frac{1}{\sin\omega}}[\nu_{i0}\sin((1-t)\omega) + \nu_{i1} \sin(t\omega)], \ t \in [0,1],
$
where $\omega = \arccos(\nu_{i0}^\intercal\nu_{i1})$.

Now, the observable noisy responses are obtained by adding a small perturbation to the random end-points on the geodesic. To this end, we represent any point on the surface of the sphere in spherical coordinates and add noise to the angle the point makes with the $z-$ axis.
A point $P$ on the surface of the sphere given by $P = (\rho \sin \phi\cos\theta, \rho \sin\phi\sin\theta, \rho \cos\phi)$, where $\rho$ is the distance from $P$ to the origin, $\theta$ is the angle between the positive x-axis and the line segment from the origin to the projection of $P$ to the $xy-$plane, and $\phi$ is the angle between the positive $z-$axis and the line segment from the origin to $P$.
A noisy observation around $P$ with a perturbation level $\alpha_n$ is generated as $P' := (\rho \sin (\phi + \eps)\cos\theta, \rho \sin (\phi  +\eps)\sin\theta, \rho \cos(\phi + \eps))$, where $\eps = \pm \alpha_n$ with equal probability $1/2$. For this perturbation scheme, the perturbed point $P'$ has norm $\rho^2$, i.e., $P'$ still lies on the surface of the sphere $S^2$. Further, $\expect{(d_g^2(P,P'))} = \arccos(P^\intercal P'/\rho^2) = \alpha_n^2 \to 0$ as $n\to \infty$. Thus, using the polar coordinate representation of every point generated on the geodesic, the noisy observations are procured as described above.

The simulation study is then carried out for different sample sizes $n=50, 400,$ and $1000$, for both sparse and dense designs, while fixing the noise level at $\alpha =0.1$. A measure of the efficacy for the fits is constructed as the Integrated Squared Error (ISE) over $500$ Monte Carlo simulation runs as
\begin{align}
	\label{sim:mise:sphe}
	\text{ISE}_r = \int_{z \in \S} \int_{t \in [0,1} d_g (Y^r(t,z), \hat{Y}^r(t,z) dt dz,
\end{align}
where $Y^r(t,z)$ and $\hat{Y}^r(t,z)$ denote, respectively, the true object on the two-dimensional sphere, lying on a geodesic (without perturbation), and the estimated object at time point $t$ and covariate value $z$ for the $r^{\text{th}}$ simulation run, where $r = 1,\dots,500.$ Here $d_g$ denotes the geodesic distance between two points on a sphere and is given by
\[
d_g(A,B) = \arccos (A^\intercal B).
\]
where $A$ and $B$ are two points on the surface of a sphere.
\begin{figure}[!htb]	
	\centering
	\includegraphics[width=.5\textwidth]{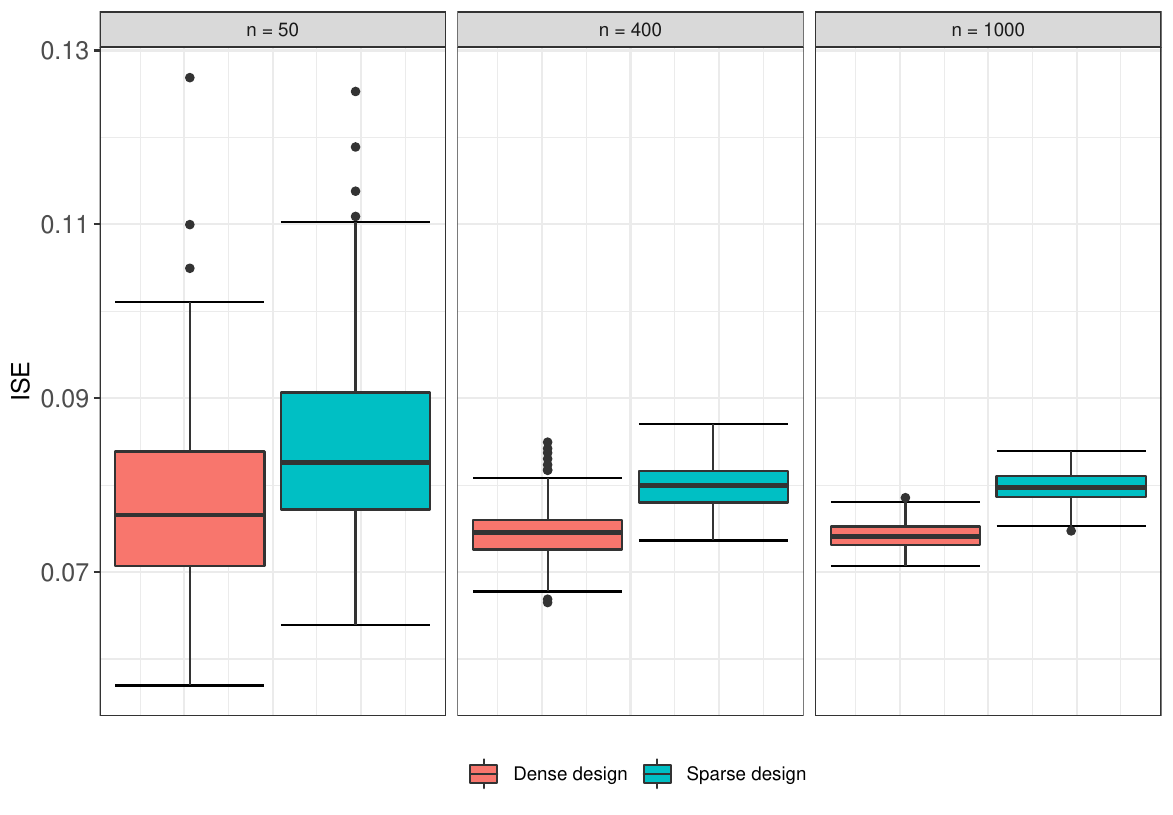}
	\centering
	\caption{Boxplots of Mean Integrated Squared Errors (MISE) calculated as per~\eqref{sim:mise:sphe}, over $500$ simulation runs and different sample sizes for object  responses situated on the surface of a 2-dimensional sphere, corresponding to setting I.}
	\label{Fig:Sim:Boxplot3}
\end{figure}	
Figure~\ref{Fig:Sim:Boxplot3} shows that with a denser design and higher sample size, the ISE reduces significantly, giving evidence for the asymptotic convergence of the estimates to the true underlying object responses.

\subsection*{S.1.2. Data analysis: Remaining life distributions as object responses }
\label{suppl:sec:data1}
The Human Mortality Database (\url{https://www.mortality.org/}) provides yearly life table data for males and females and various countries.  Here we study the time-varying association between remaining life distribution and various socioeconomic indices at the country level and consider the life tables for females over $30$ calendar years, $1990-2019$,  for  $n = 28$ countries. We consider the remaining life distribution $R(t) = P(T \le t | T \ge 75)$ as responses,  where $T$ denotes age-at-death and the remaining life distribution is  considered on the interval  $[75,120]$ (all in years).  This remaining life distribution and its density can be easily obtained from the available lifetable data that correspond to histograms with bin width one year by adding a smoothing step, for which we used the R package \emph{frechet}~\citep{fr:package}
with bandwidth 2 years. 

We then obtained  a sample of time-varying univariate probability distributions which are the responses 
for  $n = 28$ countries, where the time axis represents $30$ calendar years from $T:= [1990,2019]$ and the observation made at each calendar year for each country corresponds to the remaining life distribution over the age interval $[75,120]$.

For the first-step regression, we fit model~\eqref{gfr:estd:first_step} to obtain the estimates for the remaining life distribution at the first (time $0$, corresponding to the year $1990$) and last (time $1$, corresponding to the year $1990$) point of our time domain. The inherent assumption is that, for each country, the remaining life distributions over the years are observed around some geodesic in the Wasserstein-2 space of distributions with small error/perturbation, where the underlying geodesic connects the two distribution objects corresponding to  time $0$ and time $1$.
The fitted responses $\hmopi{0}$ and $\hmopi{1}$ are then treated  as a summary of the time-varying remaining life distributions for the $i^{\text{th}}$ country, $i=1,\dots,28,$ and are carried forward as the paired distributional response to the second-step regression as per~\eqref{model:accross:subj:est:final2}. 

For implementing the second step regression as per model~\eqref{model:accross:subj:final2}, we consider a $4-$ dimensional baseline covariate for each country, where the covariates for the  $i^{\text{th}}$ country represent (1) Unemployment rate (\% of the total labor force)  
(2) Fertility Rate (Births per women), (3) GDP per capita- International purchasing power parity, and (4) Population growth (annual \%),  measured in the calendar year $1990$. The data is obtained from the World Bank Database at \url{https://data.worldbank.org}. Our aim is to quantify the effects of this baseline/ external covariate, possibly  changing over the calendar years, on the remaining life distributions. The second-step regression with the paired object responses $(\hmopi{0},\hmopi{1})$  and Euclidean covariates $Z_i$, $i =1 \dots,n$, produces the fitted objects $\hat{\zeta}_\oplus(z) = (\hgamz,\hhgamz)^\intercal$ over varying values $Z = z$. 

\begin{figure}[!htb]	
	\centering
	\includegraphics[width=.6\textwidth]{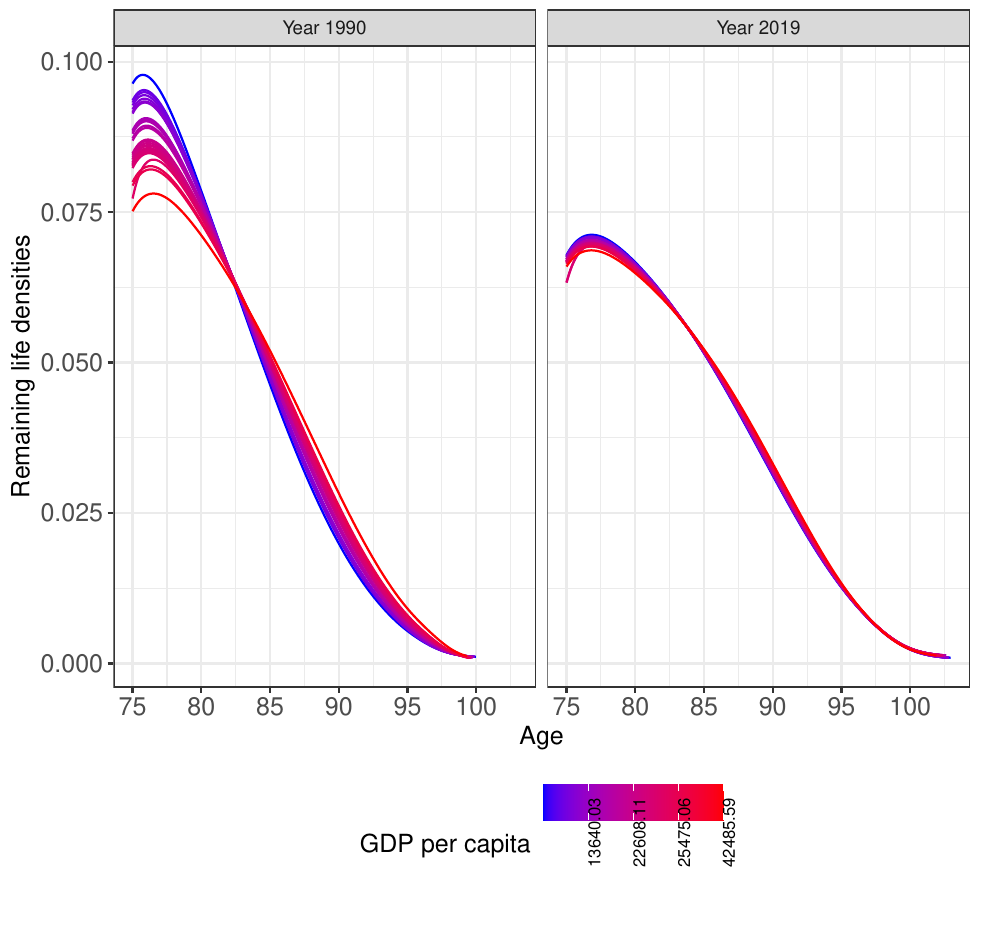}
	\centering
	\caption{Effect of the covariate GDP per Capita, at the beginning and end of the time domain. The changes in density of the remaining life distribution after age 75 as GDP per Capita rate ranges from low (blue) to high (red) are displayed when the other predictors are fixed at their mean level. The left and right panels show the fits at the calendar years $1990$ and $2019$, respectively.}
	\label{Fig:Data:Life:year01:pred1}
\end{figure}

It is of interest to see how the estimated distributions at times $0$ and $1$ given by $\hgamz$ and $\hhgamz$, respectively, change over varying levels of the baseline covariate $Z$. Here $Z$ is a $4$-dimensional predictor. To elicit the effect of each component of $Z$, we vary the levels of that component from low to high while keeping the other three components fixed at their mean level.
For example, Figure~\ref{Fig:Data:Life:year01:pred1} illustrates  how the remaining-life density changes with  increasing levels of GDP per capita, while the other three predictors are kept fixed at their mean levels. The left and right panels display the fitted densities for the calendar years $1990$ and $2019$ respectively.
The fitted densities are color coded such that blue to red indicates smaller to  larger value of GDP.
We find  that smaller values of GDP are associated with left-shifted remaining life distributions, while a larger GDP value corresponds to a shift of the mode of the age-at-death toward the right. Further, the densities for the year $1990$ are more left-skewed than the ones for $2019$, indicating an increasing right shift of the remaining life
distribution as calendar time progresses. The time effect and GDP effect are seen to be not simply additive but the GDP effect is more pronounced in 2019 than in 2010.   

\begin{figure}[!htb]	
	\centering
	\includegraphics[width=.6\textwidth]{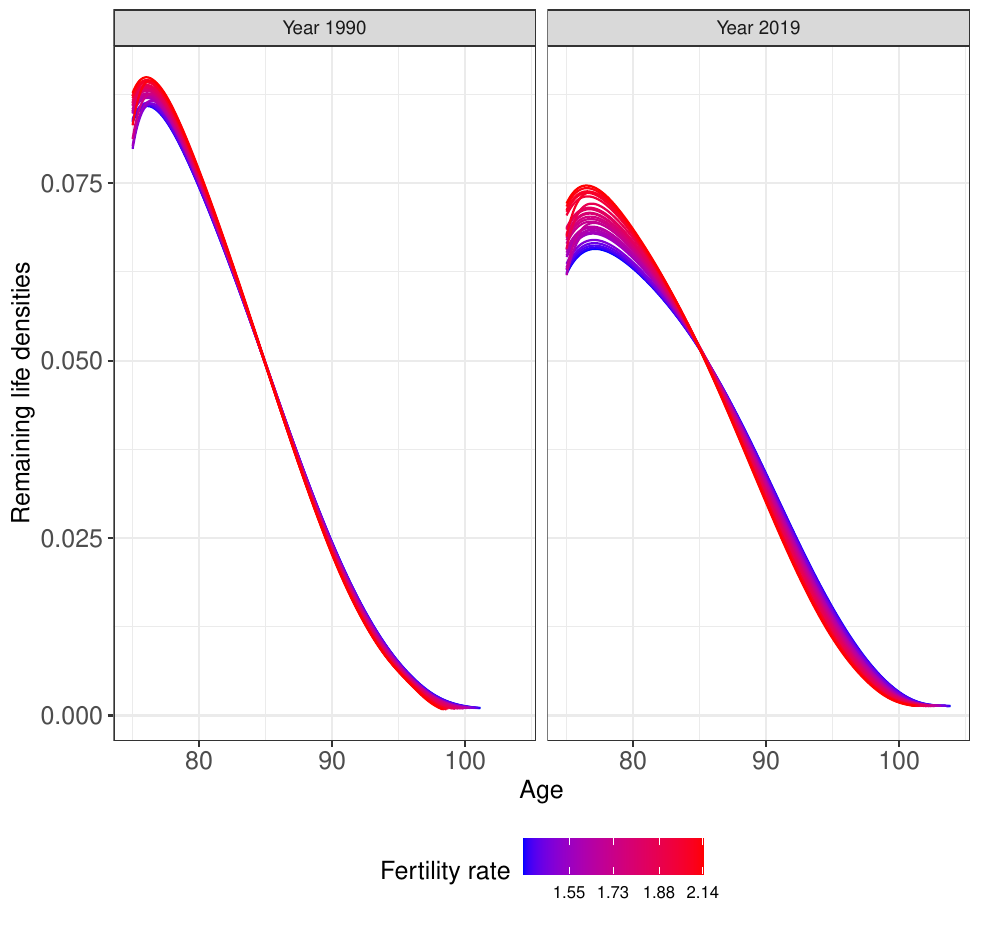}
	\centering
	\caption{Effect of the covariate Fertility rate, at the beginning and end of the time domain. The changes in density of the remaining life distribution after age 75 as Fertility rate ranges from low (blue) to high (red) are displayed when the other predictors are fixed at their mean level. The left and right panels show the fits at the calendar years $1990$ and $2019$, respectively.}
	\label{Fig:Data:Life:year01:pred2}
\end{figure}

\begin{figure}[!htb]	
	\centering
	\includegraphics[width=.6\textwidth]{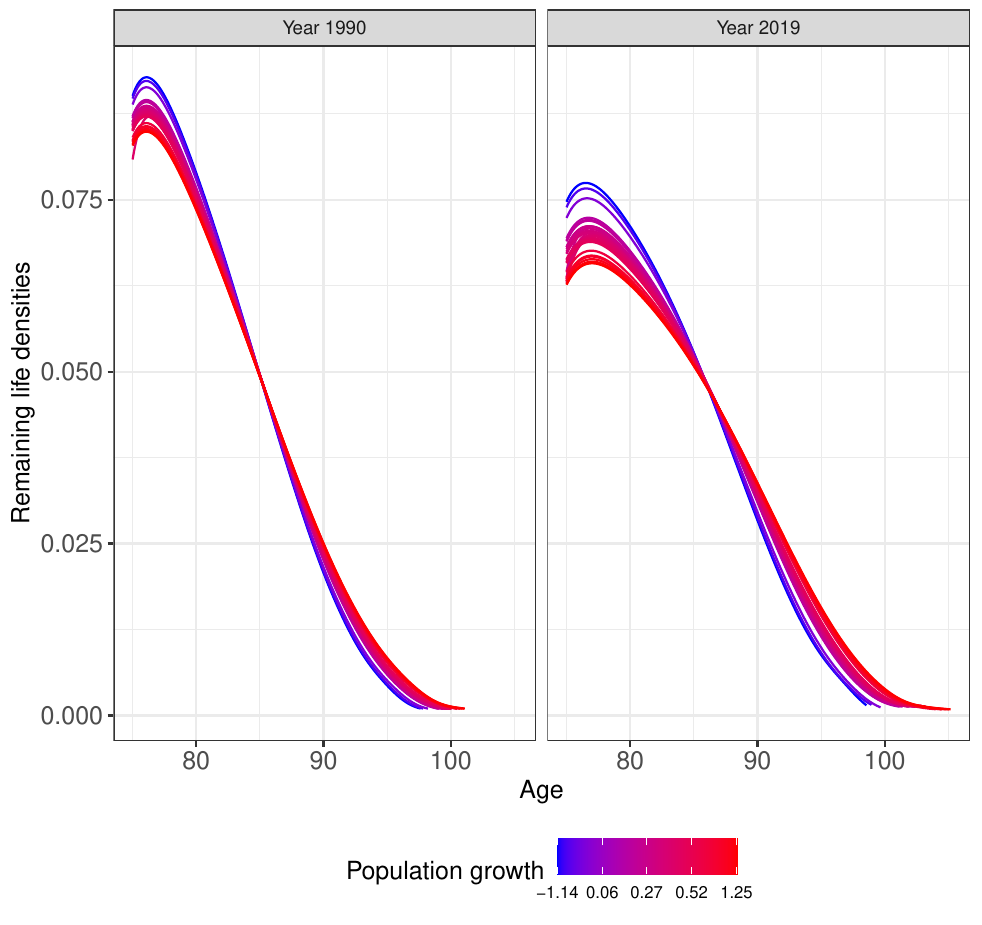}
	\centering
	\caption{Effect of the covariate Percentage of Population Growth, at the beginning and end of the time domain. The changes in density of the remaining life distribution after age 75 as Percentage of Population Growth   ranges from low (blue) to high (red) are displayed when the other predictors are fixed at their mean level. The left and right panels show the fits at the calendar years $1990$ and $2019$, respectively.}
	\label{Fig:Data:Life:year01:pred3}
\end{figure}

\begin{figure}[!htb]	
	\centering
	\includegraphics[width=.6\textwidth]{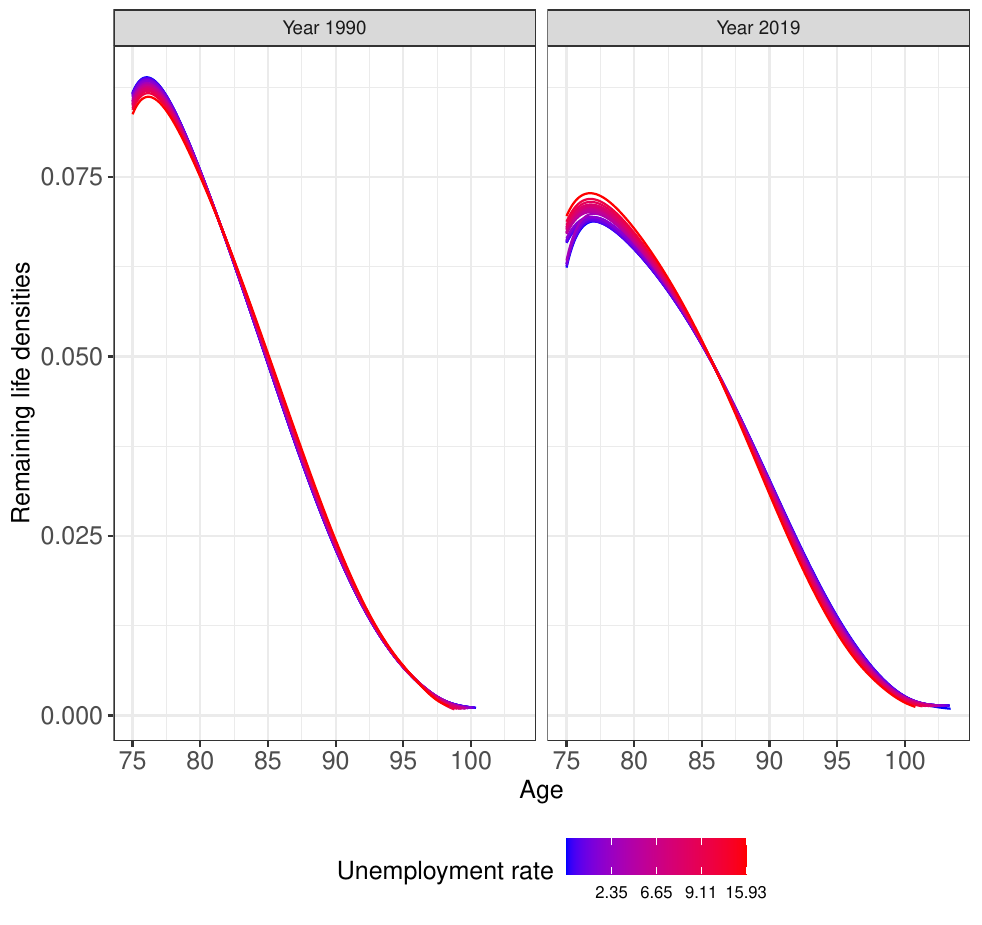}
	\centering
	\caption{Effect of the covariate Unemployment Rate, at the beginning and end of the time domain. The changes in density of the remaining life distribution after age 75 as Unemployment Rate    ranges from low (blue) to high (red) are displayed when the other predictors are fixed at their mean level. The left and right panels show the fits at the calendar years $1990$ and $2019$, respectively.}
	\label{Fig:Data:Life:year01:pred4}
\end{figure}

For increasing levels of the fertility rate, unemployment rate, and population growth percentage, similar patterns for the time-varying effect of these covariates are observed, but to a lesser extent (See Figures~\ref{Fig:Data:Life:year01:pred2},~\ref{Fig:Data:Life:year01:pred3}, and~\ref{Fig:Data:Life:year01:pred4}, respectively).

\begin{figure}[!htb]	
	\centering
	\includegraphics[width=.6\textwidth]{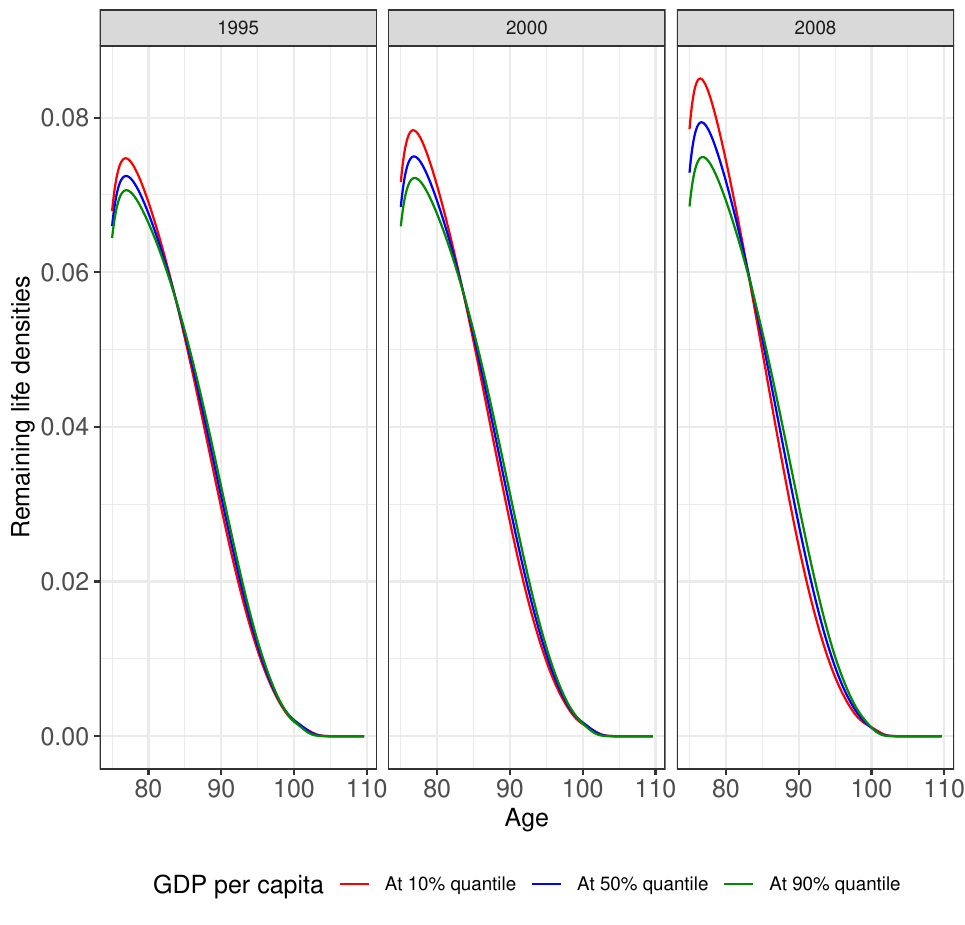}
	\centering
	\caption{Effect of the predictor GDP per Capita, evaluated at points on the fitted geodesic. The left, middle, and right panels show the fits for the years $1995, 2000$, and $2008$ respectively, where the remaining life densities are fitted at the $10\%$ (red), $50\%$ (blue)  and $90\%$ (green) quantile  levels of GDP per Capita, while the other predictors are fixed at their mean levels.}
	\label{Fig:Data:Life:years:pred1}
\end{figure}

For each country, the fitted geodesics in the Wasserstein space of distributions  summarize the time dynamics of the remaining life distributions along with the effects of the covariates.
We further demonstrate the interpretability of the proposed random effects model by displaying  the fits at the beginning and end of the time domain when varying the value of one predictor at the  $10\%, 50\%,$ and $90\%$ quantile levels, while keeping the other two predictors fixed at their mean. We then compute the estimated densities situated on the fitted geodesic in the distribution space corresponding to a grid of time points in $[1990,2019]$. The left, middle, and right panels of Figure~\ref{Fig:Data:Life:years:pred1} display the estimated densities at the calendar years $1995$, $2000$, and $2008$, respectively. For each panel, the red, blue, and green lines correspond to the $10\%$, $50\%$, and $90\%$ quantile values for GDP per Capita, while the other three predictors  
are kept fixed at their mean levels. We observe a shift in the remaining life densities towards the right  over the years, thus indicating improved remaining survival  as calendar time progresses. 

\begin{figure}[!htb]	
	\centering
	\includegraphics[width=.6\textwidth]{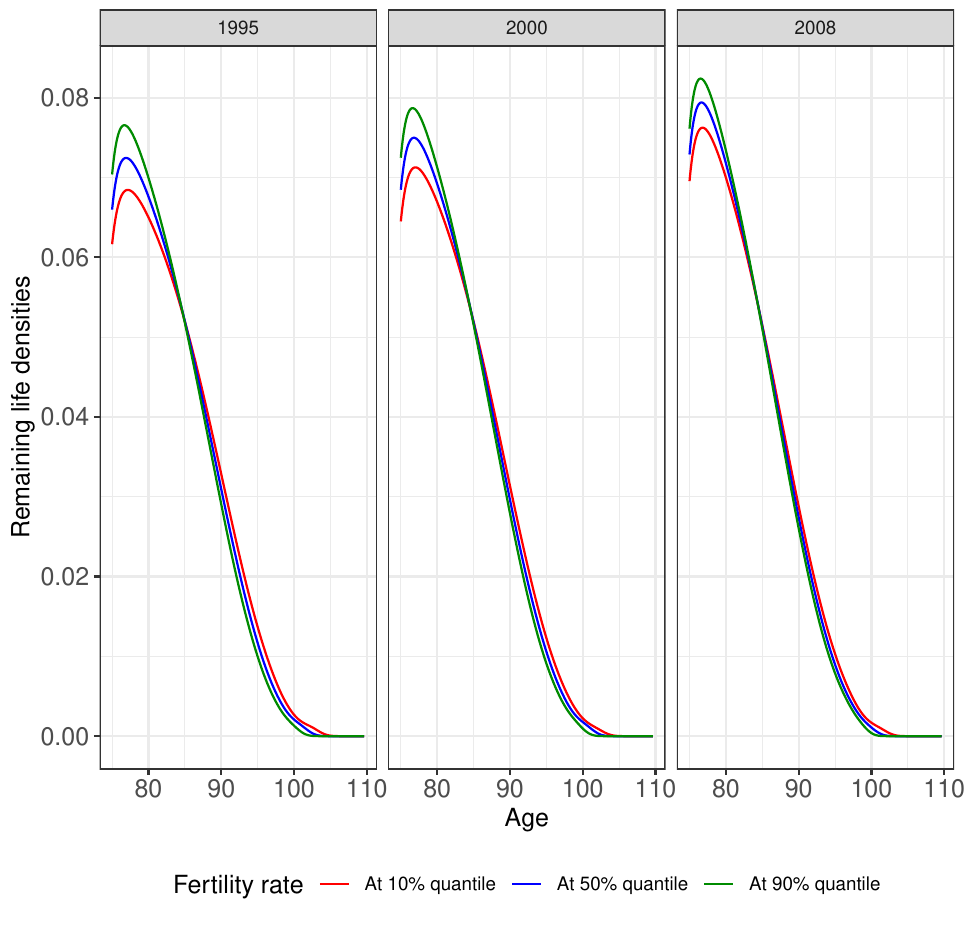}
	\centering
	\caption{Effect of the predictor Fertility Rate, evaluated at points on the fitted geodesic. The left, middle, and right panels show the fits for the years $1995, 2000$, and $2008$ respectively, where the remaining life densities are fitted at the  $10\%$ (red), $50\%$ (blue) and $90\%$ (green) quantile  levels of Fertility rate, while the other predictors are fixed at their mean levels.}
	\label{Fig:Data:Life:years:pred2}
\end{figure}

\begin{figure}[!htb]	
	\centering
	\includegraphics[width=.6\textwidth]{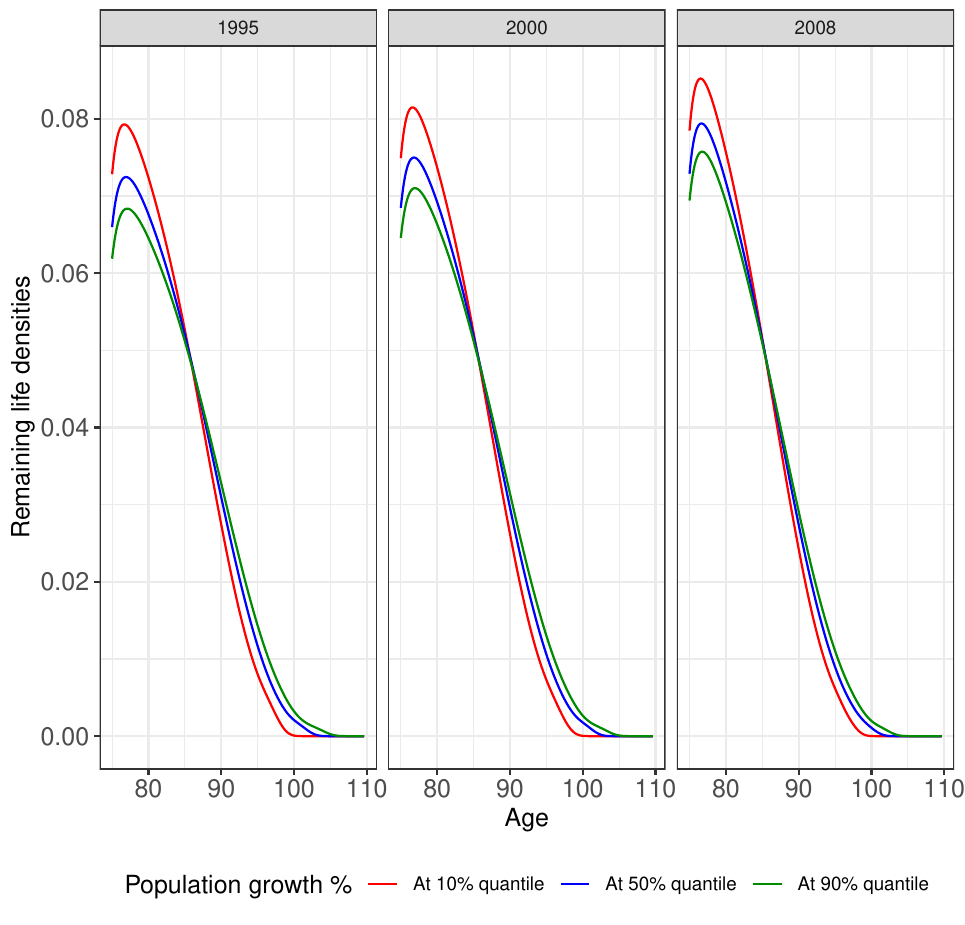}
	\centering
	\caption{Effect of the predictor Percentage of Population Growth, evaluated at points on the fitted geodesic. The left, middle, and right panels show the fits for the years $1995, 2000$, and $2008$ respectively, where the remaining life densities are fitted at the $10\%$ (red), $50\%$ (blue)  and $90\%$ (green) quantile levels of Population Growth, while the other predictors are fixed at their mean levels.}
	\label{Fig:Data:Life:years:pred3}
\end{figure}

\begin{figure}[!htb]	
	\centering
	\includegraphics[width=.6\textwidth]{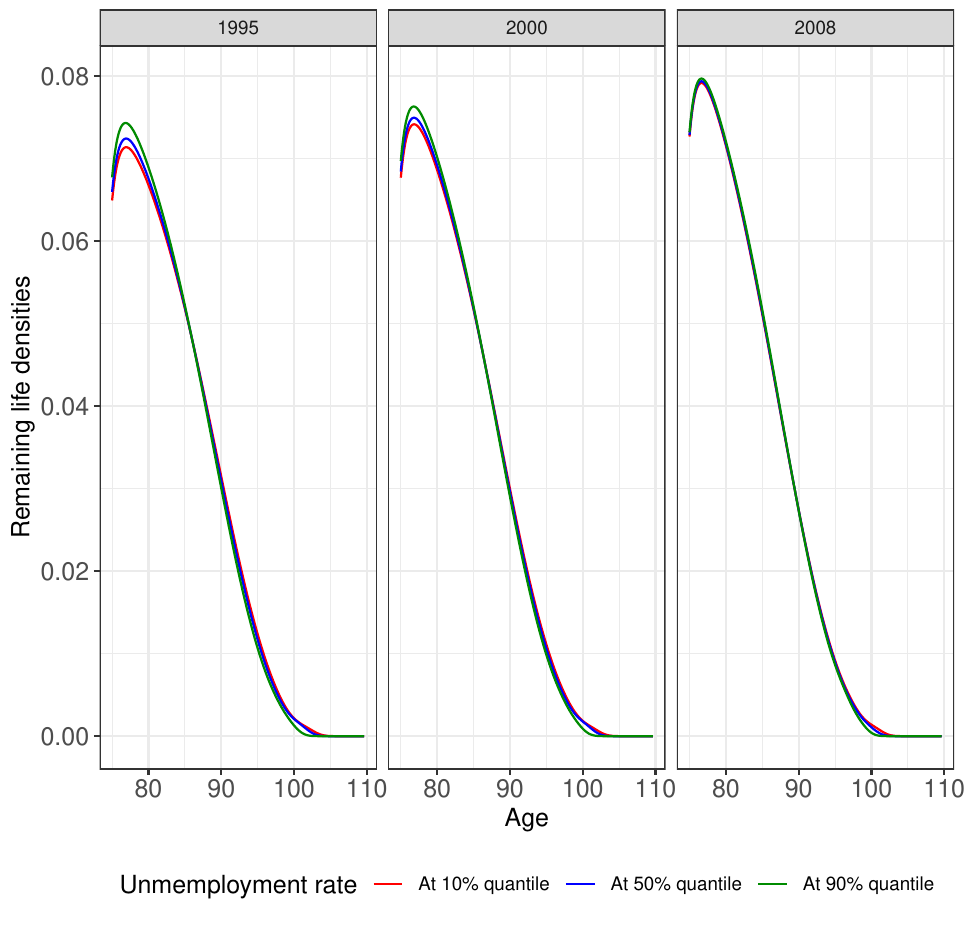}
	\centering
	\caption{Figure showing the effect of the baseline predictor- Unemployment rate, evaluated at points on the geodesic. The left, middle, and right panels show the fits for the years $1995, 2000$, and $2008$ respectively, where the remaining life densities are fitted at the $10\%, 50\%$, and $90\%$ quantile levels of the Unemployment rate (shown in red, blue, and green curves, respectively), while the other predictors are fixed at their mean levels.}
	\label{Fig:Data:Life:years:pred4}
\end{figure}

Similar interpretations emerge  for the other three predictors  from the patterns displayed in Figure~\ref{Fig:Data:Life:years:pred2},~\ref{Fig:Data:Life:years:pred3}, and~\ref{Fig:Data:Life:years:pred4}. We observe that a higher value of the covariate levels is generally associated with  right-shifted remaining life distribution,

\begin{figure}[!htb]	
	\centering
	\includegraphics[width=.8\textwidth]{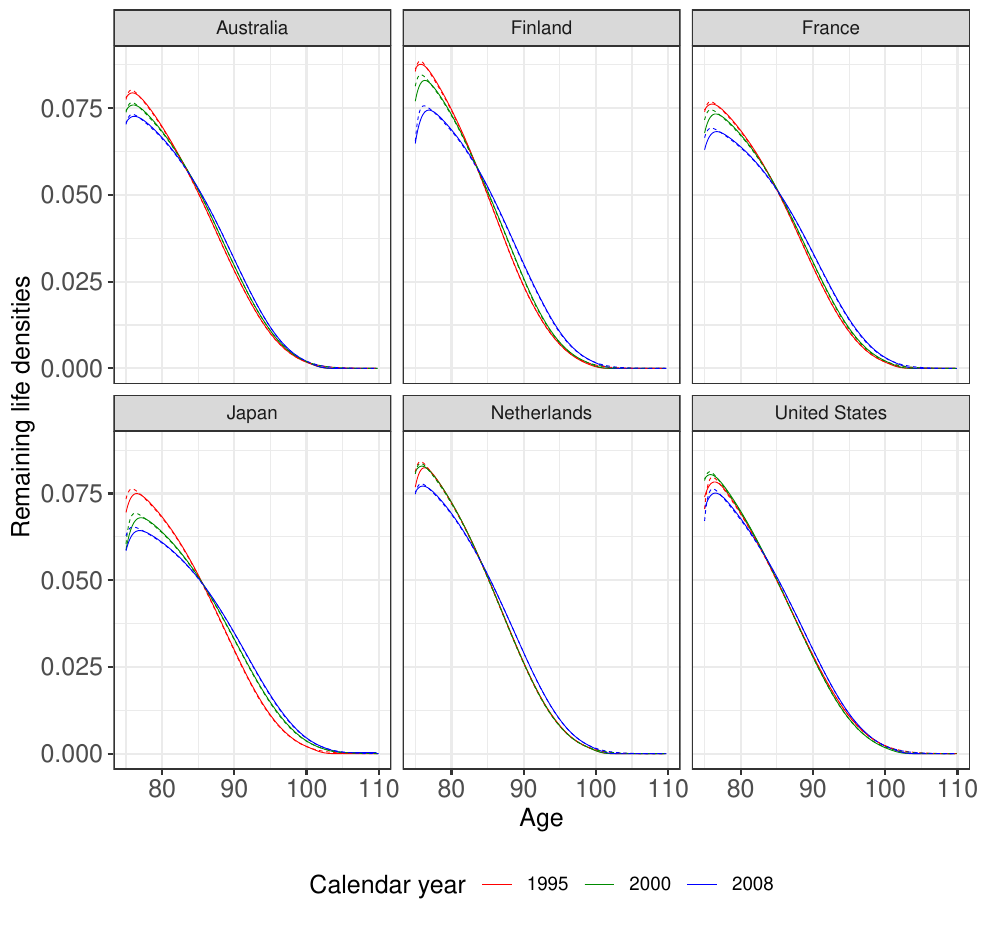}
	\centering
	\caption{Figure displaying the observed and estimated remaining life distributions, represented as densities for a select few countries over a few selected calendar years. The panels, clockwise from top-left, correspond to the countries- Australia, Finland, France, United States, Netherlands, and Japan. In each panel, the red, blue, and green lines show the densities at the calendar years $1995$, $2000$, and $2008$, respectively. The observed and predicted densities are shown in solid and dashed lines, respectively, the estimated densities being computed at the observed covariate values.}
	\label{Fig:Data:Life:fits}
\end{figure}

To summarize, in Figure~\ref{Fig:Data:Life:fits}, we illustrate the observed densities for the remaining life distributions for a few selected countries over three selected calendar years, along with the densities predicted at the observed baseline-covariate values for that country. The six panels, clockwise from top-left, correspond to Australia, Finland, France, United States, Netherlands, and Japan; while  red, green, and blue colors indicate  the calendar years $1995$, $2000$, and $2008$, respectively. The observed and predicted densities are plotted in solid and dashed lines for each country and each calendar year, and follow the same temporal pattern. The fits are close to the observations, thus giving evidence for the validity of the model. The small discrepancies in the estimated-vs-observed densities towards the beginning and end of the domain could be caused by boundary effects of the regression fits.

Finally, the performance of the fits is measured by the out-of-sample prediction performance of the proposed method. For this, we first randomly split the dataset into a training set with sample size $n_{\text{train}} = 18$ and a test set with the remaining $n_{\text{test}} =10$ subjects (countries). We then consider the fitted objects $\hat{\zeta}_\oplus(\cdot)$ obtained  from the training set and predict the responses in the test set using the covariates present in the test set. As a measure of the efficacy of the fitted model, we  compute the root mean squared prediction error as
\begin{align}
	\label{rmpe:frob}
	\text{RMPE} = \left[\frac{1}{n_{\text{test}}}\sum_{i=1}^{n_{\text{test}}} \frac{1}{n_i} \sum_{j = 1}^{n_i} d_{W}^2\p{Y^{\text{test}}_{ij}, \hat{Y}^{\text{test}}_{ij}}\right]^{-1/2},
\end{align}
where $Y^{\text{test}}_{ij} = Y^{\text{test}}_{ij}(T_{ij})$ denotes the $i^{\text{th}}$ observed response at time $T_{ij}$, $j = 1,\dots,n_i$, for the $i^{\text{th}}$ subject in the test set, $\hat{Y}^{\text{test}}_{ij} = \hat{Y}^{\text{test}}_{ij}(T_{ij})$ denotes predicted object for the second-step  fits at the covariate level $Z_i$ for the  predictors in the test set. $d_W$ denotes the Wasserstein-2 metric in the space of distribution objects. We repeat this process $500$ times and compute RMPE for each split for $n = 28$ countries. separately. The summary of the RMPE is shown in Table~\ref{Tab:rmpe:dat}.

\begin{table}[!htb]
	\centering
	\caption{Average Root Mean Prediction Error (RMPE) over $500$ repetitions, as obtained from predicted responses from the proposed two-step method.  Here, $n_{\text{train}}$ and $n_{\text{test}}$ denote the sample sizes for the  split training and testing datasets respectively.}
	\centering
	\begin{tabular}{  cccccc } 
		\hline
		$n_{\text{train}}$ &  $n_{\text{test}}$ &  First Quartile & Mean & Median & Third Quartile.\\ 
		\hline
		$18$ & $10$ &  $0.2418 $ & $0.3196$  & $0.2935$ & $0.3656$\\
		\hline
	\end{tabular}
	\label{Tab:rmpe:dat}
\end{table}

\subsection*{S.1.3. ADNI data}
\label{suppl:sec:data2}
Continuing from Section~\ref{sec:data2} in the main manuscript, we illustrate the network structure of the fitted Pearson correlation connectivity (PCC) matrices for CN and MCI subjects. The PCC matrices serve as responses residing in the space of correlation matrices equipped with the power Euclidean metric with  power $\alpha = 1/2$, coupled with baseline covariates taken as age and C-score over a time window $[0,1]$, since the first available scan. 

\begin{figure}[!htb]
	\centering 
	\begin{subfigure}{0.25\textwidth}
		\includegraphics[width=\linewidth]{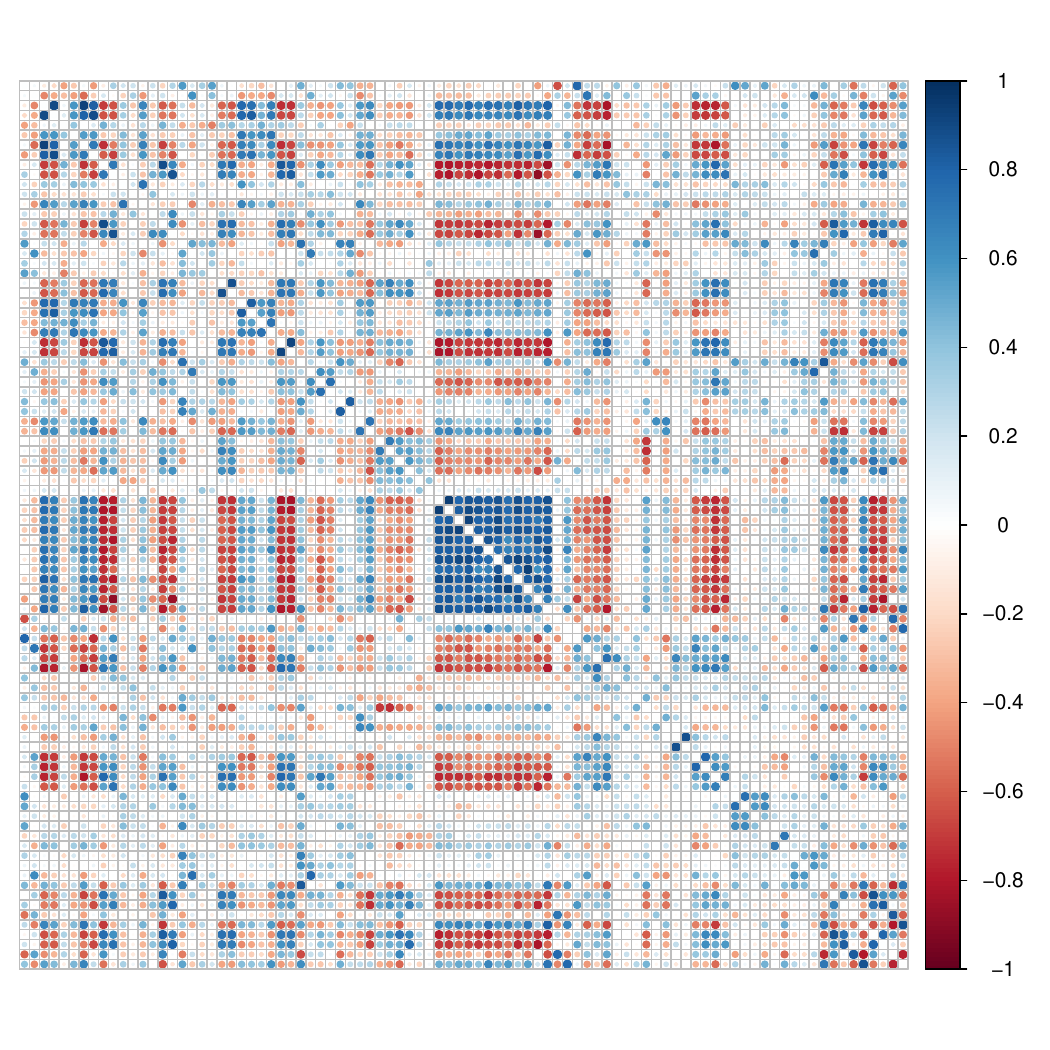}
	\end{subfigure}\hfil 
	\begin{subfigure}{0.25\textwidth}
		\includegraphics[width=\linewidth]{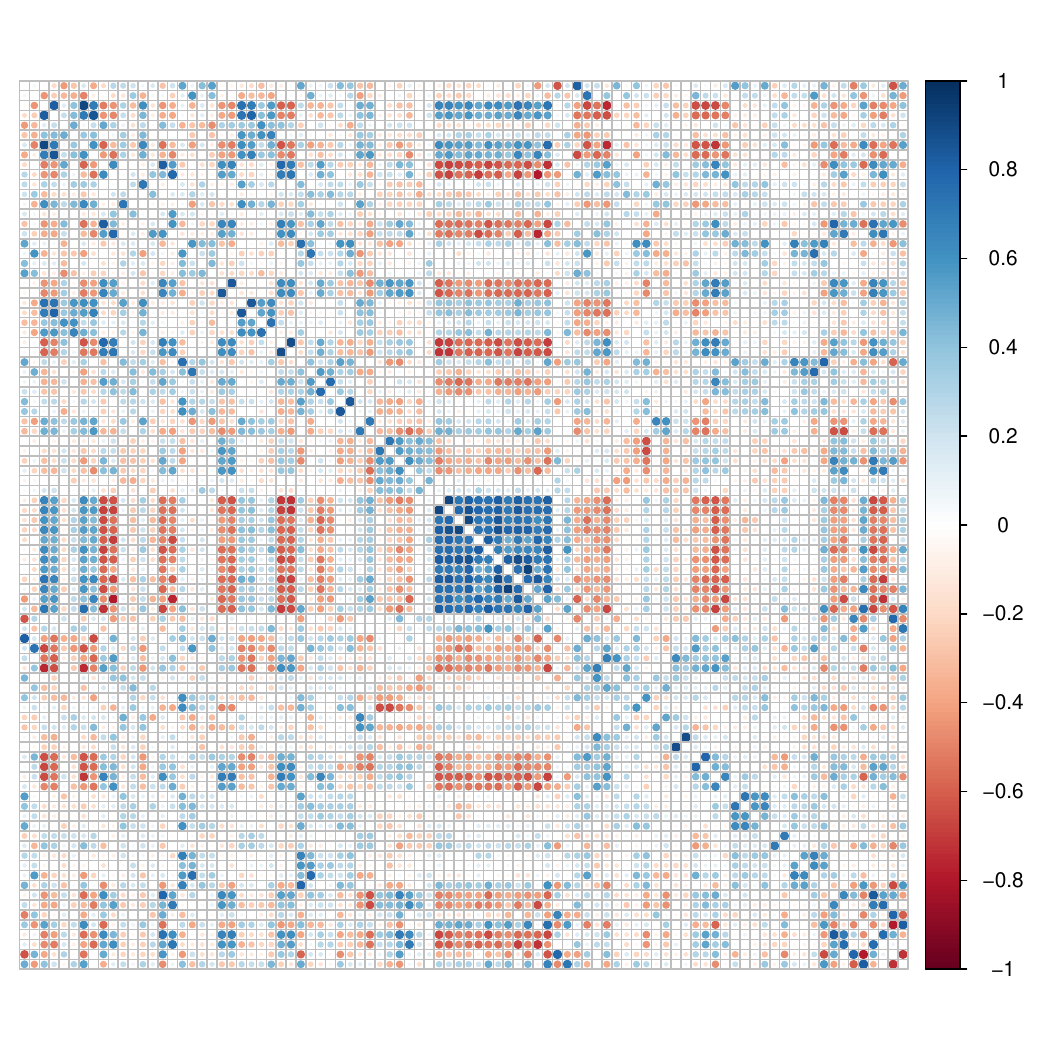}
	\end{subfigure}\hfil 
	\begin{subfigure}{0.25\textwidth}
		\includegraphics[width=\linewidth]{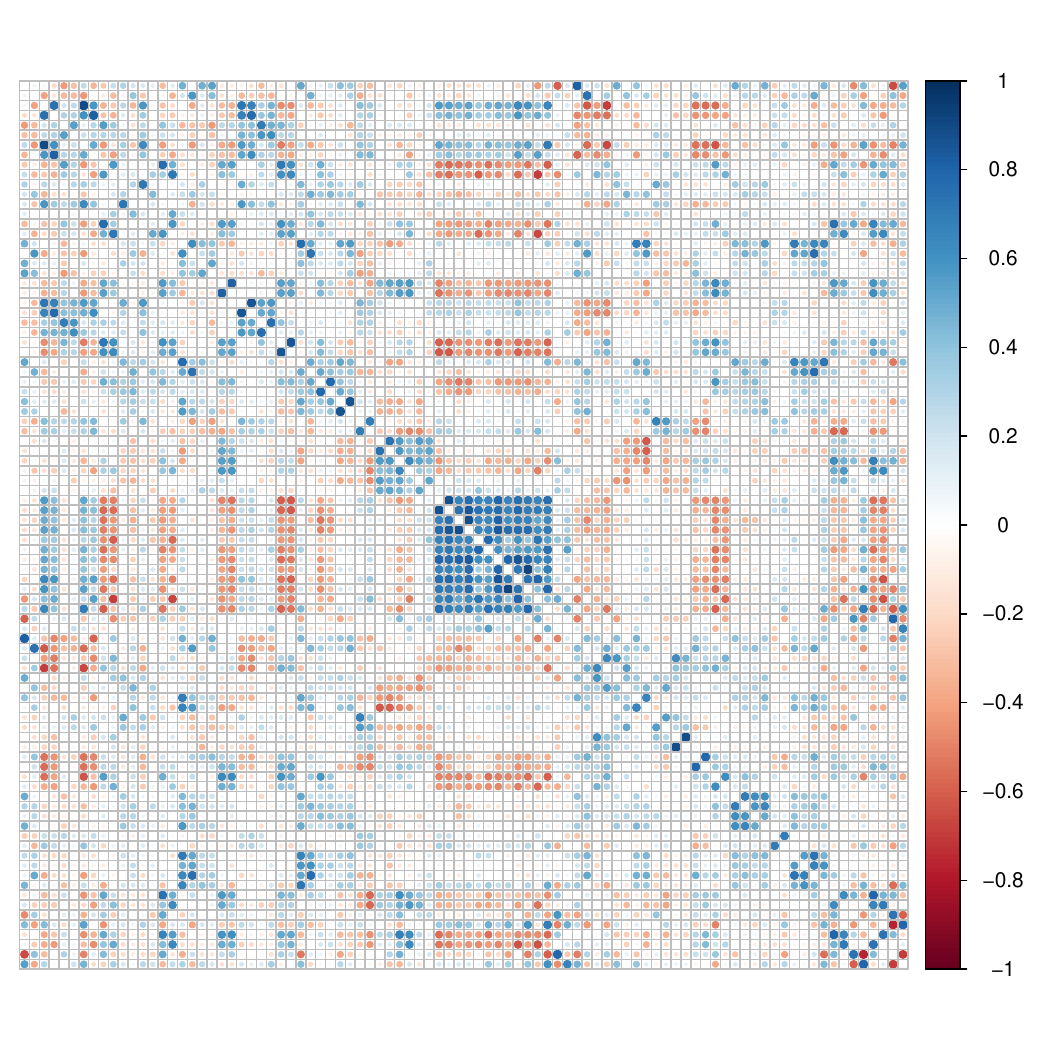}
	\end{subfigure}
	
	\medskip
	\begin{subfigure}{0.25\textwidth}
		\includegraphics[width=\linewidth]{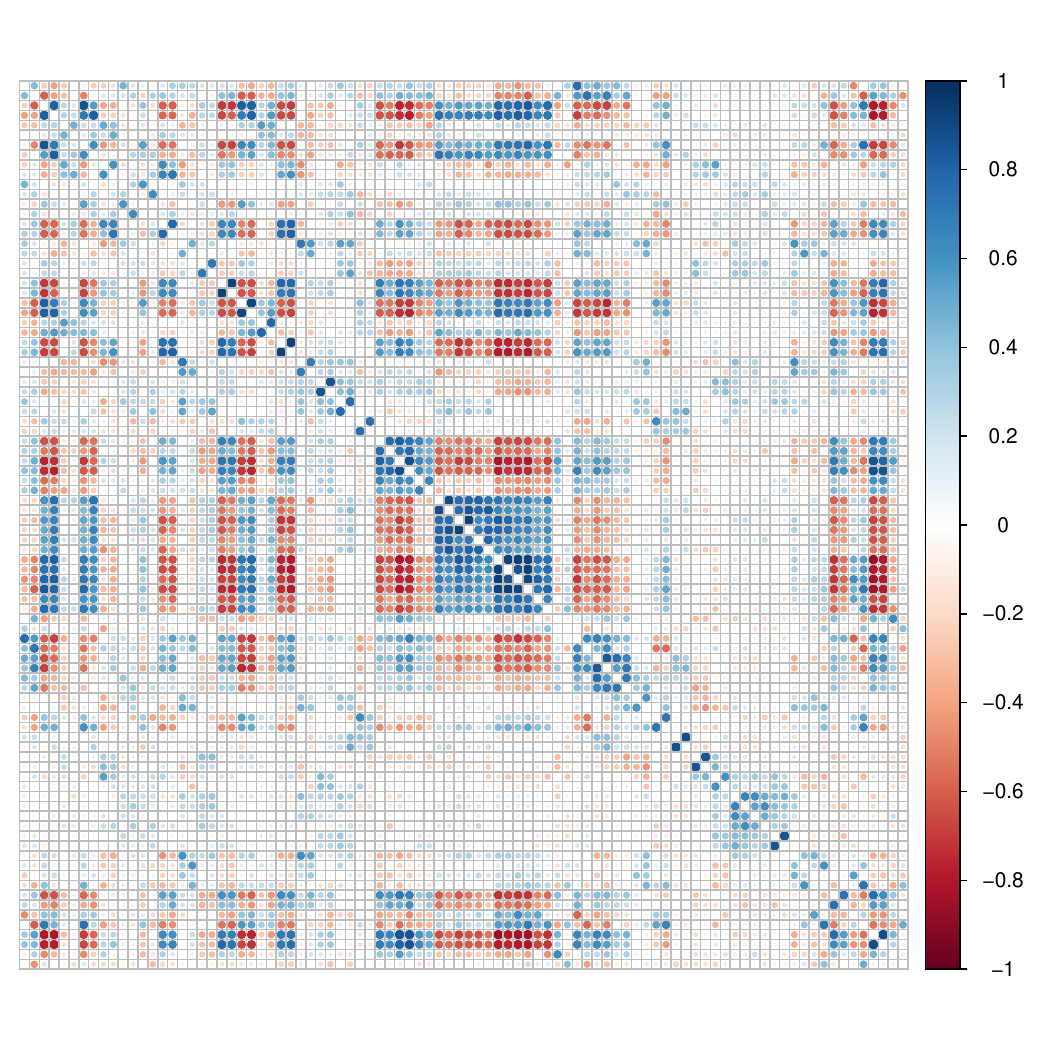}
	\end{subfigure}\hfil 
	\begin{subfigure}{0.25\textwidth}
		\includegraphics[width=\linewidth]{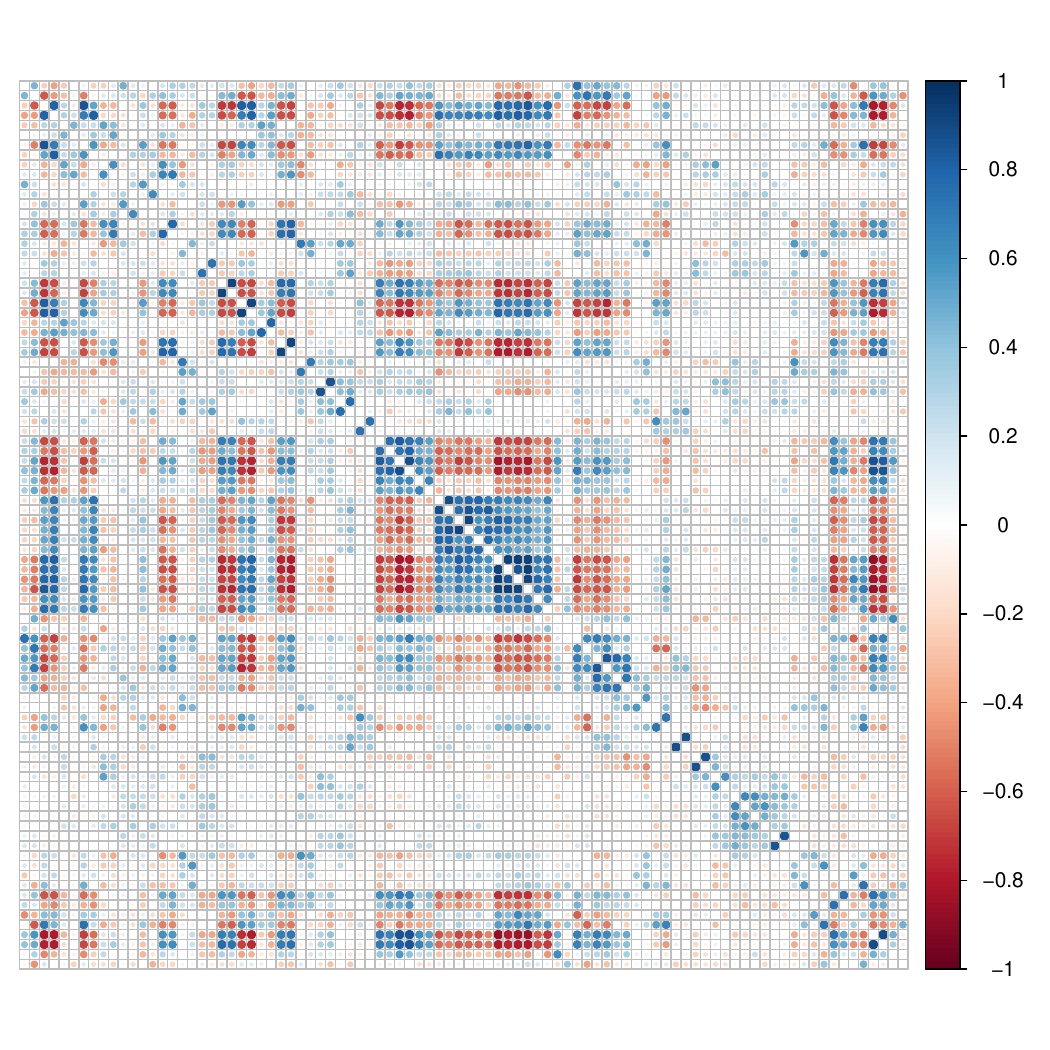}
	\end{subfigure}\hfil 
	\begin{subfigure}{0.25\textwidth}
		\includegraphics[width=\linewidth]{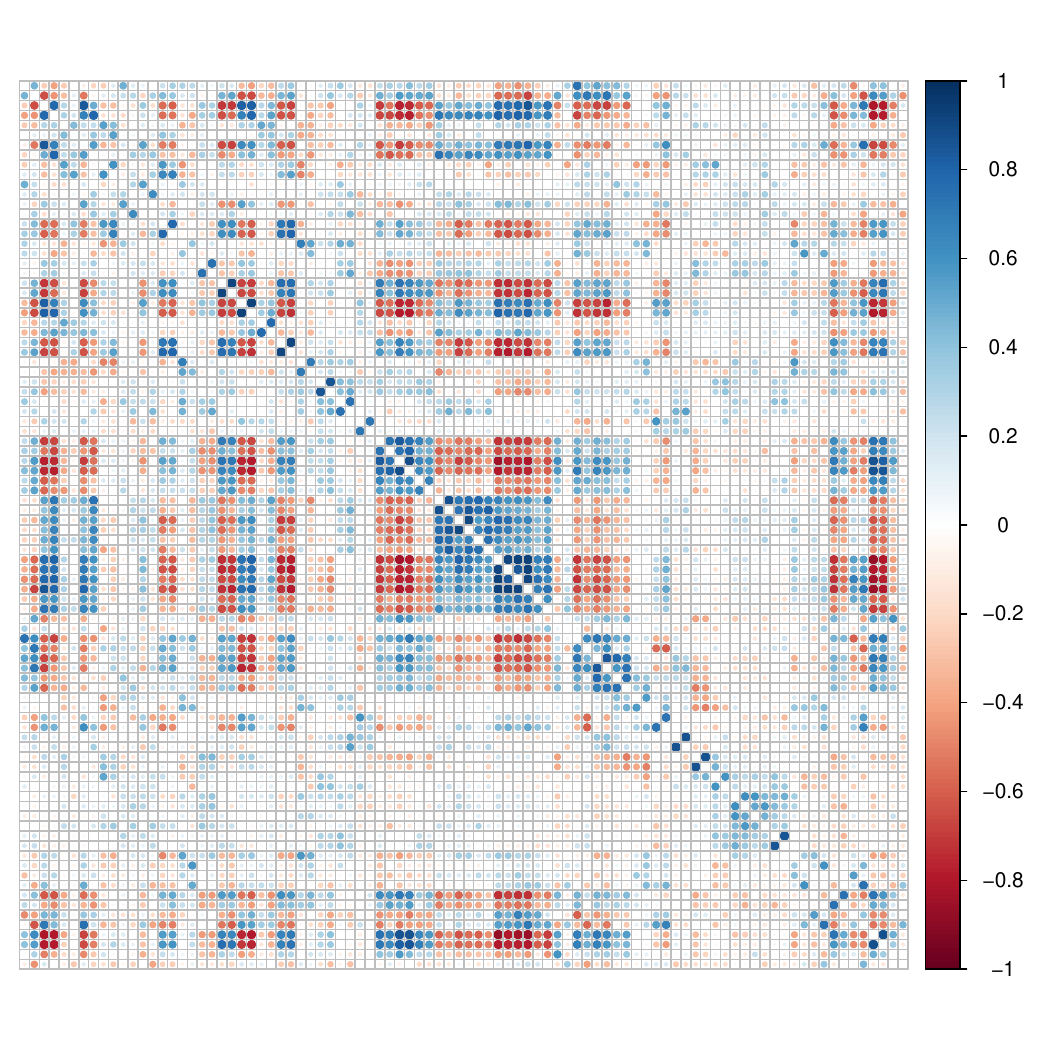}
	\end{subfigure}
	\centering
	\caption{Correlation plot of the predicted PCC matrices over varying levels of covariates at two different time points for MCI subjects. The top and bottom rows correspond to the predicted correlation matrices at times $0$ and $1$ respectively, while within each row the left, middle, and right panels depict the fits at the $10\%, 50\%$, and $90\%$  quantiles of the C-score with the other covariate age fixed at its mean level.  Positive (negative) values for correlations are drawn in red (blue). }
	\label{Fig:Data:adni:corrplot_over_score_MCI}
\end{figure}

First, the effect of the C-score for a fixed age is demonstrated for MCI subjects through correlation plots of the estimated PCC matrices. We fixed the age of the subjects at their mean level and fitted the model at varying levels of the C-score, namely, at the $10\%, 50\%$, and $90\%$  quantiles of the C-score.  Figure~\ref{Fig:Data:adni:corrplot_over_score_MCI} demonstrates the trend for the temporal correlations for varying predictor levels at different times of the study. The top and bottom rows correspond to the predicted correlation matrices (with the diagonals set to 0) at times $0$ and $1$ respectively, while within each row the left, middle, and right panels depict the fits at the $10\%, 50\%$, and $90\%$  quantiles of the C-score with the age fixed at its mean level. The overall correlation strengths decrease as  C-scores increase, reflecting the mean effects of the baseline covariates. Further, comparing the rows for each panel, we find overall weaker correlations at time $1$ compared with those at  time $0$.

\begin{figure}[!htb]
	\centering 
	\begin{subfigure}{0.25\textwidth}
		\includegraphics[width=\linewidth]{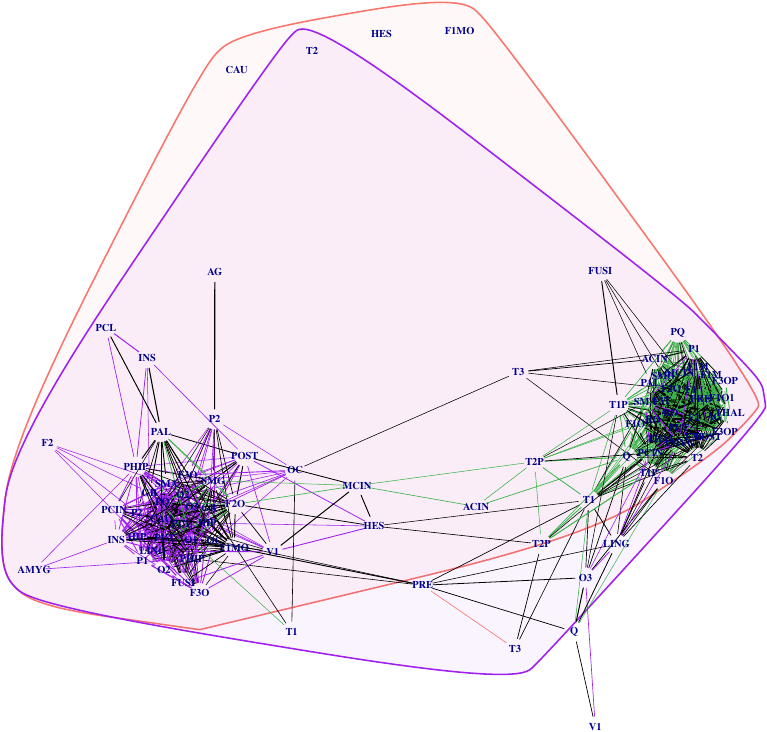}
		\caption{At time $t= 0$; $10\%$ quantile of the C-score, No. of communities = 7.}
	\end{subfigure}\hfil 
	\begin{subfigure}{0.25\textwidth}
		\includegraphics[width=\linewidth]{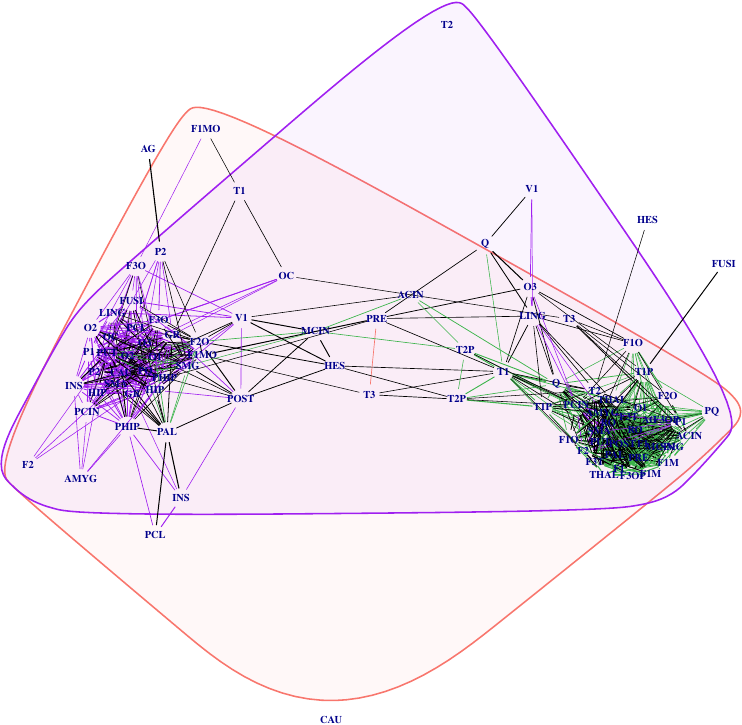}
		\caption{At time $t= 0$; $50\%$ quantile of the C-score, No. of communities = 6.}
	\end{subfigure}\hfil 
	\begin{subfigure}{0.25\textwidth}
		\includegraphics[width=\linewidth]{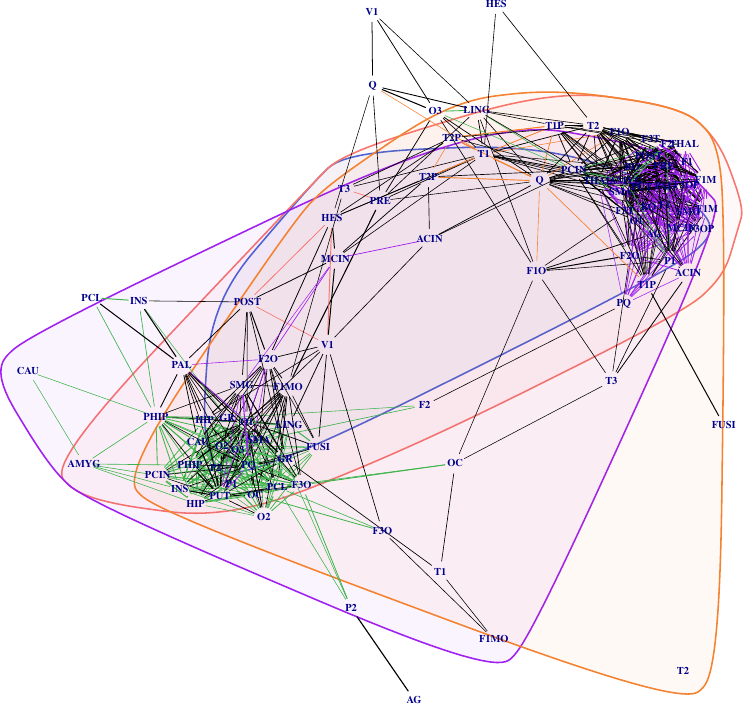}
		\caption{At time $t= 0$; $90\%$ quantile of the C-score, No. of communities = 7.}
	\end{subfigure}
	
	\medskip
	\begin{subfigure}{0.25\textwidth}
		\includegraphics[width=\linewidth]{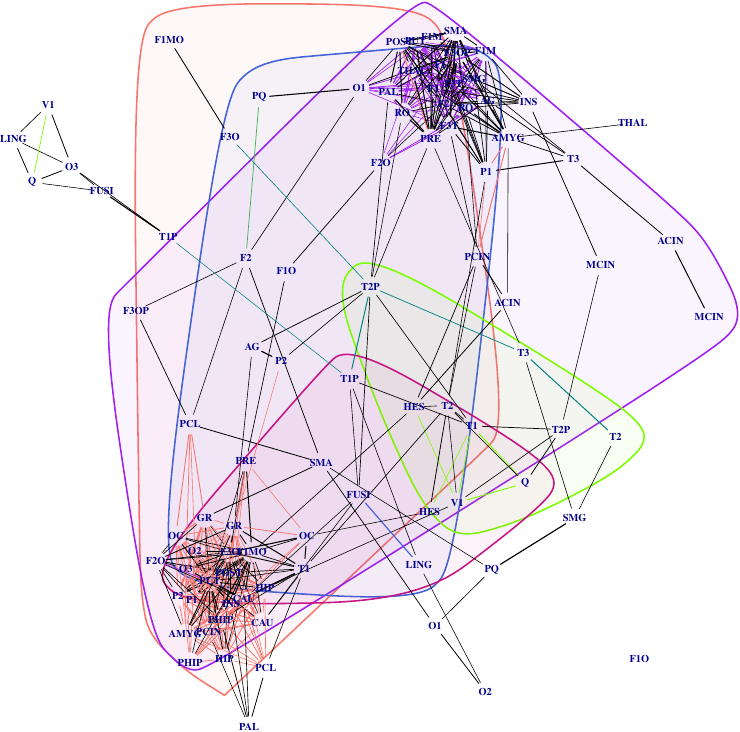}
		\caption{At time $t= 1$; $10\%$ quantile of the C-score, No. of communities = 13.}
	\end{subfigure}\hfil 
	\begin{subfigure}{0.25\textwidth}
		\includegraphics[width=\linewidth]{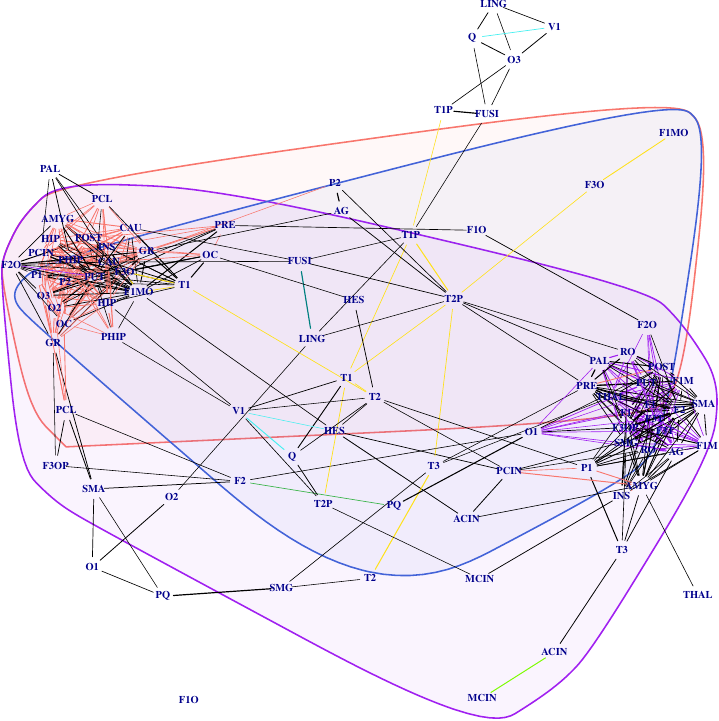}
		\caption{At time $t= 1$; $50\%$ quantile of the C-score, No. of communities = 12.}
	\end{subfigure}\hfil 
	\begin{subfigure}{0.25\textwidth}
		\includegraphics[width=\linewidth]{./plots/data_adni/network_commnities_CN_5_Tshifted2.pdf}
		\caption{At time $t= 1$; $90\%$ quantile of the C-score, No. of communities = 11.}
	\end{subfigure}
	\medskip
	\begin{subfigure}{\textwidth}
		\centering
		\includegraphics[width=.7\linewidth]{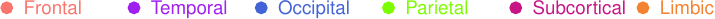}
	\end{subfigure}
	\centering
	\caption{Representation of the estimated PCC matrices at different levels of C-score as brain network, using spectral community detection method for the CN subjects. The top and bottom rows correspond to the predicted correlation matrices at times $0$ and $1$ respectively, while within each row the left, middle, and right panels depict the fits at the $10\%, 50\%$, and $90\%$  quantiles of the C-score with the other covariate age fixed at its mean level.  The communities comprising 10 or more ROIs are highlighted using colored polygons. These communities are found to be associated with different anatomical regions of the brain.}
	\label{Fig:Data:adni:network_comm_CN}
\end{figure}
\begin{figure}[!htb]
	\centering 
	\begin{subfigure}{0.25\textwidth}
		\includegraphics[width=\linewidth]{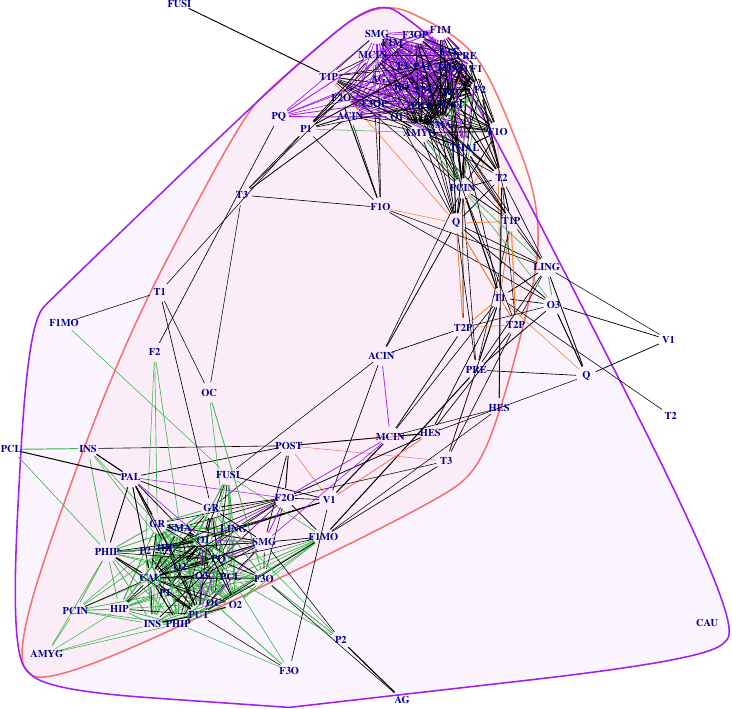}
		\caption{At time $t= 0$; $10\%$ quantile of the C-score, No. of communities = 7.}
	\end{subfigure}\hfil 
	\begin{subfigure}{0.25\textwidth}
		\includegraphics[width=\linewidth]{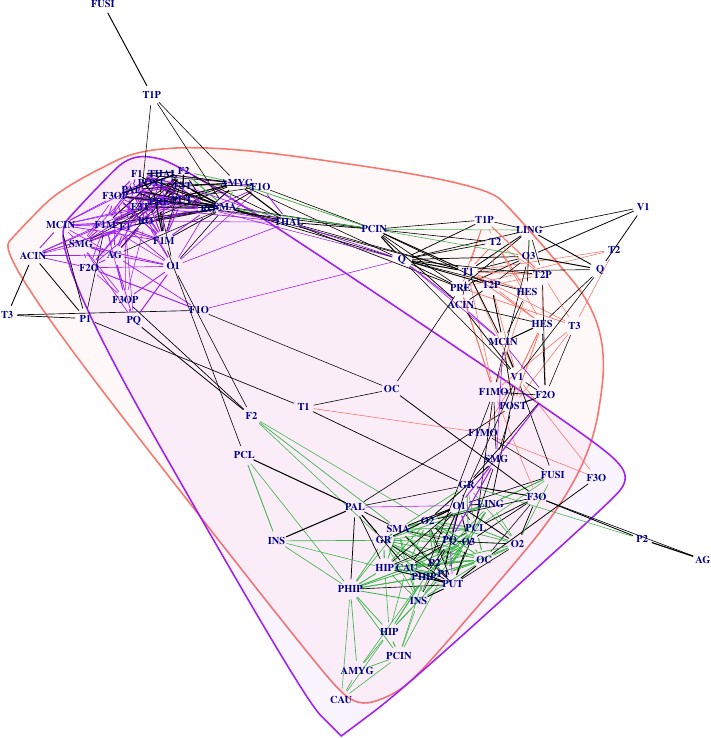}
		\caption{At time $t= 0$; $50\%$ quantile of the C-score, No. of communities = 3.}
	\end{subfigure}\hfil 
	\begin{subfigure}{0.25\textwidth}
		\includegraphics[width=\linewidth]{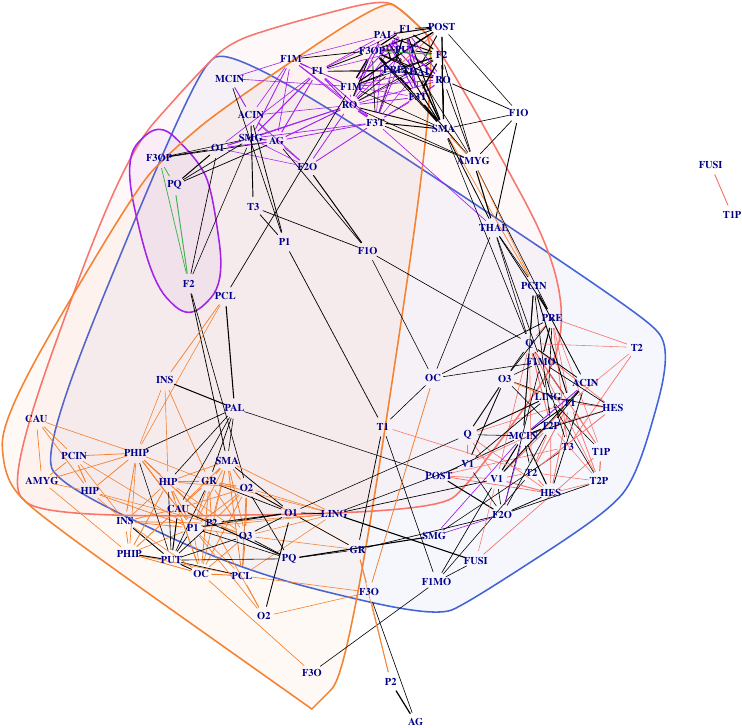}
		\caption{At time $t= 0$; $90\%$ quantile of the C-score, No. of communities = 7.}
	\end{subfigure}
	
	\medskip
	\begin{subfigure}{0.25\textwidth}
		\includegraphics[width=\linewidth]{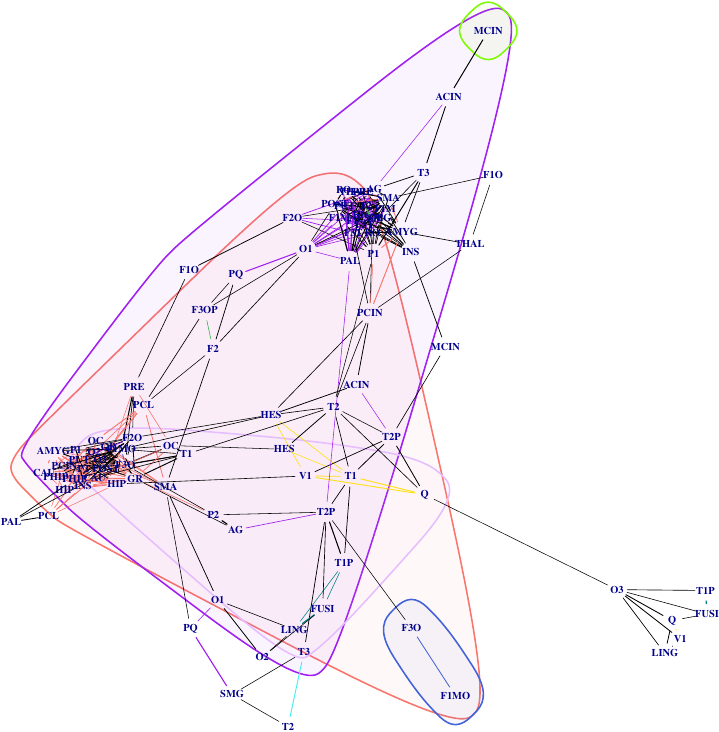}
		\caption{At time $t= 1$; $10\%$ quantile of the C-score, No. of communities = 11.}
	\end{subfigure}\hfil 
	\begin{subfigure}{0.25\textwidth}
		\includegraphics[width=\linewidth]{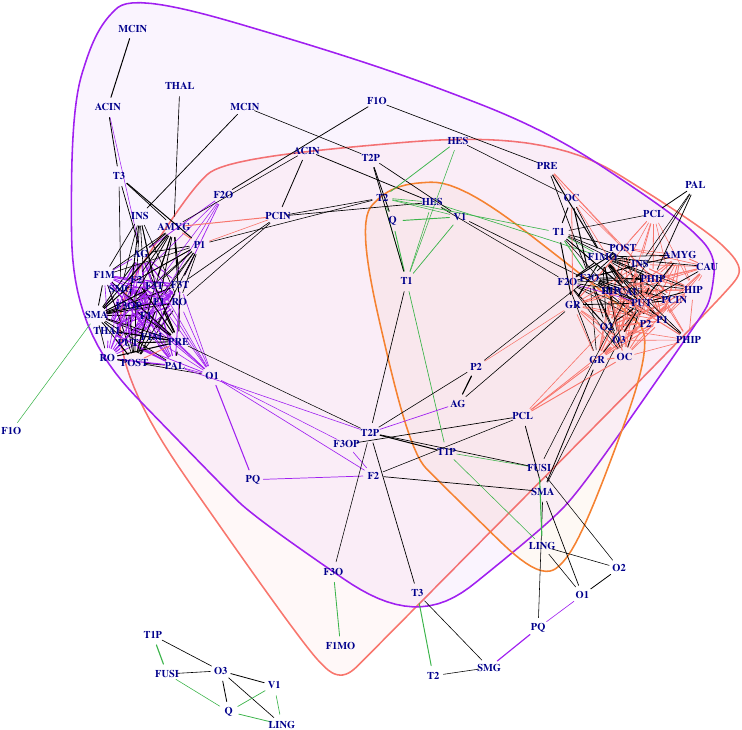}
		\caption{At time $t= 1$; $50\%$ quantile of the C-score, No. of communities = 4.}
	\end{subfigure}\hfil 
	\begin{subfigure}{0.25\textwidth}
		\includegraphics[width=\linewidth]{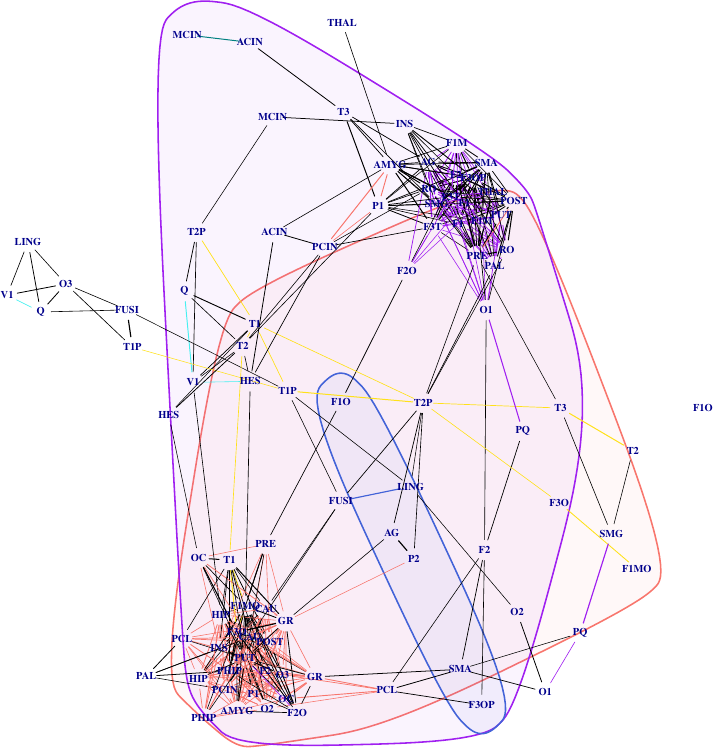}
		\caption{At time $t= 1$; $90\%$ quantile of the C-score, No. of communities = 11.}
	\end{subfigure}
	\medskip
	\begin{subfigure}{\textwidth}
		\centering
		\includegraphics[width=.7\linewidth]{./plots/data_adni/network_comm_legend_CN_Tshifted2.pdf}
	\end{subfigure}
	\centering
	\caption{Representation of the estimated PCC matrices at different levels of C-score as brain network, using spectral community detection method for the MCI subjects. The top and bottom rows correspond to the predicted correlation matrices at times $0$ and $1$ respectively, while within each row the left, middle, and right panels depict the fits at the $10\%, 50\%$, and $90\%$  quantiles of the C-score with the other covariate age fixed at its mean level.  The communities comprising 10 or more ROIs are highlighted using colored polygons. These communities are found to be associated with different anatomical regions of the brain.}
	\label{Fig:Data:adni:network_comm_MCI}
\end{figure}

We also converted  the predicted PCC matrices  into networks to better interpret and visualize  the brain structure. The predicted PCC matrices at varying levels of C-scores  and  for a fixed age were converted into weighted adjacency matrices and we explored the community detection methods for these network representations for both the CN and MCI subjects.
The predicted networks for the CN and MCI subjects are demonstrated in Figure~\ref{Fig:Data:adni:network_comm_CN} and~\ref{Fig:Data:adni:network_comm_MCI} respectively, where the nodes were placed using the Fruchterman-Reingold layout algorithm~\citep{fruc:91} for visualization. Spectral clustering~\citep{newm:06} is applied to detect the community structure in each network, where different communities are distinguished by different colors. The R package \emph{igraph} was used to find communities in graphs via directly optimizing a modularity score with a fast greedy algorithm.

The number of communities for the CN subjects at the $10\%, 50\%$, and $90\%$  quantiles of the C-score, where the other covariate age is fixed at its mean level, are $7, 6, 7$ corresponding to the fits at time $0$, respectively, and $13, 12, 11$, corresponding to the fits at time $1$, respectively. The number of communities found in the predicted networks for MCI subjects are $7,3,7$ and $11,4,11$, respectively, at time $0$ and time $1$. The communities with no less
than $10$ nodes are highlighted using colored polygons. These communities are found to be associated with different anatomical regions of the brain, where a community is identified as the anatomical region to which the majority of nodes belong. However, the communities found using the spectral clustering method overlap, especially for a higher value of the C-score, as the local interconnectivity and tendency to form a clique more locally increases. High cliquishness is known to be associated with reduced capability to rapidly combine specialized information from distributed brain regions, which may contribute to the cognitive decline of Alzheimer's subjects.

\begin{figure}[!htb]
	\centering 
	\includegraphics[width=.5\textwidth]{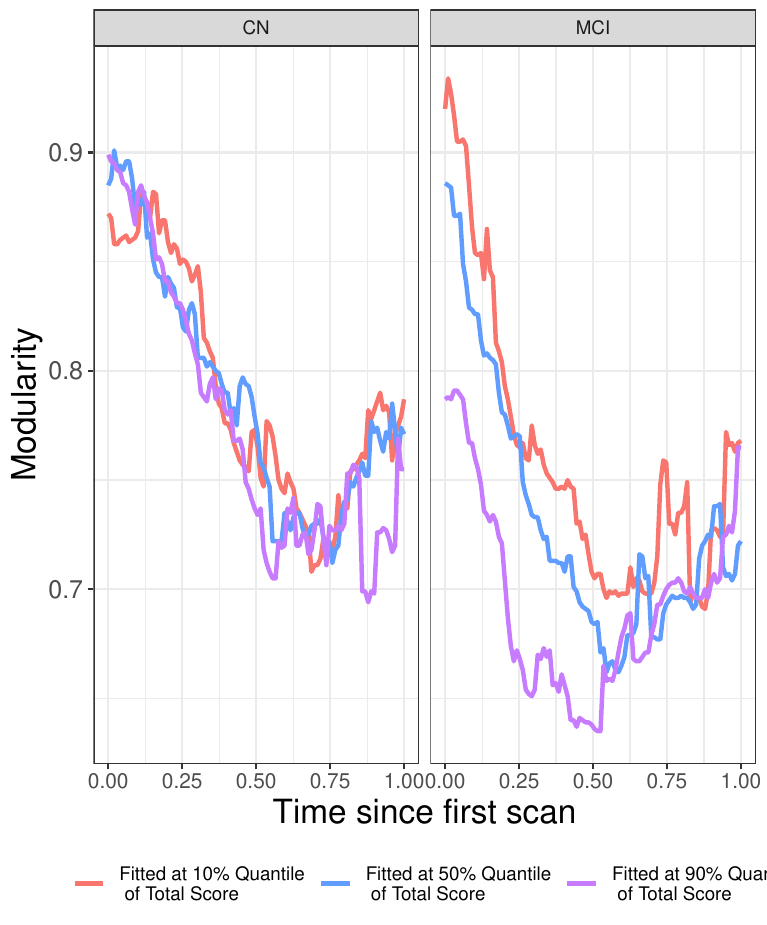}
	\centering
	\caption{Figure showing the global efficiency of the estimated brain network over time for the CN and MCI subjects (in the left and right panels, respectively). The covariate levels at which the networks are estimated are depicted in red, blue, and purple, respectively, corresponding to the $10\%, 50\%$, and $90\%$  quantiles of the C-score, with the other covariate age fixed at its mean level.}
	\label{Fig:Data:adni:globEff:CN_MCI}
\end{figure}

Finally, the global efficiency, a characteristic measure of network integration for the estimated networks evaluated at all points between time $0$ and $1$, on the underlying geodesic in the space of SPD matrices, for the CN and MCI subjects is demonstrated  in Figure~\ref{Fig:Data:adni:globEff:CN_MCI}. Global efficiency is a scaled measure of how many steps it takes when moving through the network from one node to another, where higher efficiency means that on average fewer steps are needed~\citep{alex:13, lato:01}.
In the left and right panels of Figure~\ref{Fig:Data:adni:globEff:CN_MCI}, the time-varying nature of the global efficiency of the estimated networks are illustrated for the CN and MCI subjects respectively. Each panel shows an overall decreasing trend with time. Further, for each panel, the estimated networks at the $10\%, 50\%$, and $90\%$  quantiles of the C-score are shown in red, blue, and purple, where the other covariate age is kept fixed at its mean level. The purple line is generally below the others, which suggests that higher  C-scores are associated with lower degrees of global efficiency, indicating less connectivity in the brain and a enhanced cognitive deficiency. The impairment over time looks more severe for the MCI subjects.

\end{document}